\documentclass[prd,preprint,tightenlines,floatfix,
preprintnumbers,nofootinbib,eqsecnum,superscriptaddress]{revtex4}

\usepackage[T1]{fontenc}		% fnot encoding with polish letters !or OT4

\usepackage{amsmath,amsfonts,amssymb,amstext,mathrsfs}
\usepackage{mathpazo}

\usepackage[dvips]{graphicx}
\usepackage{epsf,float}
\usepackage{revsymb}

\usepackage{dcolumn}% Align table columns on decimal point
\usepackage{braket}
\usepackage{color,xcolor}
\usepackage{graphicx}
\usepackage{subfigure}
\usepackage{multirow}
\usepackage{tabularx}
\usepackage{pstricks}
\usepackage[section]{placeins}
\usepackage{booktabs}
\usepackage{array}

\usepackage{hyperref}
\newcommand{\Pom}{\mathbb{P}}

\newcommand{\Reg}{\mathbb{R}}

\newcommand{\bdPt}{\mbox{\boldmath $dP_{t}$}}
\newcommand{\bqta}{\mbox{\boldmath $q_{t,1}$}}
\newcommand{\bqtb}{\mbox{\boldmath $q_{t,2}$}}
\newcommand{\bpta}{\mbox{\boldmath $p_{t,1}$}}
\newcommand{\bptb}{\mbox{\boldmath $p_{t,2}$}}

\begin{document}

\title{Towards a complete study of central exclusive production
of \boldmath{$K^{+}K^{-}$} pairs in proton-proton collisions
within the tensor Pomeron approach}

\vspace{0.6cm}

\author{Piotr Lebiedowicz}
 \email{Piotr.Lebiedowicz@ifj.edu.pl}
\affiliation{Institute of Nuclear Physics Polish Academy of Sciences, Radzikowskiego 152, PL-31-342 Krak\'ow, Poland}

\author{Otto Nachtmann}
 \email{O.Nachtmann@thphys.uni-heidelberg.de}
\affiliation{Institut f\"ur Theoretische Physik, Universit\"at Heidelberg,
Philosophenweg 16, D-69120 Heidelberg, Germany}

\author{Antoni Szczurek
\footnote{Also at \textit{Faculty of Mathematics and Natural Sciences, University of Rzesz\'ow, Pigonia 1, PL-35-310 Rzesz\'ow, Poland}.}}
\email{Antoni.Szczurek@ifj.edu.pl}
\affiliation{Institute of Nuclear Physics Polish Academy of Sciences, Radzikowskiego 152, PL-31-342 Krak\'ow, Poland}

\begin{abstract}
We present a study of the central exclusive production of the $K^{+} K^{-}$ pairs
in proton-proton collisions at high energies.
We consider diffractive mechanisms including
the $K^{+} K^{-}$ continuum, the dominant scalar $f_{0}(980)$, $f_{0}(1500)$, $f_{0}(1710)$
and tensor $f_{2}(1270)$, $f'_{2}(1525)$ resonances decaying
into the $K^{+} K^{-}$ pairs.
We include also photoproduction mechanisms
for the non-resonant (Drell-S\"{o}ding) and
the $\phi(1020)$ resonance contributions.
The theoretical results are calculated within 
the tensor-pomeron approach including both pomeron and reggeon exchanges.
Predictions for planned or current experiments 
at RHIC and LHC are presented.
We discuss the influence of the experimental cuts
on the integrated cross section and on 
various differential distributions for outgoing particles.
The distributions in two-kaon invariant mass, 
in a special ``glueball filter variable'',
as well as examples of angular distributions in the $K^{+}K^{-}$ rest frame
are presented.
We compare the $\phi(1020)$ and continuum photoproduction contributions
to the $f_{0}(980)$ and continuum diffractive contributions
and discuss whether the $\phi(1020)$ resonance could be extracted experimentally.
For the determination of some model parameters we also
include a discussion of $K$-nucleon scattering,
in particular total cross sections, and of $\phi(1020)$ photoproduction.
\end{abstract}

%\pacs{}

\maketitle

%-------------------------------------------------------------------
\section{Introduction}
%-------------------------------------------------------------------

Diffractive exclusive production of light mesons
mediated by double pomeron exchange
is expected to be an ideal process
for the investigation of gluonic bound states (glueballs)
due to the gluonic nature of the pomeron.
Such processes were studied extensively at CERN starting from 
the Intersecting Storage Rings (ISR) experiments
%in proton-proton collisions at $\sqrt{s} = 62$~GeV
\cite{Akesson:1983jz,Waldi:1983sc,Akesson:1985rn,Breakstone:1986xd,
Breakstone:1989ty,Breakstone:1990at},
later at the Super Proton Synchrotron (SPS) in fixed-target experiments
by the WA76 and WA102 collaborations
\cite{Armstrong:1989rv,Armstrong:1991ch,Barberis:1996iq,Barberis:1999am,French:1999tm,
Barberis:1999cq,Barberis:1999zh,Barberis:2000cd},
and more recently by the COMPASS collaboration 
\cite{Austregesilo:2013vky,Austregesilo:2016sss}.
For reviews of experimental results see for instance \cite{Kirk:2000ws, Kirk:2014nwa, Albrow:2010yb}.
The measurement of two charged pions in $p \bar{p}$ collisions %production 
was performed by the CDF collaboration at Tevatron \cite{Aaltonen:2015uva}.
Exclusive reactions are of particular interest
since they can be studied in current experiments at the LHC
by the ALICE, ATLAS, CMS \cite{Khachatryan:2017xsi}, and LHCb collaborations,
as well as by the STAR collaboration at RHIC \cite{Adamczyk:2014ofa,Sikora:2016evz}.
In such experiments it is of great advantage for the theoretical analysis
if the leading outgoing protons can be measured.
There are several efforts to complete installation of forward proton detectors.  
The CMS collaboration combines efforts with the TOTEM collaboration 
while the ATLAS collaboration may use the ALFA sub-detectors.
Also the STAR experiment at RHIC is equipped with detectors of similar type. 

On the theoretical side, the main contribution to
the central diffractive exclusive production at high energies
can be understood as being due to the exchange of two pomerons
between the external nucleons and the centrally produced hadronic system.
We believe that the soft pomeron exchange can be effectively treated 
as an effective rank-2 symmetric-tensor exchange as introduced in \cite{Ewerz:2013kda}.
In \cite{Ewerz:2016onn} it was shown that the tensor-pomeron model 
is consistent with the experimental data 
on the helicity structure of proton-proton elastic scattering at
$\sqrt{s} = 200$~GeV and small $|t|$ from the STAR experiment \cite{Adamczyk:2012kn}.
The paper \cite{Ewerz:2016onn} also contains some remarks on the history 
of the views of the pomeron spin structure.
In \cite{Lebiedowicz:2013ika} the central exclusive production 
of several scalar and pseudoscalar mesons 
in the reaction $p p \to p p M$ was studied for the relatively low WA102 energy.
Then, in \cite{Lebiedowicz:2016ioh}, the model was applied
to the reaction $p p \to p p \pi^+ \pi^-$ at high energies
including the $\pi^+ \pi^-$ continuum, the dominant scalar
$f_{0}(500)$, $f_{0}(980)$ and tensor $f_{2}(1270)$ resonances 
decaying into the $\pi^+ \pi^-$ pairs.
The resonant $\rho^0$ and non-resonant (Drell-S\"oding)
$\pi^{+}\pi^{-}$ photoproduction was studied in \cite{Lebiedowicz:2014bea}.
In \cite{Bolz:2014mya}, an extensive study of the reaction 
$\gamma p \to \pi^+ \pi^- p$ was presented.
The $\rho^{0}$ meson production associated with 
a very forward/backward $\pi N$ system
in the $pp \to pp \rho^{0} \pi^{0}$ and $pp \to pn \rho^{0} \pi^{+}$ processes
was discussed in \cite{Lebiedowicz:2016ryp}.
Also the central exclusive $\pi^+ \pi^-\pi^+ \pi^-$ production 
via the intermediate $\sigma\sigma$ and $\rho^0\rho^0$ states in $pp$ collisions
was studied in \cite{Lebiedowicz:2016zka}.
Recently, in \cite{Lebiedowicz:2018sdt}, the central exclusive production of 
the $p \bar{p}$ in the continuum and via scalar resonances in $pp$ collisions was studied.

Some time ago two of us considered the exclusive $pp \to pp K^{+} K^{-}$ reaction
in a simple Regge-like model \cite{Lebiedowicz:2011tp}.
The Born approximation is usually not sufficient and 
absorption corrections have to be taken into account,
see e.g. \cite{Harland-Lang:2013dia,Lebiedowicz:2015eka}.
In \cite{Lebiedowicz:2011tp} the production of the diffractive $K^{+} K^{-}$ continuum 
and of the scalar $\chi_{c0}$ meson decaying
via $\chi_{c0} \to K^{+} K^{-}$ was studied.
For other related works see \cite{Lebiedowicz:2009pj}
for the $pp \to pp \pi^{+} \pi^{-}$ reaction,
\cite{Szczurek:2009yk} for the exclusive $f_{0}(1500)$,
and \cite{Lebiedowicz:2011nb} for $\chi_{c0}$ meson production.

In \cite{Fiore:2017xnx} a model for the exclusive diffractive meson production 
in $pp$ collisions was discussed
based on the convolution of the Donnachie-Landshoff parametrization 
of the pomeron distribution in the proton 
with the pomeron-pomeron-meson total cross section.
In this approach the cross section is calculated 
by summing over the direct-channel contributions 
from the pomeron and two different $f_{1}$ and $f_{2}$ trajectories
associated to the glueball candidate $f_{0}(980)$ and the $f_{2}(1270)$ resonances, respectively.
Also the $f_{0}(500)$ resonance contribution dominating the small mass region
and a slowly increasing background were taken into account.
%In such a model, the strength of the resonance contributions needs to be extracted 
%from experimental data.
The absolute contribution of resonances, e.g. of the $f_{0}(980)$ and the $f_{2}(1270)$,
to the total cross section cannot be derived within this approach, 
and must hence be deduced from experimental data.
But the relative weights of the various resonances on one trajectory
are correlated by the duality argument made in \cite{Fiore:2017xnx}.

The aim of the study presented here is 
%the formulation of a model for simulating such differential distributions.
%In the following we extend 
the application of the tensor-pomeron model 
to central exclusive production of $K^{+}K^{-}$ pairs in $pp$ collisions.
We wish to show first predictions in the tensor-pomeron approach
for the production of the diffractive $K^{+}K^{-}$ continuum, 
of the scalar $f_{0}(980)$, $f_{0}(1500)$, $f_{0}(1710)$,
and the tensor $f_{2}(1270)$, $f'_{2}(1525)$ resonances 
decaying into $K^+ K^-$ pairs.
This model, being formulated at the amplitude level,
allows us also to calculate interference effects
of the various contributions.
In the following we wish to show differential distributions
which can be helpful in the investigation of scalar and tensor resonance parameters.
Therefore, we shall treat each resonance in its own right and shall not a priori suppose
any correlations of the coupling parameters of different resonances.
%by a Partial Wave Analysis (PWA).}
In addition the resonant $\phi(1020)$ and non-resonant (Drell-S\"oding) 
$K^+ K^-$ photoproduction mechanisms will be discussed.
So far the cross sections for the exclusive $pp \to pp \phi(1020)$ reaction
were calculated within a pQCD $k_{t}$-factorization approach \cite{Cisek:2010jk},
and in a color dipole approach \cite{Santos:2014vwa,Goncalves:2017wgg}.

%--------------------------------------
\section{Exclusive $K^{+}K^{-}$ production}
\label{sec:excl_prod}
%--------------------------------------

We study central exclusive production of $K^+ K^-$ 
in proton-proton collisions at high energies
\begin{eqnarray}
p(p_{a},\lambda_{a}) + p(p_{b},\lambda_{b}) \to
p(p_{1},\lambda_{1}) + K^{+}(p_{3}) + K^{-}(p_{4}) + p(p_{2},\lambda_{2}) \,,
\label{2to4_reaction}
\end{eqnarray}
where $p_{a,b}$, $p_{1,2}$ and $\lambda_{a,b}$, 
$\lambda_{1,2} \in \lbrace +1/2, -1/2 \rbrace$,
indicated in brackets,
denote the four-momenta and helicities of the protons, 
and $p_{3,4}$ denote the four-momenta of the charged kaons, respectively.

\begin{figure}
\includegraphics[width=7.cm]{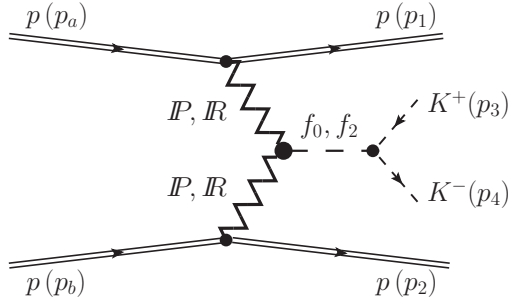}      
  \caption{\label{fig:resonant_Born}
  \small The Born diagram for double-pomeron/reggeon
  central exclusive scalar and tensor resonances production 
  and their subsequent decays into $K^+ K^-$ in proton-proton collisions.
}
\end{figure}

The full amplitude of $K^{+} K^{-}$ production 
is a sum of the continuum amplitude 
and the amplitudes with the $s$-channel resonances:
\begin{equation}
\begin{split}
{\cal M}_{pp \to pp K^{+} K^{-}} =
{\cal M}^{KK{\rm-continuum}}_{pp \to pp K^{+} K^{-}} + 
{\cal M}^{KK{\rm-resonances}}_{pp \to pp K^{+} K^{-}}\,.
\end{split}
\label{amplitude}
\end{equation}
The amplitude for exclusive resonant $K^+ K^-$ production
via the pomeron-pomeron fusion,
shown by the diagram of Fig.~\ref{fig:resonant_Born},
can be written as
\begin{equation}
\begin{split}
{\cal M}^{KK{\rm-resonances}}_{pp \to pp K^{+} K^{-}}
= {\cal M}^{(\Pom \Pom \to f_{0} \to K^{+}K^{-})}_{pp \to pp K^{+} K^{-}}
 + {\cal M}^{(\Pom \Pom \to f_{2} \to K^{+}K^{-})}_{pp \to pp K^{+} K^{-}}\,.
\end{split}
\label{amplitude_f0f2_pomTpomT}
\end{equation}
As indicated in Fig.~\ref{fig:resonant_Born} also contributions 
involving non-leading reggeons $\Reg$: 
$\rho_{\Reg}$ ($\rho$~reggeon), 
$\omega_{\Reg}$ ($\omega$~reggeon), 
$f_{2 \Reg}$ ($f_{2}$~reggeon),
$a_{2 \Reg}$ ($a_{2}$~reggeon) can contribute.
The relevant production modes via $(C_{1},C_{2})$ fusion
\footnote{Here $C_{1}$ and $C_{2}$ are the charge-conjugation
quantum numbers of the exchange objects and $C_{1},C_{2} \in \{+1,-1 \}$.}
giving resonances
are listed in Table~II of \cite{Lebiedowicz:2016ioh}.
However, in the present paper we shall consider only resonance production
by pomeron-pomeron fusion in order not to be swamped by too many,
essentially unknown, coupling parameters.

Turning now to continuum diffractive $K^{+} K^{-}$ production shown in
Fig.~\ref{fig:diagrams_Born} we have again pomeron and reggeon contributions.
Here we will be able to extract all relevant coupling parameters
from the kaon-nucleon total cross section data.
Therefore, we shall include in the calculation pomeron and reggeon exchanges.
In this way we will also get an estimate of the possible importance of the latter exchanges.
In the following we treat the $C=+1$ pomeron 
and the reggeons $\Reg_{+} = f_{2 \Reg}, a_{2 \Reg}$ 
as effective tensor exchanges
while the $C=-1$ reggeons 
$\Reg_{-} = \omega_{\Reg}, \rho_{\Reg}$ are treated as effective vector exchanges.
%------------------------------------------------------------------
%-------------------------------------------------------------------

In Table~\ref{table:table} we have listed intermediate resonances
that contribute to the $pp \to pp K^{+} K^{-}$ 
and/or $pp \to pp \pi^{+} \pi^{-}$ reactions.
%~~~~~~~~~~~~~~~~~~~~~~~~~~~~~~~~~~~~~~~~~~~~~~~~~~~~~~~~~~~~~~~~~~~~~
\begin{table}[!h]
\caption{A list of resonances, up to a mass of 1800~MeV,
that decay into $K^{+} K^{-}$ and/or $\pi^{+} \pi^{-}$.
The meson masses, their total widths $\Gamma$ and branching fractions
are taken from PDG~\cite{Olive:2016xmw}.
}
\begin{tabular}{|l|l|l|l|l|l|l|}
\hline
Meson		&$I^{G}J^{PC}$& $m$ (MeV) 	& $\Gamma$ (MeV) & $\Gamma_{K \overline{K}}/\Gamma$	& $\Gamma_{\pi \pi}/\Gamma$ & Other decay modes\\ \hline
$f_0(500)$	&$0^{+}0^{++}$&	$400 - 550$ 		& $400 - 700$	& - & dominant  & $\gamma \gamma$ \\
$\rho(770)$&$1^{+}1^{--}$&	$769.0 \pm 1.0$ 	& $151.7 \pm 2.6$ 	& - & $\frac{\Gamma_{\pi^{+}\pi^{-}}}{\Gamma}$ = 1 &  \\
$f_0(980)$	&$0^{+}0^{++}$&	$990 \pm 20$ 		& $10 - 100$	& seen & dominant  & $\gamma \gamma$ \\
$a_0(980)$	&$1^{-}0^{++}$&	$980 \pm 20$        & $50 - 100$	& seen  & - & $\eta \pi$, $\gamma \gamma$ \\ 
$\phi(1020)$&$0^{-}1^{--}$&	$1019.460 \pm 0.016$ 	& $4.247 \pm 0.016$ 	&  $\frac{\Gamma_{K^{+}K^{-}}}{\Gamma}$ = $0.489 \pm 0.005$  & - & 
$K^{0}_{L}K^{0}_{S}$, $3 \pi$, $\eta \gamma$\\
$f_{2}(1270)$&$0^{+}2^{++}$&	$1275.5 \pm 0.8$ 	& $186.7^{+2.2}_{-2.5}$ 	& $0.046^{+0.005}_{-0.004}$ & $0.842^{+0.029}_{-0.009}$ & $4 \pi$, $\gamma \gamma$ \\
$a_{2}(1320)$&$1^{-}2^{++}$&	$1318.1 \pm 0.7$ & $109.8 \pm 2.4$ 	& $0.049 \pm 0.008$ & - & $3 \pi$, $\eta \pi$, $\omega \pi \pi$, $\gamma \gamma$ \\ 
$f_{0}(1370)$&$0^{+}0^{++}$&	$1200 - 1500$ & $200 - 500$ 	 & seen & seen & $4 \pi$ ($\rho \rho$), $\eta \eta$, $\gamma \gamma$ \\
$a_{0}(1450)$&$1^{-}0^{++}$&	$1474 \pm 19$ & $265 \pm 13$ 	 & $0.082 \pm 0.028$ & - & $\pi \eta$, $\pi \eta'(958)$, $\gamma \gamma$ \\ 
$f_{0}(1500)$&$0^{+}0^{++}$&	$1504 \pm 6$ & $109 \pm 7$ 	 & $0.086 \pm 0.010$ & 
$0.349 \pm 0.023$ & $4 \pi$, $\eta \eta$, $\eta \eta'(958)$ \\ 
$f'_{2}(1525)$&$0^{+}2^{++}$&	$1525 \pm 5$ & $73^{+6}_{-5}$ 	 & $0.887 \pm 0.022$ & 
$(8.2 \pm 1.5) \times 10^{-3}$ & $\eta \eta$, $\gamma \gamma$ \\
$f_{2}(1640)$&$0^{+}2^{++}$&	$1639 \pm 6$ & $99^{+60}_{-40}$ 	 & seen & - & $4 \pi$, $\omega \omega$ \\ 
$\phi(1680)$&$0^{-}1^{--}$&	$1680 \pm 20$ & $150 \pm 50$ 	 & seen & - & $K \overline{K}^{*}(892)$ \\ 
$\rho_{3}(1690)$&$1^{+}3^{--}$&	$1696 \pm 4$ & $204 \pm 18$ 	 & $0.0158 \pm 0.0026$ & 
$0.236 \pm 0.013$ & $4 \pi$, $K \overline{K} \pi$ \\
$\rho(1700)$&$1^{+}1^{--}$&	$1740.8 \pm 22.2$ & $187.2 \pm 26.7$ 	 & seen & seen & 
$4 \pi$ ($\rho \pi \pi$)\\
$a_{2}(1700)$&$1^{-}2^{++}$&	$1732 \pm 16$ & $194 \pm 40$ 	 & seen & - & $\eta \pi$ \\
$f_{0}(1710)$&$0^{+}0^{++}$&	$1723^{+6}_{-5}$ & $139 \pm 8$ 	 & seen & seen & $\eta \eta$, $\omega \omega$ \\ \hline
\end{tabular}
\label{table:table}
\end{table}
%~~~~~~~~~~~~~~~~~~~~~~~~~~~~~~~~~~~~~~~~~~~~~~~~~~~~~~~~~~~~~~~~~~~~~

%--------------------------------------
\section{Diffractive contributions}
\label{sec:diff_mech}
%--------------------------------------

%--------------------------------------
\subsection{$K^{+}K^{-}$ continuum central production}
\label{sec:diff_continuum}
%--------------------------------------

%------------------------------------------------------------------
\begin{figure}
\includegraphics[width=5.5cm]{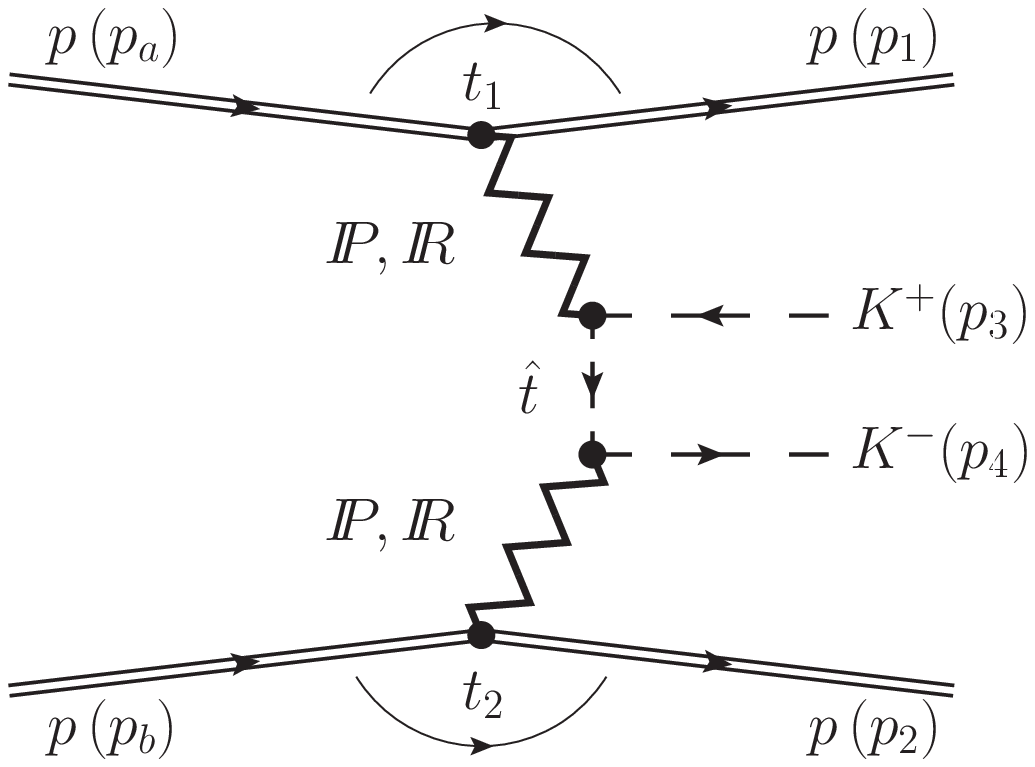}
\includegraphics[width=5.5cm]{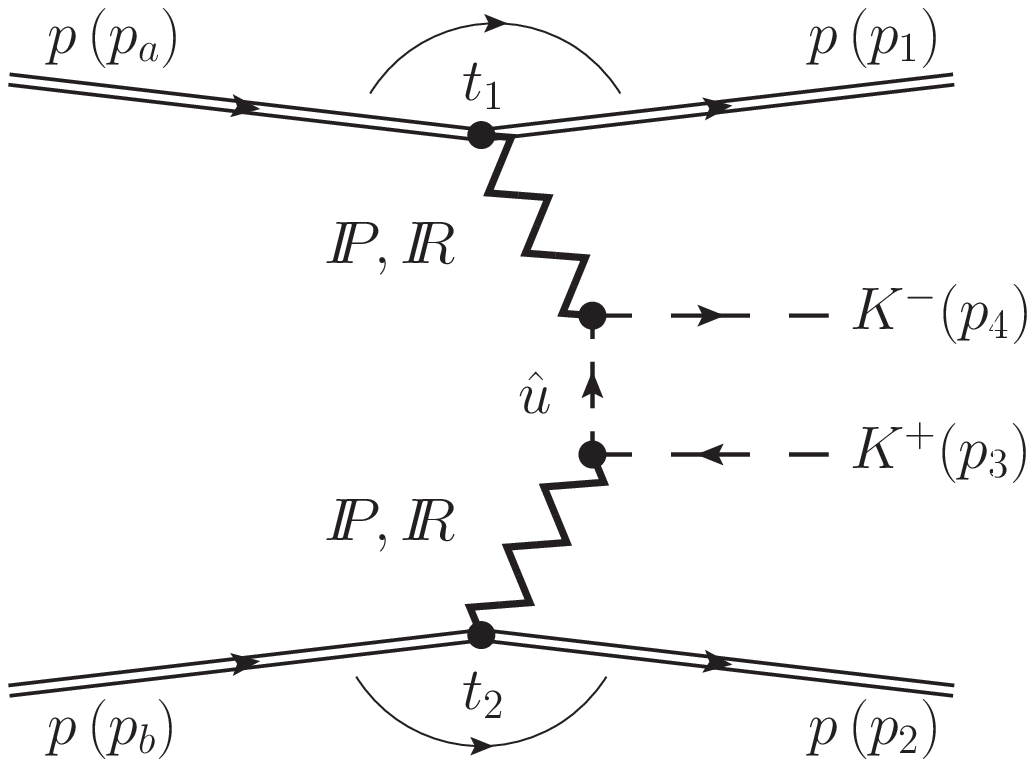}      
  \caption{\label{fig:diagrams_Born}
  \small The Born diagrams for double-pomeron/reggeon
  central exclusive $K^+ K^-$ continuum production in proton-proton collisions.
}
\end{figure}
%-------------------------------------------------------------------

The generic diagrams for diffractive exclusive $K^+ K^-$ continuum production
are shown in Fig.~\ref{fig:diagrams_Born}.
At high energies the exchange objects to be considered are
the pomeron $\Pom$ and the reggeons $\Reg$.
%Their charge conjugation and $G$-parity quantum numbers
%are listed in Table~I of \cite{Lebiedowicz:2016ioh}.
The amplitude can be written as the following sum:
\begin{eqnarray}
{\cal M}^{KK{\rm-continuum}}_{pp \to pp K^{+} K^{-}} &=&
{\cal M}^{(\Pom \Pom \to K^{+}K^{-})} +
{\cal M}^{(\Pom \Reg \to K^{+}K^{-})} +
{\cal M}^{(\Reg \Pom \to K^{+}K^{-})} +
{\cal M}^{(\Reg \Reg \to K^{+}K^{-})}. \; \qquad
\label{cont_diff}
\end{eqnarray}

The $\Pom\Pom$-exchange amplitude on the Born level 
can be written as the sum:
\begin{eqnarray}
{\cal M}^{(\Pom \Pom \to K^{+}K^{-})} =
{\cal M}^{({\rm \hat{t}})}_{\lambda_{a} \lambda_{b} \to \lambda_{1} \lambda_{2} K^{+}K^{-}}+
{\cal M}^{({\rm \hat{u}})}_{\lambda_{a} \lambda_{b} \to \lambda_{1} \lambda_{2} K^{+}K^{-}}
\,,
\label{pompom_amp}
\end{eqnarray}
where
\begin{equation}
\begin{split}
& {\cal M}^{({\rm \hat{t}})}_{\lambda_{a} \lambda_{b} \to \lambda_{1} \lambda_{2} \pi^{+}\pi^{-}} 
= \\
& \quad (-i)
\bar{u}(p_{1}, \lambda_{1}) 
i\Gamma^{(\Pom pp)}_{\mu_{1} \nu_{1}}(p_{1},p_{a}) 
u(p_{a}, \lambda_{a})\,
i\Delta^{(\Pom)\, \mu_{1} \nu_{1}, \alpha_{1} \beta_{1}}(s_{13},t_{1}) \,
i\Gamma^{(\Pom KK)}_{\alpha_{1} \beta_{1}}(p_{t},-p_{3}) \,
i\Delta^{(K)}(p_{t}) \\
& \quad \times  i\Gamma^{(\Pom KK)}_{\alpha_{2} \beta_{2}}(p_{4},p_{t})\,
i\Delta^{(\Pom)\, \alpha_{2} \beta_{2}, \mu_{2} \nu_{2}}(s_{24},t_{2}) \,
\bar{u}(p_{2}, \lambda_{2}) 
i\Gamma^{(\Pom pp)}_{\mu_{2} \nu_{2}}(p_{2},p_{b}) 
u(p_{b}, \lambda_{b}) \,,
\end{split}
\label{amplitude_t}
\end{equation}
\begin{equation} 
\begin{split}
& {\cal M}^{({\rm \hat{u}})}_{\lambda_{a} \lambda_{b} \to \lambda_{1} \lambda_{2} \pi^{+}\pi^{-}} 
= \\ 
& \quad (-i)\,
\bar{u}(p_{1}, \lambda_{1}) 
i\Gamma^{(\Pom pp)}_{\mu_{1} \nu_{1}}(p_{1},p_{a}) 
u(p_{a}, \lambda_{a}) \,
i\Delta^{(\Pom)\, \mu_{1} \nu_{1}, \alpha_{1} \beta_{1}}(s_{14},t_{1}) \,
i\Gamma^{(\Pom KK)}_{\alpha_{1} \beta_{1}}(p_{4},p_{u}) 
\,i\Delta^{(K)}(p_{u})  \\
& \quad \times  
i\Gamma^{(\Pom KK)}_{\alpha_{2} \beta_{2}}(p_{u},-p_{3})\, 
i\Delta^{(\Pom)\, \alpha_{2} \beta_{2}, \mu_{2} \nu_{2}}(s_{23},t_{2}) \,
\bar{u}(p_{2}, \lambda_{2}) 
i\Gamma^{(\Pom pp)}_{\mu_{2} \nu_{2}}(p_{2},p_{b}) 
u(p_{b}, \lambda_{b}) \,.
\end{split}
\label{amplitude_u}
\end{equation}
Here $p_{t} = p_{a} - p_{1} - p_{3}$ and 
$p_{u} = p_{4} - p_{a} + p_{1}$, $s_{ij} = (p_{i} + p_{j})^{2}$.
The normal kaon propagator is $i\Delta^{(K)}(k) = i/(k^{2}-m_{K}^{2})$.
Furthermore $\Delta^{(\Pom)}$ and $\Gamma^{(\Pom pp)}$ 
denote the effective propagator and proton vertex function, respectively, 
for the tensorial pomeron.
%For the explicit expressions, see section~3 of \cite{Ewerz:2013kda}.
%In a similar way the $\Pom f_{2 \Reg}$, $f_{2 \Reg} \Pom$ 
%and $f_{2 \Reg} f_{2 \Reg}$ amplitudes can be written.
The propagator of the tensor-pomeron exchange is written as
(see Eq.~(3.10) of \cite{Ewerz:2013kda}):
\begin{eqnarray}
i \Delta^{(\Pom)}_{\mu \nu, \kappa \lambda}(s,t) = 
\frac{1}{4s} \left( g_{\mu \kappa} g_{\nu \lambda} 
                  + g_{\mu \lambda} g_{\nu \kappa}
                  - \frac{1}{2} g_{\mu \nu} g_{\kappa \lambda} \right)
(-i s \alpha'_{\Pom})^{\alpha_{\Pom}(t)-1}
\label{pomeron_propagator}
\end{eqnarray}
and fulfils the following relations
\begin{equation} 
\begin{split}
&\Delta^{(\Pom)}_{\mu \nu, \kappa \lambda}(s,t) = 
\Delta^{(\Pom)}_{\nu \mu, \kappa \lambda}(s,t) =
\Delta^{(\Pom)}_{\mu \nu, \lambda \kappa}(s,t) =
\Delta^{(\Pom)}_{\kappa \lambda, \mu \nu}(s,t) \,, \\
&g^{\mu \nu} \Delta^{(\Pom)}_{\mu \nu, \kappa \lambda}(s,t) = 0, \quad 
g^{\kappa \lambda} \Delta^{(\Pom)}_{\mu \nu, \kappa \lambda}(s,t) = 0 \,.
\end{split}
\label{pomeron_propagator_aux}
\end{equation}
%
%For the $f_{2 \Reg}$ reggeon exchange a similar form of the effective propagator 
%and the $f_{2 \Reg}pp$ and $f_{2 \Reg} \pi \pi$ effective vertices is assumed, 
%see (3.12) and (3.49), (3.53) of \cite{Ewerz:2013kda}.
Here the pomeron trajectory $\alpha_{\Pom}(t)$
is assumed to be of standard linear form, see e.g. \cite{Donnachie:1992ny,Donnachie:2002en},
\begin{eqnarray}
\alpha_{\Pom}(t) = \alpha_{\Pom}(0)+\alpha'_{\Pom}\,t, \quad 
\alpha_{\Pom}(0) = 1.0808,\quad \alpha'_{\Pom} = 0.25 \; \mathrm{GeV}^{-2}\,.
%\\
%&&\alpha_{\Reg}(t) = \alpha_{\Reg}(0)+\alpha'_{\Reg}\,t, \quad 
%\alpha_{\Reg}(0) = 0.5475,\quad \alpha'_{\Reg} = 0.9 \; \mathrm{GeV}^{-2}\,.
\label{trajectory}
\end{eqnarray}
The pomeron-proton vertex function is written as
%is supplemented by a vertex form-factor, 
%taken here to be the Dirac electromagnetic form factor of the proton for simplicity
(see Eq.~(3.43) of \cite{Ewerz:2013kda})
\begin{eqnarray}
&&i\Gamma_{\mu \nu}^{(\Pom pp)}(p',p)= 
i\Gamma_{\mu \nu}^{(\Pom \bar{p} \bar{p})}(p',p)
\nonumber\\
&& \qquad =-i 3 \beta_{\Pom NN} F_{1}\bigl((p'-p)^2\bigr)
\left\lbrace 
\frac{1}{2} 
\left[ \gamma_{\mu}(p'+p)_{\nu} 
     + \gamma_{\nu}(p'+p)_{\mu} \right]
- \frac{1}{4} g_{\mu \nu} ( p\!\!\!/' + p\!\!\!/ )
\right\rbrace , \qquad \;\;\;
%\nonumber\\
\label{vertex_pomNN}
\end{eqnarray}
where 
%$p\!\!\!/ = \gamma_{\kappa} p^{\kappa}$ and 
$\beta_{\Pom NN} = 1.87$~GeV$^{-1}$.
%$\Delta^{(\Pom)}$ denotes the effective (spin-2) pomeron propagator.
The $\Pom KK$ vertices in the amplitudes (\ref{amplitude_t}) and (\ref{amplitude_u})
can be written in analogy to the $\Pom \pi \pi$ vertices (see (3.45) of \cite{Ewerz:2013kda}) 
but with the replacement $\beta_{\Pom \pi \pi} \to \beta_{\Pom KK}$,
%and (B.69) of \cite{Bolz:2014mya},
%
\begin{eqnarray}
i\Gamma_{\mu \nu}^{(\Pom KK)}(k',k)=
-i 2 \beta_{\Pom KK} 
\left[ (k'+k)_{\mu}(k'+k)_{\nu} - \frac{1}{4} g_{\mu \nu} (k' + k)^{2} \right] \, F_{M}((k'-k)^2)\,.
\label{vertex_pomKK}
\end{eqnarray}
The form factors, 
%$F^{(i)}_{\pi N}(t)$ ($i = \Pom, \Reg$) 
taking into account that the hadrons are extended objects,
%(see section~3.2 of \cite{Donnachie:2002en})
are chosen as 
%
%\begin{equation}
%\begin{split}
%F^{(i)}_{\pi N}(t) = F_1(t) F_M(t)\,,
%\end{split}
%\label{nachtmann_ff}
%\end{equation}
%
%where 
%
\begin{eqnarray}
F_{1}(t)= \frac{4 m_{p}^{2}-2.79\,t}{(4 m_{p}^{2}-t)(1-t/m_{D}^{2})^{2}}\,, \qquad
F_{M}(t)= \frac{1}{1-t/\Lambda_{0}^{2}}\,,
\label{Fpion}
\end{eqnarray}
where $m_{p}$ is the proton mass and $m_{D}^{2} = 0.71$~GeV$^{2}$
is the dipole mass squared and $\Lambda_{0}^{2} = 0.5$~GeV$^{2}$; 
see Eqs.~(3.29) and (3.34) of \cite{Ewerz:2013kda}, respectively.

The off-shellness of the intermediate kaons
is taken into account by the inclusion of form factors. 
%is unknown in particular 
%at higher values of $p_{t}^{2}$ or $p_{u}^{2}$.
The form factors are normalized to unity at
the on-shell point $\hat{F}_{K}(m_{K}^{2}) = 1$
and parametrised here in the monopole form
\begin{eqnarray} 
%\hat{F}_{K}(p^{2})=
%\exp\left(\frac{p^{2}-m_{K}^{2}}{\Lambda^{2}_{off,E}}\right) \,,
%\label{off-shell_form_factors_exp}
\hat{F}_{K}(k^{2})=
\dfrac{\Lambda^{2}_{off,M} - m_{K}^{2}}{\Lambda^{2}_{off,M} - k^{2}} \,,
\label{off-shell_form_factors_mon} 
%&&F_{M^{*}}(\hat{t}/\hat{u})=
%\exp\left(- \sqrt{-(\hat{t}/\hat{u}-m_{M^{*}}^{2})} / \Lambda_{off}\right) \,.
%\label{off-shell_form_factors_other}
\end{eqnarray}
where $\Lambda_{off,M}$ could be adjusted to experimental data.
We take $\Lambda_{off,M} = 0.7$~GeV,
that is, the same value as for the pion off-shell form factor
in the reaction $pp \to pp \pi^{+} \pi^{-}$ discussed in \cite{Lebiedowicz:2016ioh}.
In \cite{Lebiedowicz:2016ioh} we fixed a parameter of the form factor for off-shell pion
and a few parameters of the pomeron-pomeron-meson coupling constants 
to describe the CDF data \cite{Aaltonen:2015uva};
see Fig.~9 of \cite{Lebiedowicz:2016ioh}.
%It was shown in Fig.~9 of \cite{Lebiedowicz:2015eka} that
%the monopole form (\ref{off-shell_form_factors_mon})
%is supported by the preliminary CDF results \cite{Albrow_Project_new}
%particularly at higher values of two-pion invariant mass, $M_{\pi\pi} > 1.5$~GeV.
%Thus, in the numerical calculations below,
%see section~\ref{sec:section_4},
%we used the monopole form of the off-shell pion form factors.

In our calculations we include both the tensor-pomeron and the reggeon $\Reg_{+}$ and $\Reg_{-}$ exchanges. 
In the following we collect the expressions for reggeon effective propagators and vertex functions
in order to make our present paper self contained.
For extensive discussions motivating the following expressions we refer to 
section~3 of \cite{Ewerz:2013kda}.

The ansatz for the $C=+1$ reggeons $\Reg_{+} = f_{2 \Reg}, a_{2 \Reg}$
is similar to (\ref{pomeron_propagator}) - (\ref{vertex_pomNN}).
The $\Reg_{+}$ propagator is obtained from (\ref{pomeron_propagator}) with the replacements
\begin{eqnarray}
&&\alpha_{\Pom}(t) \to \alpha_{\Reg_{+}}(t) = \alpha_{\Reg_{+}}(0)+\alpha'_{\Reg_{+}}\,t\,, \nonumber \\
&&\alpha_{\Reg_{+}}(0) = 0.5475\,, \nonumber \\
&&\alpha'_{\Reg_{+}} = 0.9 \; \mathrm{GeV}^{-2}\,.
\label{A5}
\end{eqnarray}
In (\ref{A5}) and in the following the parameters of the reggeon trajectories 
are taken from \cite{Donnachie:2002en}.
The $f_{2 \Reg}$- and $a_{2 \Reg}$-proton vertex functions are obtained from (\ref{vertex_pomNN})
with the replacements
% ($M_{0} = 1$~GeV)
%
\begin{eqnarray}
&&3 \beta_{\Pom NN} \to \frac{g_{f_{2 \Reg} pp}}{M_{0}}\,, \nonumber \\
&&g_{f_{2 \Reg} pp} = 11.04\,,
\label{A6}
\end{eqnarray}
and
\begin{eqnarray}
&&3 \beta_{\Pom NN} \to \frac{g_{a_{2 \Reg} pp}}{M_{0}}\,, \nonumber \\
&&g_{a_{2 \Reg} pp} = 1.68\,,
\label{A7}
\end{eqnarray}
respectively.
In (\ref{A6}), (\ref{A7}) and in the following $M_{0} = 1$~GeV 
is used in various places for dimensional reasons.
The $f_{2 \Reg}$- and $a_{2 \Reg}$-kaon vertex functions are obtained from (\ref{vertex_pomKK})
with the replacements
\begin{eqnarray}
&&2 \beta_{\Pom KK} \to \frac{g_{f_{2 \Reg} KK}}{2 M_{0}}\,,
\label{A7a}\\
&&2 \beta_{\Pom KK} \to \frac{g_{a_{2 \Reg} KK}}{2 M_{0}}\,,
\label{A7b}
\end{eqnarray}
respectively.
For the $C=-1$ reggeons $\Reg_{-} = \omega_{\Reg}, \rho_{\Reg}$
we assume an effective vector propagator
(see Eqs.~(3.14) - (3.15) of \cite{Ewerz:2013kda})
\begin{eqnarray}
i \Delta^{(\Reg_{-})}_{\mu \nu}(s,t) = 
i g_{\mu \nu} \frac{1}{M_{-}^{2}} (-i s \alpha'_{\Reg_{-}})^{\alpha_{\Reg_{-}}(t)-1}\,,
\label{A8}
\end{eqnarray}
with
\begin{eqnarray}
&&\alpha_{\Reg_{-}}(t) = \alpha_{\Reg_{-}}(0)+\alpha'_{\Reg_{-}}\,t\,, \nonumber \\
&&\alpha_{\Reg_{-}}(0) = 0.5475\,, \nonumber \\
&&\alpha'_{\Reg_{-}} = 0.9 \; \mathrm{GeV}^{-2}\,,
\label{A9}\\
&&M_{-} = 1.41 \; \mathrm{GeV}\,.
\label{A9a}
\end{eqnarray}
The value of (\ref{A9a}) is taken from \cite{Ewerz:2013kda}
as default value for the parameter of the propagators 
for $\omega_{\Reg}$ and $\rho_{\Reg}$ exchanges.

For the $\Reg_{-}$-proton vertices we have
(see Eqs.~(3.59) - (3.62) of \cite{Ewerz:2013kda})
\begin{eqnarray}
i\Gamma_{\mu}^{(\omega_{\Reg} pp)}(p',p)
&=& i\Gamma_{\mu}^{(\omega_{\Reg} nn)}(p',p)
= -i\Gamma_{\mu}^{(\omega_{\Reg} \bar{p} \bar{p})}(p',p) \nonumber\\
&=& -i g_{\omega_{\Reg} pp} F_{1}\bigl((p'-p)^2\bigr) \gamma_{\mu}\,,
\label{A10a} \\
i\Gamma_{\mu}^{(\rho_{\Reg} pp)}(p',p)
&=& -i\Gamma_{\mu}^{(\rho_{\Reg} nn)}(p',p)
= -i\Gamma_{\mu}^{(\rho_{\Reg} \bar{p} \bar{p})}(p',p) \nonumber\\
&=& -i g_{\rho_{\Reg} pp} F_{1}\bigl((p'-p)^2\bigr) \gamma_{\mu}\,,
\label{A10b}
\end{eqnarray}
with
\begin{eqnarray}
&&g_{\omega_{\Reg} pp} = 8.65\,, \nonumber \\
&&g_{\rho_{\Reg} pp} = 2.02\,,
\label{A11}
\end{eqnarray}
respectively.
Note that in (\ref{A10b}) the vertex function for the isospin 1 $\rho_{\Reg}$ reggeon changes sign
when we replace protons by neutrons.
This is also the case for the isospin 1 $a_{2\Reg}$ reggeon exchange; see (3.51) of \cite{Ewerz:2013kda}.
The $\Reg_{-}$-kaon vertex ($\Reg_{-} = \omega_{\Reg}, \rho_{\Reg}$) 
can be written in analogy to the $\rho_{\Reg}$-pion vertex 
(see (3.63) of \cite{Ewerz:2013kda})
\begin{eqnarray}
i\Gamma_{\mu}^{(\Reg_{-} K^{+}K^{+})}(k',k)
&=& -i\Gamma_{\mu}^{(\Reg_{-} K^{-}K^{-})}(k',k) \nonumber\\
&=& -\frac{i}{2} g_{\Reg_{-} KK} F_{M}\bigl((k'-k)^2\bigr) (k'+k)_{\mu}\,.
\label{A12}
\end{eqnarray}

To obtain the pomeron/reggeon-kaon coupling constants 
we consider the following elastic scattering processes at high energies
\begin{eqnarray}
%&&\pi^\pm (p_1) + p(p_2,\lambda_2) \to \pi^\pm (p'_1) + p(p'_2,\lambda'_2) 
%\label{pip_pip_reaction} \\
&&K^\pm (p_1) + p(p_2,\lambda_2) \to K^\pm (p_3) + p(p_4,\lambda_4) \,,
\label{Kp_Kp_reaction} \\
&&K^\pm (p_1) + n(p_2,\lambda_2) \to K^\pm (p_3) + n(p_4,\lambda_4) \,.
\label{Kn_Kn_reaction}
\end{eqnarray}
We treat (\ref{Kp_Kp_reaction}) and (\ref{Kn_Kn_reaction}) in analogy 
to the elastic $\pi^{\pm} p$ scattering; see section~7 of \cite{Ewerz:2013kda}.
For the case of the elastic kaon-nucleon scattering amplitudes
we set for $p$ and $n$ also $N(I_{3})$ with $I_{3} = +1/2$ and $I_{3} = -1/2$, respectively.
We obtain
\begin{equation}
\begin{split}
\langle K^\pm(p_3), N(I_{3},p_4,\lambda_4) | {\cal T} &| K^\pm(p_1), N(I_{3},p_2,\lambda_2) \rangle
\\
=i\, 2 s\, \delta_{\lambda_{4} \lambda_{2}} F_{1}(t) F_{M}(t)
\bigg\{ &\,
6 \beta_{{\Pom}KK} \beta_{{\Pom}NN}
(- i s \alpha_{\Pom}' )^{\alpha_{\Pom}(t) -1} %\,{F}^{(\Pom)}_{K N}(t)
\\
&
+
\frac{1}{2} \left[ g_{f_{2\Reg}KK}\, g_{f_{2\Reg}pp} 
                 + (-1)^{I_{3}-\frac{1}{2}}\, g_{a_{2\Reg}KK}\, g_{a_{2\Reg}pp} \right]
M_0^{-2}
(- i s \alpha_{{\Reg}_+}')^{\alpha_{{\Reg}_+}(t) -1} %\,{F}^{(\Reg)}_{K N}(t)
\\
&
\pm
\frac{i}{2} \left[ g_{\omega_{\Reg}KK}\, g_{\omega_{\Reg}pp}
                 + (-1)^{I_{3}-\frac{1}{2}}\,g_{\rho_{\Reg}KK}\, g_{\rho_{\Reg}pp} \right]
M_-^{-2} 
(- i s \alpha_{{\Reg}_-}')^{\alpha_{{\Reg}_-}(t) -1} %\,{F}^{(\Reg)}_{K N}(t)
\bigg\}\,.
\end{split}
\label{Kp_Kp_el}
\end{equation}
Here we have $s = (p_1 + p_2)^2$ and $t = (p_1 - p_3)^2$
and we work in the approximation $s \gg |t|$, $m_{p}^{2}$.
%{\color{blue}Note that in (\ref{Kp_Kp_el}) under $p \leftrightarrow n$
%odd-isospin $\rho_{\Reg}$ and $a_{2\Reg}$ change sign of contribution.}

For the total cross sections we obtain from the optical theorem for large $s$
\begin{equation}
\begin{split}
\sigma_{\rm tot}& (K^\pm, N(I_{3})) =
\frac{1}{2s} 
\sum_{\lambda_2}
\text{Im}\,
\langle K^\pm(p_1), N(I_{3},p_2,\lambda_2) | {\cal T} | K^\pm(p_1), N(I_{3},p_2,\lambda_2) \rangle 
\\
= &\,
2 \bigg\{
6\beta_{{\Pom}KK}\,\beta_{{\Pom}NN} \cos\left[\frac{\pi}{2}(\alpha_{\Pom}(0) -1 )\right]
(s \alpha_{\Pom}' )^{\alpha_{\Pom}(0) -1}
\\
& \hspace*{.5cm}
+ 
\frac{1}{2} \left[ g_{f_{2 \Reg}KK}\, g_{f_{2 \Reg}pp} 
                 + (-1)^{I_{3}-\frac{1}{2}}\, g_{a_{2\Reg}KK}\, g_{a_{2\Reg}pp} \right]
M_0^{-2} \cos\left[\frac{\pi}{2}(\alpha_{{\Reg}_+}(0) -1 )\right]
(s \alpha_{{\Reg}_+}')^{\alpha_{{\Reg}_+}(0) -1} 
\\
&\hspace*{.5cm}
\mp
\frac{1}{2} \left[ g_{\omega_{\Reg}KK}\, g_{\omega_{\Reg}pp}
                 + (-1)^{I_{3}-\frac{1}{2}}\,g_{\rho_{\Reg}KK}\, g_{\rho_{\Reg}pp} \right]
M_-^{-2} \cos\left[\frac{\pi}{2}\alpha_{{\Reg}_-}(0)\right]
(s \alpha_{{\Reg}_-}')^{\alpha_{{\Reg}_-}(0) -1}
\bigg\}.
\end{split}
\label{Kp_Kp_tot}
\end{equation}

Following Donnachie and Landshoff \cite{Donnachie:1992ny} 
we use a two component parametrisation for the total cross sections
of kaon-nucleon scattering
\begin{eqnarray}
\sigma_{\rm tot}(a,b) = X_{ab} \left(s\,M_{0}^{-2}\right)^{0.0808}
                       +Y_{ab} \left(s\,M_{0}^{-2}\right)^{-0.4525} \,.
\label{Kp_tot_fit}
\end{eqnarray}
Here $(a,b)$ = $(K^{+}, p)$, $(K^{-}, p)$, $(K^{+}, n)$, $(K^{-}, n)$, and $M_{0} = 1$~GeV.
The numbers $X_{ab} \equiv X$ and $Y_{ab}$ are 
\begin{equation}
\begin{split}
&X = 11.93\;{\rm mb} \;\widehat{=}\; 30.64\;{\rm GeV}^{-2} \,,\\
&Y_{K^{+}p} = 7.58\;{\rm mb} \;\widehat{=}\; 19.47\;{\rm GeV}^{-2}\,,\quad
Y_{K^{-}p}= 25.33\;{\rm mb} \;\widehat{=}\; 65.05\;{\rm GeV}^{-2}\,,\\ 
&Y_{K^{+}n}=  9.08\;{\rm mb} \;\widehat{=}\; 23.32\;{\rm GeV}^{-2}\,,\quad
Y_{K^{-}n}= 19.09\;{\rm mb} \;\widehat{=}\; 49.03\;{\rm GeV}^{-2} \,,
\end{split}
\label{Kp_tot_parameters}
\end{equation}
where the values for the $X$, $Y_{K^{+} p}$ and 
$Y_{K^{-} p}$ are taken from Fig.~3.2 of \cite{Donnachie:2002en}
and the values for the $Y_{K^{+} n}$ and $Y_{K^{-} n}$ 
are from our fit to the world data from \cite{Olive:2016xmw}.

We compare now (\ref{Kp_Kp_tot}) with (\ref{Kp_tot_fit}) taking into account
the parameters of the pomeron and reggeon trajectories
and of their vertices from \cite{Ewerz:2013kda}
%and take the parameters of the pomeron and reggeon trajectories
%from \cite{Ewerz:2013kda}
%and of the nucleon vertices 
quoted above in Eqs.~(\ref{pomeron_propagator}) to (\ref{A11}).
We get then the following results for the couplings
\begin{eqnarray}
&&\beta_{\Pom KK} = 1.54\;{\rm GeV}^{-1}\,, 
\label{Kp_couplings_pom}\\
&&g_{f_{2\Reg} KK} = 4.47\,, \,
g_{a_{2\Reg} KK} = 2.28\,, \,
g_{\omega_{\Reg} KK} = 5.99\,, \,
g_{\rho_{\Reg} KK} = 7.15\,.
\label{Kp_couplings_reg}
\end{eqnarray}
%

%--------------------------------------
\subsection{Scalar mesons central production}
\label{sec:diff_f0}
%--------------------------------------

The $K^{+} K^{-}$ production amplitude through the $s$-channel exchange of scalar mesons,
such as $f_{0}(980)$, $f_{0}(1370)$, $f_{0}(1500)$, and $f_{0}(1710)$, 
via the $\Pom \Pom$ fusion can be written as
\begin{equation}
\begin{split}
{\cal M}^{(\Pom \Pom \to f_{0} \to K^{+}K^{-})}_{\lambda_{a} \lambda_{b} \to \lambda_{1} \lambda_{2} K^{+}K^{-}} 
= & (-i)\,
\bar{u}(p_{1}, \lambda_{1}) 
i\Gamma^{(\Pom pp)}_{\mu_{1} \nu_{1}}(p_{1},p_{a}) 
u(p_{a}, \lambda_{a})\;
i\Delta^{(\Pom)\, \mu_{1} \nu_{1}, \alpha_{1} \beta_{1}}(s_{1},t_{1}) \\
& \times 
i\Gamma^{(\Pom \Pom f_{0})}_{\alpha_{1} \beta_{1},\alpha_{2} \beta_{2}}(q_{1},q_{2}) \;
i\Delta^{(f_{0})}(p_{34})\;
i\Gamma^{(f_{0} KK)}(p_{3},p_{4})\\
& \times 
i\Delta^{(\Pom)\, \alpha_{2} \beta_{2}, \mu_{2} \nu_{2}}(s_{2},t_{2}) \;
\bar{u}(p_{2}, \lambda_{2}) 
i\Gamma^{(\Pom pp)}_{\mu_{2} \nu_{2}}(p_{2},p_{b}) 
u(p_{b}, \lambda_{b}) \,,
\end{split}
\label{amplitude_f0_pomTpomT}
\end{equation}
where
$s_{1} = (p_{a} + q_{2})^{2} = (p_{1} + p_{34})^{2}$,
$s_{2} = (p_{b} + q_{1})^{2} = (p_{2} + p_{34})^{2}$, and
$p_{34} = p_{3} + p_{4}$.
The effective Lagrangians and the vertices 
for the fusion of two tensor pomerons into the $f_{0}$ meson 
were discussed in appendix~A of \cite{Lebiedowicz:2013ika}.
The $\Pom \Pom f_{0}$ vertex, including a form factor, 
reads as follows ($p_{34} = q_{1} + q_{2}$)
\begin{eqnarray}
i\Gamma_{\mu \nu,\kappa \lambda}^{(\Pom \Pom f_{0})} (q_{1},q_{2}) =
\left( i\Gamma_{\mu \nu,\kappa \lambda}'^{(\Pom \Pom f_{0})}\mid_{bare} +
       i\Gamma_{\mu \nu,\kappa \lambda}''^{(\Pom \Pom f_{0})} (q_{1}, q_{2})\mid_{bare} \right)
%i \, g_{\Pom \Pom M} \, M_{0} \,  %F^{M}_{\Pom \Pom M}(t_{1},t_{2})
%\left( g_{\mu \kappa} g_{\nu \lambda} + g_{\mu \lambda} g_{\nu \kappa}
%-\frac{1}{2} g_{\mu \nu} g_{\kappa \lambda} \right) 
\tilde{F}^{(\Pom \Pom f_{0})}(q_{1}^{2},q_{2}^{2},p_{34}^{2}) \,;
\label{vertex_pompomS}
\end{eqnarray}
see (A.21) of \cite{Lebiedowicz:2013ika}.
The vertex (\ref{vertex_pompomS}) contains
two independent $\Pom \Pom f_{0}$ couplings
corresponding to the lowest allowed values of $(l, S)$, 
that is $(l, S) = (0,0)$ and $(2,2)$.

We take the factorized form for the $\Pom \Pom f_{0}$ form factor
\begin{eqnarray}
\tilde{F}^{(\Pom \Pom f_{0})}(q_{1}^{2},q_{2}^{2},p_{34}^{2}) = 
F_{M}(q_{1}^{2}) F_{M}(q_{2}^{2}) F^{(\Pom \Pom f_{0})}(p_{34}^{2})\,
\label{Fpompommeson}
\end{eqnarray}
normalised to $\tilde{F}^{(\Pom \Pom f_{0})}(0,0,m_{f_{0}}^{2}) = 1$.
In practical calculations we take
\begin{eqnarray}
F^{(\Pom \Pom f_{0})}(p_{34}^{2}) = 
\exp{ \left( \frac{-(p_{34}^{2}-m_{f_{0}}^{2})^{2}}{\Lambda_{f_{0}}^{4}} \right)}\,,
\quad \Lambda_{f_{0}} = 1\;{\rm GeV}\,.
\label{Fpompommeson_ff}
\end{eqnarray}

There has been a long history of uncertainty about 
the properties of the $f_{0}(1710)$ meson,
one of the earliest glueball candidates.
%In addition to the well known $f_{2}(1270)/a_{2}(1320)$ and $f'_{2}(1525)$ mesons
%The WA76 collaboration found evidence for a further tensor state
%in the region around 1.7~GeV \cite{Armstrong:1989rv},
%so called $\theta/f_{J}(1720)$ state.
This state was observed in the WA76 experiment 
at $\sqrt{s} = 23.8$~GeV \cite{Armstrong:1989rv}
in both the $K^{+}K^{-}$ and $K_{S}^{0}K_{S}^{0}$ channels
in the dikaon invariant mass region around 1.7~GeV.
%shows that it must have $J^{PC} = (even)^{++}$
By studying the $K^{+}K^{-}$ angular distributions
%in the Gottfried-Jackson frame 
the authors of \cite{Armstrong:1989rv}
found that the so called $\theta/f_{J}(1720)$ state has $J^{PC} = 2^{++}$.
In \cite{French:1999tm} a reanalysis of the $K^{+}K^{-}$ channel 
from the WA76 experiment was performed.
A partial wave analysis of the centrally produced $K^{+}K^{-}$ system,
as performed in \cite{French:1999tm} (see Fig.~4 there),
shows in the $S$-wave a threshold enhancement and a structure
in the 1.5 - 1.7~GeV mass interval which has been interpreted as 
being due to the $f_{0}(1500)$ and $f_{J}(1710)$ with $J=0$.
The $D$-wave shows peaks in the 1.3~GeV and 1.5~GeV mass regions,
presumably due to the $f_{2}(1270)/a_{2}(1320)$ and $f'_{2}(1525)$ resonances.
In the $D$-wave at higher masses
there is no evidence for any significant structure
in the 1.7~GeV mass region and only a wide structure around 2.2~GeV 
is seen that may be due to the $f_{2}(2150)$ meson. 
In the $P$-wave ($P_{1}^{-}$) a peak corresponding to the $\phi(1020)$ is observed.
These results are compatible with those
coming from WA102 experiment \cite{Barberis:1999am} at $\sqrt{s} = 29$~GeV.
The $f_{J}(1710)$ with $J =2$ state has been observed 
also in radiative $J/\psi$ decays \cite{Bai:1996dc}.
However, a new analysis of $J/\psi \to \gamma K^{+} K^{-}$
and $\gamma K_{S}^{0} K_{S}^{0}$ \cite{Bai:2003ww} strongly
demonstrates that the mass region around 1.7~GeV
is predominantly $0^{++}$ from the $f_{0}(1710)$.
\footnote{It is mentioned in \cite{Bai:2003ww} that
the amount of the possible $2^{++}$ component
in the 1.7~GeV mass region is of the order of a few percent.}
This conclusion is consistent with the latest central production data 
of WA76 and WA102 \cite{French:1999tm,Barberis:1999am,Barberis:1999cq}.

An important variable characterising the production mechanisms
of the various $f_{0}$ mesons is the azimuthal angle $\phi_{pp}$
between the outgoing protons, $p (p_{1})$ and $p (p_{2})$ in (\ref{2to4_reaction}).
As can be seen from the experimental results presented in
\cite{Barberis:1999cq,Barberis:1999zh,Kirk:2000ws} 
for the $f_{0}(980)$, $f_{0}(1500)$, and $f_{0}(1710)$ states
the cross sections peak at $\phi_{pp} = 0$
in contrast to the $f_{0}(1370)$ meson.
%The preference of the $f_{0}(1370)$ meson for
%the azimuthal angle $\phi_{pp} \approx \pi$ domain in contrast to the more enigmatic
%$f_{0}(980)$ and $f_{0}(1500)$ scalars has been observed by the WA102 Collaboration.
It was shown in \cite{Lebiedowicz:2013ika} that the appropriate angular shapes 
for the central production of $f_{0}(980)$ and $f_{0}(1500)$ mesons 
could be obtained with the $\Pom \Pom f_{0}$ vertices 
corresponding to the sum of the two lowest values of $(l,S)$ couplings,
$(l,S) = (0,0)$ and $(2,2)$,
with appropriate coupling constants $g_{\Pom \Pom M}'$ and $g_{\Pom \Pom M}''$.
For the production of $f_{0}(1370)$ meson the $(l,S) = (0,0)$ coupling alone
already describes the azimuthal angular correlation reasonably well.
In \cite{Lebiedowicz:2013ika} we determined
the corresponding (dimensionless) $\Pom \Pom f_{0}$ coupling constants 
by approximately fitting the theoretical results to the WA102 data
for the angular distributions and the total cross sections
given in Table~1 of \cite{Kirk:2000ws}.
The following ``preferred'' values for the couplings were obtained,
see Table~3 of \cite{Lebiedowicz:2013ika},
%For the convenience of the reader we write in the following
%their numerical values:
$(g'_{\Pom \Pom f_{0}(980)}, g''_{\Pom \Pom f_{0}(980)}) = (0.788,4.0)$,
$(g'_{\Pom \Pom f_{0}(1500)}, g''_{\Pom \Pom f_{0}(1500)}) = (1.22,6.0)$,
and $(g'_{\Pom \Pom f_{0}(1370)},g''_{\Pom \Pom f_{0}(1370)}) = (0.81,0)$.

%--------------------------------------------------------
\begin{figure}[!ht]
\includegraphics[width=0.49\textwidth]{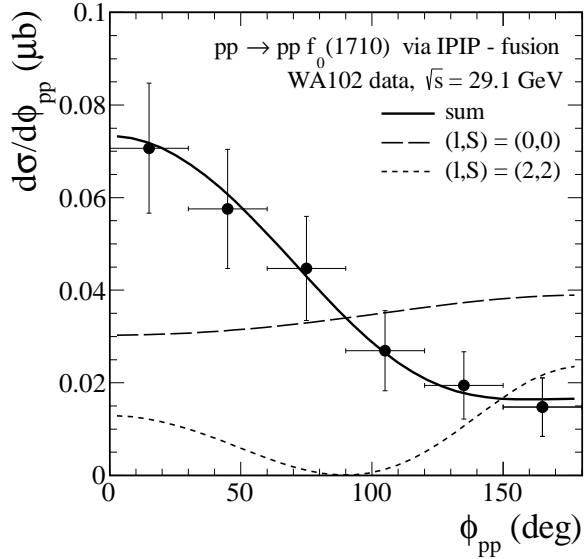}
  \caption{\label{fig:dsig_dphi12_f01710}
  \small
The distribution in azimuthal angle $\phi_{pp}$ between the outgoing protons for
central exclusive production of the $f_{0}(1710)$ meson 
by the fusion of two tensor pomerons at $\sqrt{s} = 29.1$~GeV.
The experimental data points from \cite{Barberis:1999cq} have been
normalized to the total cross section 
$\sigma = 245$~nb from \cite{Kirk:2000ws}.
Plotted is the cross section $d\sigma/d\phi_{pp}$ for $0<\phi_{pp}<\pi$.
We show the individual contributions to the cross section with 
$(l,S) = (0,0)$ (the long-dashed line), $(l,S) = (2,2)$ (the short-dashed line),
and their coherent sum (the solid line).}
\end{figure}
%--------------------------------------------------------
In Fig.~\ref{fig:dsig_dphi12_f01710} we show the distribution in 
azimuthal angle $\phi_{pp}$ between the outgoing protons for the
central exclusive production of the $f_{0}(1710)$ meson
at $\sqrt{s} =29.1$~GeV with the data measured by the WA102 collaboration
in \cite{Kirk:2000ws}.
Similarly as for the $f_{0}(980)$ and $f_{0}(1500)$ mesons 
(see Figs.~5 and 6 in \cite{Lebiedowicz:2013ika}, respectively)
also for the $f_{0}(1710)$ meson both $(l,S)$ contributions
are necessary to describe the $\phi_{pp}$ distribution accurately.
For the $f_{0}(1710)$ we obtain the coupling constants as
$(g'_{\Pom \Pom f_{0}(1710)}, g''_{\Pom \Pom f_{0}(1710)}) = (0.45,2.6)$.

The scalar-meson propagator in (\ref{amplitude_f0_pomTpomT}) 
is parametrized as
\begin{eqnarray}
i\Delta^{(f_{0})}(p_{34}) = \dfrac{i}{p_{34}^{2}-m_{f_{0}}^2+i m_{f_{0}} \Gamma_{f_{0}}}
%(p_{34}^{2})}\,,
\label{prop_scalar}
\end{eqnarray}
with a constant decay width.
%where the running (energy-dependent) width is parametrized as
%
%\begin{eqnarray}
%\Gamma_{f_{0}}(p_{34}^{2}) = \Gamma_{f_{0}}
%\left( \frac{p_{34}^{2} - 4 m_{K}^2}{m_{f_{0}}^2 - 4 m_{K}^2} \right)^{1/2}
%%\frac{m_{f_{0}}}{k} \, 
%\theta(p_{34}^{2}-4 m_{K}^2)\,.
%\end{eqnarray}
%

For the $f_{0} KK$ vertex we have ($M_{0} \equiv 1$~GeV)
\begin{eqnarray}
i\Gamma^{(f_{0} KK)}(p_{3},p_{4}) = i g_{f_{0} K^{+} K^{-}} M_{0}\, F^{(f_{0} KK)}(p_{34}^{2})\,,
\label{vertex_f0KK}
\end{eqnarray}
where the dimensionless coupling constant
$g_{f_{0} K^{+} K^{-}}$ is related to the partial decay width of the $f_{0}$ meson
(for an 'on-shell' $f_{0}$ state $p_{34}^{2} = m_{f_{0}}^{2}$)
\begin{eqnarray}
\Gamma(f_{0} \to K^{+} K^{-}) = 
\frac{M_{0}^{2}}{16 \pi m_{f_{0}}}
|g_{f_{0} K^{+} K^{-}}|^{2} \left( 1-\frac{4 m_{K}^{2}}{m_{f_{0}}^{2}} \right)^{1/2}\,.
\label{gamma_f0KK}
\end{eqnarray}
The analogous relation for $f_{0} \to \pi^{+} \pi^{-}$ reads
\begin{eqnarray}
\Gamma(f_{0} \to \pi^{+} \pi^{-}) = 
\frac{M_{0}^{2}}{16 \pi m_{f_{0}}}
|g_{f_{0} \pi^{+} \pi^{-}}|^{2} \left( 1-\frac{4 m_{\pi}^{2}}{m_{f_{0}}^{2}} \right)^{1/2}\,.
\label{gamma_f0pipi}
\end{eqnarray}
In (\ref{vertex_f0KK}) we assume that $F^{(f_{0} KK)}(p_{34}^{2})$
has the same form as $F^{(\Pom \Pom f_{0})}(p_{34}^{2})$, 
see (\ref{Fpompommeson_ff}).

In order to estimate the coupling constants $g_{f_{0} K^{+} K^{-}}$
for the various $f_{0}$ states from (\ref{gamma_f0KK}) we need data for
the partial decay rates $\Gamma(f_{0} \to K^{+}K^{-})$.
Since the Particle Data Group \cite{Olive:2016xmw}
does not give these decay rates explicitly we shall estimate them
in the following using the available information.

The $f_{0}$ states have isospin $I = 0$. 
Assuming isospin invariance in the decays we get
\begin{eqnarray}
&&\Gamma(f_{0} \to K^{+} K^{-}) = \Gamma(f_{0} \to K^{0} \overline{K}^{0}) = 
                       \frac{1}{2}\Gamma(f_{0} \to K \overline{K})\,,
\label{gamma_f0KK_1}\\
&&\Gamma(f_{0} \to \pi^{+} \pi^{-}) = 2 \Gamma(f_{0} \to \pi^{0} \pi^{0}) = 
                             \frac{2}{3}\Gamma(f_{0} \to \pi \pi)\,.
\label{gamma_f0pipi_1}
\end{eqnarray}
Let us now consider the various $f_{0}$ states in turn. 
The $f_{0}(980)$ has only the $\pi \pi$, $K \overline{K}$ and the electromagnetic
$\gamma \gamma$ decays.
Therefore we have, to very good approximation, for the total decay rate
%The total width is given by the sum of the partial decay width
%
%\begin{eqnarray}
%\Gamma_{f_{0}} &=& \Gamma(f_{0} \to K \overline{K}) + \Gamma(f_{0} \to \pi \pi) \nonumber \\
%               &=& \Gamma(f_{0} \to K^{+} K^{-}) + \Gamma(f_{0} \to K^{0} \overline{K}^{0}) +
%                   \Gamma(f_{0} \to \pi^{+} \pi^{-}) + \Gamma(f_{0} \to \pi^{0} \pi^{0})\,.
%\label{gamma_f0KK_1}
%\end{eqnarray}
%
%We assume that
%
%\begin{eqnarray}
%&&\Gamma(f_{0} \to K^{+} K^{-}) = \Gamma(f_{0} \to K^{0} \overline{K}^{0})\,,
%\label{gamma_f0KK_2a}\\
%&&\Gamma(f_{0} \to \pi^{+} \pi^{-}) = 2 \Gamma(f_{0} \to \pi^{0} \pi^{0})\,.
%\label{gamma_f0KK_2b}
%\end{eqnarray}
%
%From (\ref{gamma_f0KK_1}) - (\ref{gamma_f0KK_2b}) we get
%
\begin{eqnarray}
\Gamma_{f_{0}(980)} = \Gamma(f_{0}(980) \to \pi \pi) + \Gamma(f_{0}(980) \to K \overline{K}) \,.
\label{gamma_f0KK_2}
\end{eqnarray}
%
%Particle Data Group \cite{Olive:2016xmw} does not 
%give explicitly $\Gamma(f_{0} \to K^{+} K^{-})$.
%except for $f_{0}(1500)$ meson.

In \cite{Aubert:2006nu} the ratio
\begin{eqnarray}
\Gamma(f_{0}(980) \to K^{+} K^{-})/\Gamma(f_{0}(980) \to \pi^{+} \pi^{-}) = 0.69 \pm 0.32
\label{ratio_f0980}
\end{eqnarray}
was found from the $B$ meson decays.
To obtain $g_{f_{0}(980) K^{+} K^{-}}$ we assume the approximate relation
\footnote{
We cannot use formula (\ref{gamma_f0KK}) for the $f_{0}(980) \to K^{+} K^{-}$ decay
as the threshold for the decay is above the $f_{0}(980)$ resonance position.
}
\begin{eqnarray}
\sigma(f_{0}(980) \to K^{+} K^{-})/\sigma(f_{0}(980) \to \pi^{+} \pi^{-}) = 0.69 \pm 0.32\,,
\label{sigma_f0980}
\end{eqnarray}
where $\sigma(f_{0}(980) \to \pi^{+} \pi^{-}, K^{+} K^{-})$
are the integrated cross sections for the $pp \to pp (f_{0}(980) \to \pi^{+} \pi^{-}, K^{+} K^{-})$ processes
via the $\Pom \Pom$ fusion at $\sqrt{s} =13$~TeV.
We get from (\ref{gamma_f0KK_1}) - (\ref{sigma_f0980}), and with
$m_{f_{0}(980)} = 980$~MeV, $\Gamma_{f_{0}(980)} = 50$~MeV, 
assuming $g_{f_{0}(980) K^{+} K^{-}} > 0$ and $g_{f_{0}(980) \pi^{+} \pi^{-}} > 0$,
\begin{eqnarray}
&&g_{f_{0}(980) K^{+} K^{-}} = 2.88^{+0.60}_{-0.77} \,,
\label{g_f0980_KK}\\
&&g_{f_{0}(980) \pi^{+} \pi^{-}} = 0.95^{+0.13}_{-0.09}\,.
\label{g_f0980_pipi}
\end{eqnarray}
%If we take $\Gamma_{f_{0}(980)} = 70$~MeV
%we get $g_{f_{0}(980) K^{+} K^{-}} = 3.89$ and $g_{f_{0}(980) \pi^{+} \pi^{-}} = 1.12$.
%with $\Gamma_{f_{0}(980)} = 20$~MeV
%we get $g_{f_{0}(980) K^{+} K^{-}} = 2.80$ and $g_{f_{0}(980) \pi^{+} \pi^{-}} = 0.60$, respectively.}
The error bars in (\ref{g_f0980_KK}) were obtained using only error bars in (\ref{sigma_f0980}).
Uncertainties of the rather poorly known $\Gamma_{f_{0}(980)}$ are similar.

For the $f_{0}(1370)$ meson we take the following input:
\begin{eqnarray}
m_{f_{0}(1370)} = 1370~{\rm MeV}, \; \Gamma_{f_{0}(1370)} = 350~{\rm MeV}\,,
\label{g_f01370_input}
\end{eqnarray}
from \cite{Olive:2016xmw}, and
\begin{eqnarray}
&&\Gamma(f_{0}(1370) \to K \overline{K})/ \Gamma_{f_{0}(1370)} = 0.35 \pm 0.13 \,, \nonumber \\
&&\Gamma(f_{0}(1370) \to \pi \pi)/ \Gamma_{f_{0}(1370)} = 0.26 \pm 0.09\,,
\label{BR_f01370}
\end{eqnarray}
from \cite{Bugg:1996ki}. 
From (\ref{gamma_f0KK_1}) and (\ref{gamma_f0KK_2}) we get then,
assuming again $g_{f_{0}(1370) K^{+} K^{-}} > 0$ and $g_{f_{0}(1370) \pi^{+} \pi^{-}} > 0$,
\begin{eqnarray}
&&g_{f_{0}(1370) K^{+} K^{-}} = 2.47 \,,
\label{g_f01370_KK} \\
&&g_{f_{0}(1370) \pi^{+} \pi^{-}} = 2.07 \,.
\label{g_f01370_pipi}
\end{eqnarray}

For the $f_{0}(1500)$ we have from \cite{Olive:2016xmw}
\begin{eqnarray}
&&m_{f_{0}(1500)} = 1504~{\rm MeV}, \; \Gamma_{f_{0}(1500)} = 109~{\rm MeV}\,,\nonumber \\
&&\Gamma(f_{0}(1500) \to K \overline{K})/ \Gamma_{f_{0}(1500)} = 0.086 \pm 0.010 \,, \nonumber \\
&&\Gamma(f_{0}(1500) \to \pi \pi)/ \Gamma_{f_{0}(1500)} = 0.349 \pm 0.023 \,.
\label{g_f01500_input}
\end{eqnarray}
With (\ref{gamma_f0KK_1}) and (\ref{gamma_f0pipi_1}) we get,
assuming positive coupling constants
\begin{eqnarray}
&&g_{f_{0}(1500) K^{+} K^{-}} = 0.69 \,,
\label{g_f01500_KK}\\
&&g_{f_{0}(1500) \pi^{+} \pi^{-}} = 1.40 \,.
\label{g_f01500_pipi}
\end{eqnarray}

For the $f_{0}(1710)$, finally, we have from \cite{Olive:2016xmw}
\begin{eqnarray}
m_{f_{0}(1710)} = 1723~{\rm MeV}, \; \Gamma_{f_{0}(1710)} = 139~{\rm MeV}\,,
\label{g_f01710_input}
\end{eqnarray}
and from \cite{Albaladejo:2008qa}
\begin{eqnarray}
&&\Gamma(f_{0}(1710) \to K \overline{K})/ \Gamma_{f_{0}(1710)} = 0.36 \pm 0.12 \,,
\label{BR_f01710}\\
&&\Gamma(f_{0}(1710) \to \pi\pi)/\Gamma(f_{0}(1710) \to K \overline{K}) = 0.32 \pm 0.14 \,.
\label{ratio_f01710}
\end{eqnarray}
The ratio (\ref{ratio_f01710}) is consistent 
with $0.41^{+0.11}_{-0.17}$ from \cite{Ablikim:2006db} 
and with $0.31^{+0.05}_{-0.03}$ from \cite{Dobbs:2015dwa}
within the errors.
This gives, for positive couplings, from (\ref{gamma_f0KK_1}) - (\ref{gamma_f0pipi_1})
\begin{eqnarray}
&&g_{f_{0}(1710) K^{+} K^{-}} = 2.36 \,,
\label{g_f01710_KK}\\
&&g_{f_{0}(1710) \pi^{+} \pi^{-}} = 1.40 \,.
\label{g_f01710_pipi}
\end{eqnarray}
%

%--------------------------------------
\subsection{$f_{2}(1270)$ and $f'_{2}(1525)$ mesons central production}
\label{sec:diff_f2}
%--------------------------------------
For diffractive $K^{+}K^{-}$ production through 
the $s$-channel $f_{2}$-meson exchange 
the amplitude is more complicated to treat.
The $f_{2}(1270)$ and $f'_{2}(1525)$ mesons could be considered 
as potential candidates.
The amplitude for the $\Pom \Pom$ fusion is given by
\begin{equation}
\begin{split}
{\cal M}^{(\Pom \Pom \to f_{2} \to K^{+}K^{-})}_{\lambda_{a} \lambda_{b} \to \lambda_{1} \lambda_{2} K^{+}K^{-}} 
= & (-i)\,
\bar{u}(p_{1}, \lambda_{1}) 
i\Gamma^{(\Pom pp)}_{\mu_{1} \nu_{1}}(p_{1},p_{a}) 
u(p_{a}, \lambda_{a})\;
i\Delta^{(\Pom)\, \mu_{1} \nu_{1}, \alpha_{1} \beta_{1}}(s_{1},t_{1}) \\
& \times 
i\Gamma^{(\Pom \Pom f_{2})}_{\alpha_{1} \beta_{1},\alpha_{2} \beta_{2}, \rho \sigma}(q_{1},q_{2}) \;
i\Delta^{(f_{2})\,\rho \sigma, \alpha \beta}(p_{34})\;
i\Gamma^{(f_{2} KK)}_{\alpha \beta}(p_{3},p_{4})\\
& \times 
i\Delta^{(\Pom)\, \alpha_{2} \beta_{2}, \mu_{2} \nu_{2}}(s_{2},t_{2}) \;
\bar{u}(p_{2}, \lambda_{2}) 
i\Gamma^{(\Pom pp)}_{\mu_{2} \nu_{2}}(p_{2},p_{b}) 
u(p_{b}, \lambda_{b}) \,.
\end{split}
\label{amplitude_f2_pomTpomT}
\end{equation}

The $\Pom \Pom f_{2}$ vertex can be written as 
\begin{eqnarray}
i\Gamma_{\mu \nu,\kappa \lambda,\rho \sigma}^{(\Pom \Pom f_{2})} (q_{1},q_{2}) =
\left( i\Gamma_{\mu \nu,\kappa \lambda,\rho \sigma}^{(\Pom \Pom f_{2})(1)} \mid_{bare}
+ \sum_{j=2}^{7}i\Gamma_{\mu \nu,\kappa \lambda,\rho \sigma}^{(\Pom \Pom f_{2})(j)}(q_{1},q_{2}) \mid_{bare} 
\right)
%i \, g_{\Pom \Pom M} \, M_{0} \,  %F^{M}_{\Pom \Pom M}(t_{1},t_{2})
%\left( g_{\mu \kappa} g_{\nu \lambda} + g_{\mu \lambda} g_{\nu \kappa}
%-\frac{1}{2} g_{\mu \nu} g_{\kappa \lambda} \right) 
\tilde{F}^{(\Pom \Pom f_{2})}(q_{1}^{2},q_{2}^{2},p_{34}^{2}) \,.\nonumber\\
\label{vertex_pompomT}
\end{eqnarray}
A possible choice for the 
$i\Gamma_{\mu \nu,\kappa \lambda,\rho \sigma}^{(\Pom \Pom f_{2})(j)}\mid_{bare}$
terms $j = 1, ..., 7$ is given in appendix~A of \cite{Lebiedowicz:2016ioh}.
In \cite{Lebiedowicz:2016ioh} we found that the $j=2$ coupling
for $g^{(2)}_{\Pom \Pom f_{2}(1270)} = 9.0$ is optimal
to describe the main characteristics measured in the WA102 and ISR experiments
and by the CDF collaboration \cite{Aaltonen:2015uva}
including e.g. a gap survival factor $\langle S^{2} \rangle = 0.1$ for the CDF.
%Neglecting the reggeon exchanges for the $f_{2}(1270)$ production
%infers that our predictions for the LHC energies should be treated rather as the upper limit.

The $f_{2}(1270)$ and $f'_{2}(1525)$ have similar $\phi_{pp}$ and $dP_{t}$
dependences \cite{Barberis:1999cq}.
$dP_{t}$ is the so-called ``glueball-filter variable'' \cite{Close:1997pj}
defined by the difference of the transverse momentum vectors
of the outgoing protons in (\ref{2to4_reaction})
\begin{eqnarray}
\bdPt = \bqta - \bqtb = \bptb - \bpta \,, \quad dP_{t} = |\bdPt|\,.
\label{dPt_variable}
\end{eqnarray}
It has been observed in Ref.~\cite{Barberis:1996iq}
that all the undisputed $q\bar{q}$ states 
(i.e. $\eta$, $\eta'$, $f_{1}(1285)$ etc.)
are suppressed when $dP_{t} \to 0$, whereas the glueball candidates, 
e.g. $f_{0}(1500)$, survive.
As can be seen in Refs.~\cite{Barberis:1996iq,Barberis:1999cq} 
the $f_{2}(1270)$ and $f_{2}'(1525)$ states
have larger $dP_{t}$ and their cross sections peak at $\phi_{pp} = \pi$
in contrast to the ``enigmatic'' $f_{0}(980)$, $f_{0}(1500)$ and $f_{0}(1710)$ states.
Note, that at $\sqrt{s} = 29.1$~GeV the experimental cross section 
for the production of the $f_{2}(1270)$ meson, 
whose production has been found to
be consistent with double pomeron/reggeon exchange, 
is more than 48 times greater 
than the cross section of the $f'_{2}(1525)$ meson \cite{Kirk:2000ws}.
For the $f'_{2}(1525)$ we assume also only the $j = 2$ coupling
with $g^{(2)}_{\Pom \Pom f'_{2}(1525)} = 2.0$
fixed to the experimental total cross section from \cite{Kirk:2000ws}.
With this we roughly reproduced the shapes of the differential distributions 
of the WA102 data \cite{Barberis:1999cq}.
In the future the corresponding $\Pom \Pom f_{2}$ coupling constants 
could be adjusted by comparison with precise experimental data.

In (\ref{vertex_pompomT}) $\tilde{F}^{(\Pom \Pom f_{2})}$ is a form factor 
for which we take
\begin{eqnarray}
&&\tilde{F}^{(\Pom \Pom f_{2})}(q_{1}^{2},q_{2}^{2},p_{34}^{2}) = 
F_{M}(q_{1}^{2}) F_{M}(q_{2}^{2}) F^{(\Pom \Pom f_{2})}(p_{34}^{2})\,,
\label{Fpompommeson_tensor}\\
&&F^{(\Pom \Pom f_{2})}(p_{34}^{2}) = 
\exp{ \left( \frac{-(p_{34}^{2}-m_{f_{2}}^{2})^{2}}{\Lambda_{f_{2}}^{4}} \right)}\,,
\quad \Lambda_{f_{2}} = 1\;{\rm GeV}\,.
\label{Fpompommesonf2_ff}
\end{eqnarray}

Here, for qualitative calculations only, one may use 
the tensor-meson propagator with the simple Breit-Wigner form
%, setting $p_{34}^{2} = s_{34}$
\begin{eqnarray}
i\Delta_{\mu \nu, \kappa \lambda}^{(f_{2})}(p_{34})&=&
\frac{i}{p_{34}^{2}-m_{f_{2}}^2+i m_{f_{2}} \Gamma_{f_{2}}}
\left[ 
\frac{1}{2} 
%\left( -g_{\mu \kappa} + \frac{k_{\mu} k_{\kappa}}{m_{f_{2}}^2} \right)
%\left( -g_{\nu \lambda} + \frac{k_{\nu} k_{\lambda}}{m_{f_{2}}^2} \right)
%\right. \nonumber \\ 
%&& +  \left.
%\frac{1}{2} 
%\left( -g_{\mu \lambda} + \frac{k_{\mu} k_{\lambda}}{m_{f_{2}}^2} \right)
%\left( -g_{\nu \kappa} + \frac{k_{\nu} k_{\kappa}}{m_{f_{2}}^2} \right)
( \hat{g}_{\mu \kappa} \hat{g}_{\nu \lambda}  + \hat{g}_{\mu \lambda} \hat{g}_{\nu \kappa} )
-\frac{1}{3} 
%\left( -g_{\mu \nu} + \frac{k_{\mu} k_{\nu}}{m_{f_{2}}^2} \right)
%\left( -g_{\kappa \lambda} + \frac{k_{\kappa} k_{\lambda}}{m_{f_{2}}^2} \right)
\hat{g}_{\mu \nu} \hat{g}_{\kappa \lambda}
%\right.
% \nonumber \\ 
%&& +  \left.
%\frac{1}{24} \frac{k^{2}-m_{f_{2}}^{2}}{m_{f_{2}}^{4}}
%\left[ 
%g_{\mu \nu} g_{\kappa \lambda} (k^{2}-2 m_{f_{2}}^{2})
%-4 (g_{\mu \nu} k_{\kappa} k_{\lambda} + g_{\kappa \lambda} k_{\mu} k_{\nu})
%\right]
\right] \,, 
\label{prop_f2}
\end{eqnarray}
where $\hat{g}_{\mu \nu} = -g_{\mu \nu} + p_{34 \mu} p_{34 \nu} / p_{34}^2$.
In (\ref{prop_f2}) $\Gamma_{f_{2}}$ is the total decay width of the $f_{2}$ resonance
and $m_{f_{2}}$ its mass.

The $f_{2} KK$ vertex can be written as (see Eq.~(3.37) of \cite{Ewerz:2013kda}
for the analogous $f_{2} \pi \pi$ vertex)
\begin{eqnarray}
i\Gamma_{\mu \nu}^{(f_{2} KK)}(p_{3},p_{4})=
-i \,\frac{g_{f_{2} K^{+}K^{-}}}{2 M_{0}} \,
\left[ (p_{3}-p_{4})_{\mu} (p_{3}-p_{4})_{\nu}
- \frac{1}{4} g_{\mu \nu} (p_{3}-p_{4})^{2} \right] \, F^{(f_{2} KK)}(p_{34}^{2})\,,
\nonumber \\
\label{vertex_f2pipi_N}
\end{eqnarray}
where $g_{f_{2} K^{+}K^{-}}$ 
can be obtained from the corresponding partial decay width.
We assume that
\begin{eqnarray}
F^{(f_{2} KK)}(p_{34}^{2}) = F^{(\Pom \Pom f_{2})}(p_{34}^{2}) = 
\exp{ \left( \frac{-(p_{34}^{2}-m_{f_{2}}^{2})^{2}}{\Lambda_{f_{2}}^{4}} \right)}\,,
\quad \Lambda_{f_{2}} = 1\;{\rm GeV}\,.
\label{Fpompommeson_ff_tensor}
\end{eqnarray}

In analogy to the $f_{2} \to \pi \pi$ decay,
treated in section~5.1 of \cite{Ewerz:2013kda}, we can write
\begin{eqnarray}
\Gamma(f_{2} \to K^{+} K^{-}) = 
\frac{m_{f_{2}}}{480 \pi}\,
|g_{f_{2} K^{+} K^{-}}|^{2} 
\left( \frac{m_{f_{2}}}{M_{0}} \right)^{2}
\left( 1-\frac{4 m_{K}^{2}}{m_{f_{0}}^{2}} \right)^{5/2}\,.
\label{gamma_f2KK}
\end{eqnarray}
We assume further that isospin symmetry holds, that is,
\begin{eqnarray}
\Gamma(f_{2} \to K^{+} K^{-}) = \frac{1}{2} \, \Gamma(f_{2} \to K \overline{K}) \,.
\label{gamma_f2KK_aux}
\end{eqnarray}
With $\Gamma(f_{2} \to K\overline{K})/\Gamma_{f_{2}}$ from \cite{Olive:2016xmw} 
(see Table~\ref{table:table}) we get,
assuming $g_{f_{2} K^{+} K^{-}} > 0$,
\begin{eqnarray}
&&g_{f_{2}(1270) K^{+} K^{-}} = 5.54 \,,
\label{g_f21270_KK}\\
&&g_{f'_{2}(1525) K^{+} K^{-}} = 7.32 \,.
\label{g_f21525_KK}
\end{eqnarray}
For the $\pi^{+} \pi^{-}$ decay channel,
based on (5.6) of \cite{Ewerz:2013kda} 
and the numerical values from Table~\ref{table:table},
we have
\begin{eqnarray}
&&g_{f_{2}(1270) \pi^{+} \pi^{-}} = 9.28 \,,\\
\label{g_f21270_pipi}
&&g_{f'_{2}(1525) \pi^{+} \pi^{-}} = 0.43 \,.
\label{g_f21525_pipi}
\end{eqnarray}
%

%--------------------------------------
\section{Photoproduction contributions}
\label{sec:photoprod_mech}
%--------------------------------------

\begin{figure}[!ht]
(a)\includegraphics[width=0.3\textwidth]{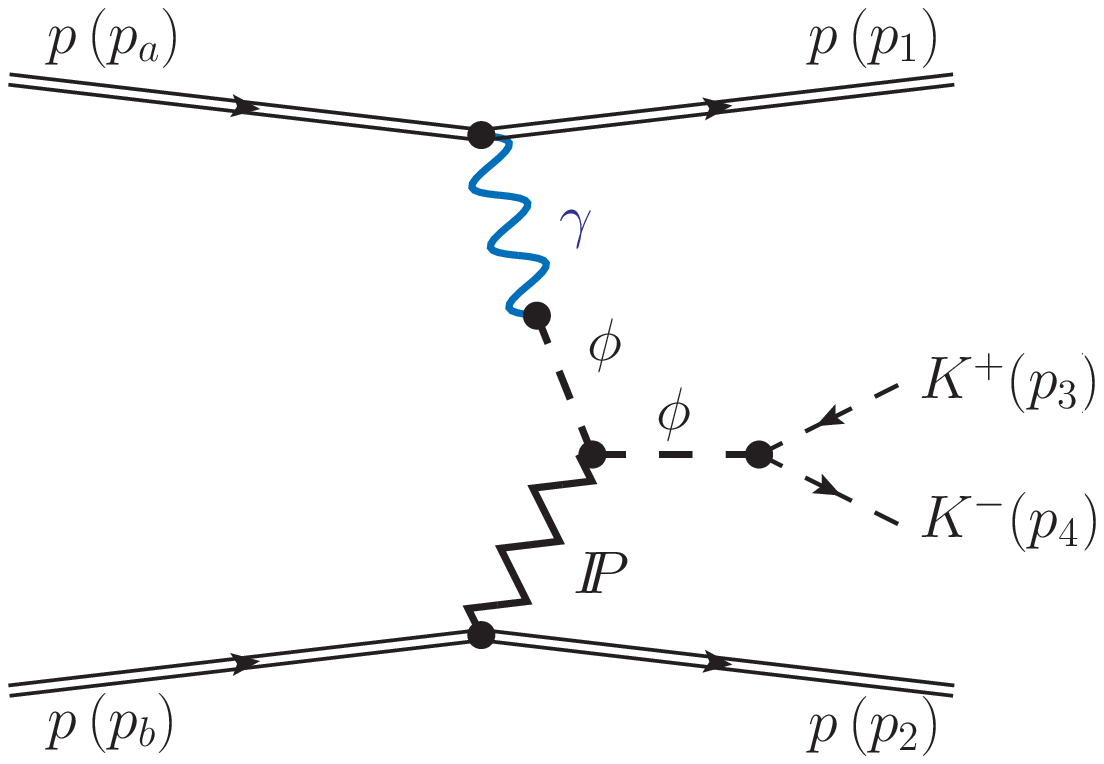}
(b)\includegraphics[width=0.3\textwidth]{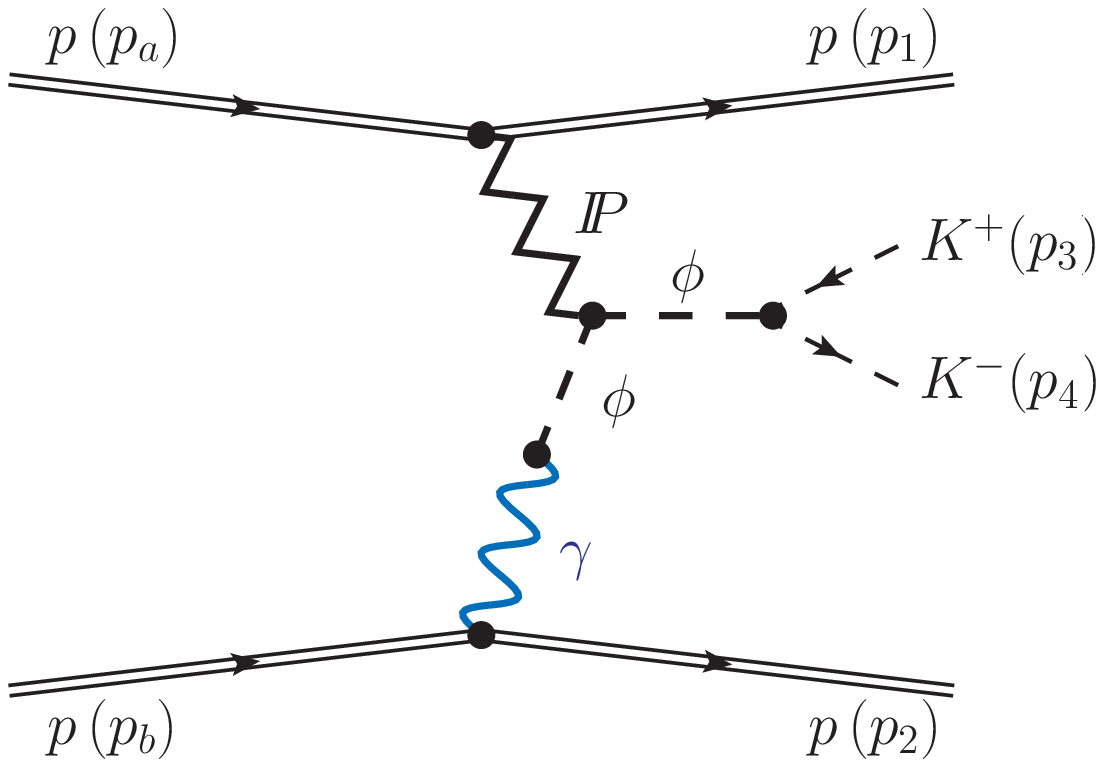}
  \caption{\label{fig:gampom_pomgam_s}
  \small
The central exclusive $\phi$ meson production and its subsequent decay 
into $P$-wave $K^+ K^-$ in proton-proton collisions.}
\end{figure}

\begin{figure}[!ht]
(a)\includegraphics[width=0.28\textwidth]{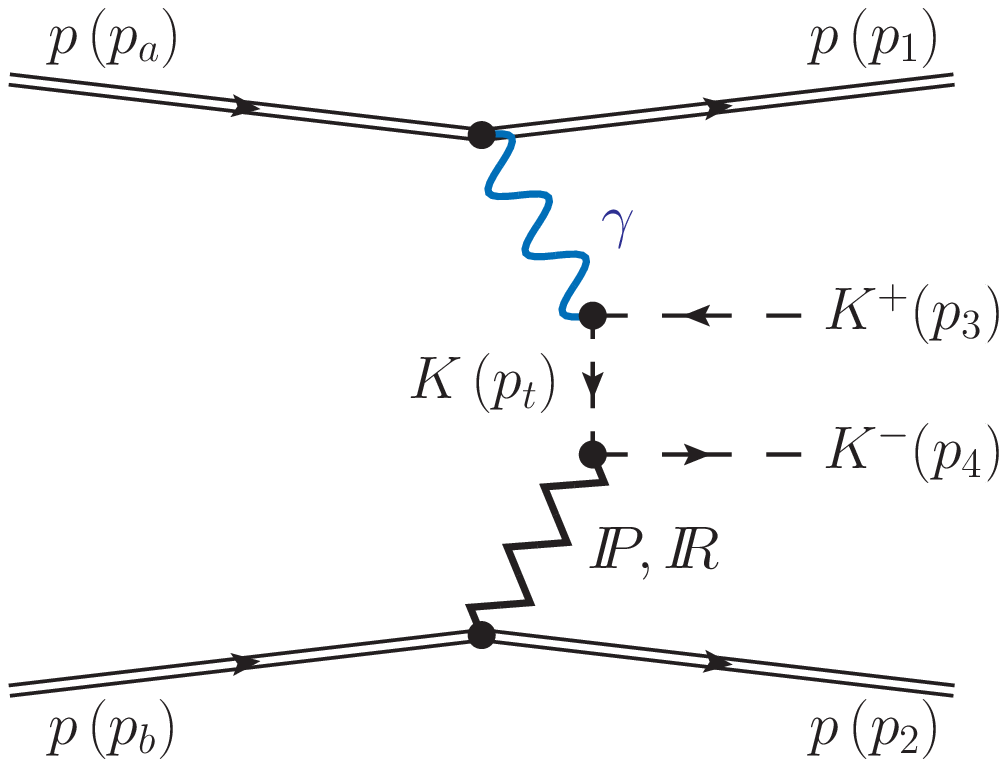}
(b)\includegraphics[width=0.28\textwidth]{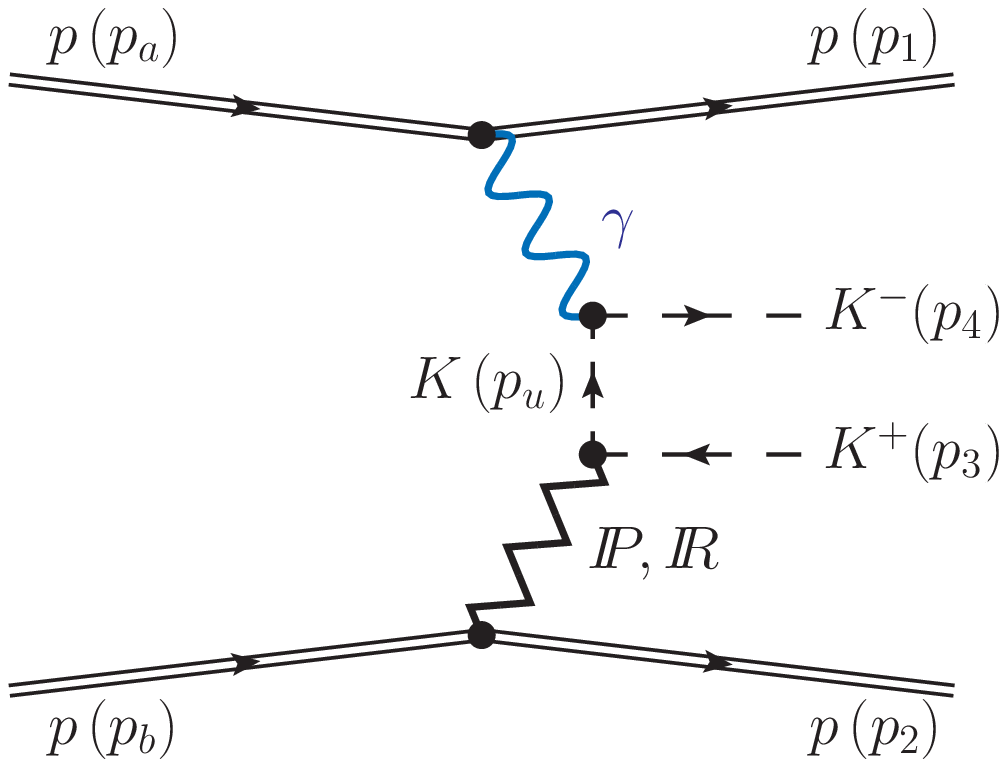}
(c)\includegraphics[width=0.28\textwidth]{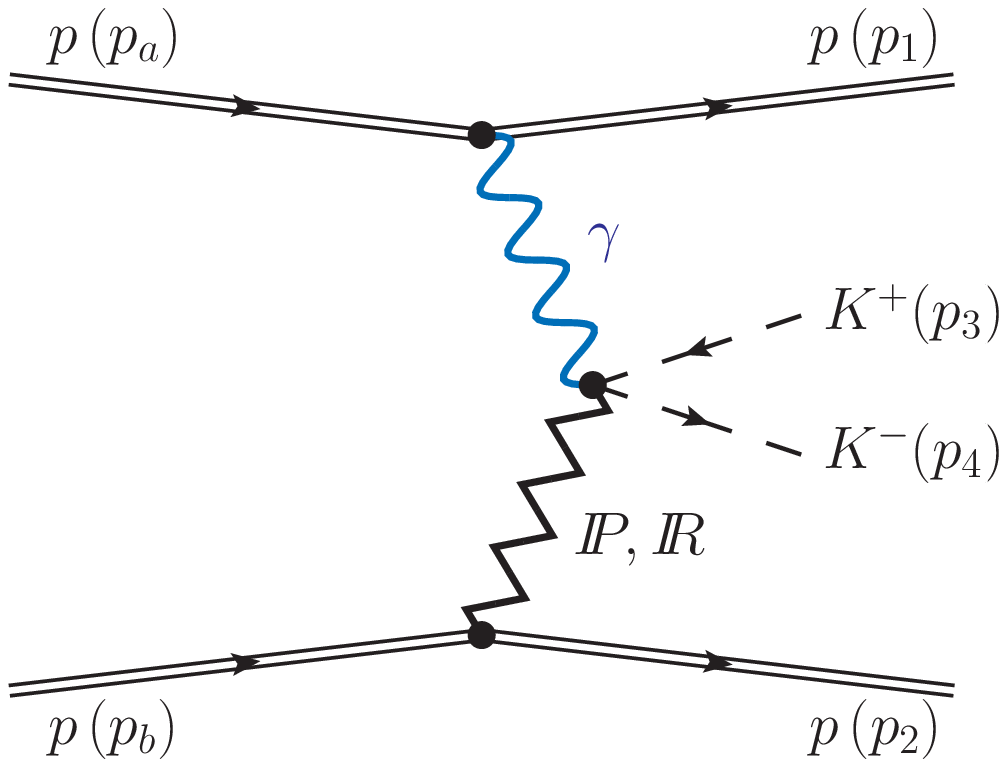}
  \caption{\label{fig:gampom_pomgam_b}
  \small
The diagrams for photon-induced central exclusive continuum
$K^+ K^-$ production in proton-proton collisions.
There are also 3 additional diagrams with the role of
$(p(p_{a}), p(p_{1}))$ and $(p(p_{b}), p(p_{2}))$ exchanged.
}
\end{figure}

For the $\phi$ resonance production we consider the diagrams shown 
in Fig.~\ref{fig:gampom_pomgam_s}.
In these diagrams all vertices and propagators will be taken here
according to Ref.~\cite{Ewerz:2013kda}.
The diagrams to be considered for the non-resonant (Drell-S\"oding)
contribution are shown in Fig.~\ref{fig:gampom_pomgam_b}.
In the following we collect formulae for the amplitudes for the $p p \to p p K^+ K^-$
reaction within the tensor-pomeron approach \cite{Ewerz:2013kda}.
%--------------------------------------------------------
%--------------------------------------------------------

%--------------------------------------------------------
%--------------------------------------------------------

%--------------------------------------
\subsection{$K^{+}K^{-}$ continuum central production}
\label{sec:photoprod_continuum}
%--------------------------------------
The amplitude for photoproduction of the $K^+ K^-$ continuum
can be written as the following sum:
\begin{eqnarray}
{\cal M}^{KK{\rm-continuum}}_{pp \to pp K^{+} K^{-}} &=&
{\cal M}^{(\gamma \Pom \to K^{+}K^{-})} +
{\cal M}^{(\Pom \gamma \to K^{+}K^{-})} +
{\cal M}^{(\gamma \Reg \to K^{+}K^{-})} +
{\cal M}^{(\Reg \gamma \to K^{+}K^{-})} . \; \qquad
\label{cont_photoprod}
\end{eqnarray}

The $\gamma \Pom$-exchange amplitude 
can be written as the sum:
\begin{eqnarray}
{\cal M}^{(\gamma \Pom \to K^{+}K^{-})} =
{\cal M}^{({a})}_{\lambda_{a} \lambda_{b} \to \lambda_{1} \lambda_{2} K^{+}K^{-}}+
{\cal M}^{({b})}_{\lambda_{a} \lambda_{b} \to \lambda_{1} \lambda_{2} K^{+}K^{-}}+
{\cal M}^{({c})}_{\lambda_{a} \lambda_{b} \to \lambda_{1} \lambda_{2} K^{+}K^{-}}\,,
\label{gampom_amp}
\end{eqnarray}
where
\begin{eqnarray}
&&{\cal M}^{(a)}_{\lambda_{a} \lambda_{b} \to \lambda_{1} \lambda_{2} K^{+}K^{-}} 
= (-i)
\bar{u}(p_{1}, \lambda_{1}) 
i\Gamma^{(\gamma pp)}_{\mu}(p_{1},p_{a}) 
u(p_{a}, \lambda_{a})\,
i\Delta^{(\gamma)\,\mu \nu}(q_{1})\,
i\Gamma^{(\gamma KK)}_{\nu}(p_{t},-p_{3})\nonumber \\
&&\qquad \times  
i\Delta^{(K)}(p_{t}) \,
i\Gamma^{(\Pom KK)}_{\alpha \beta}(p_{4},p_{t})\,
%i\Gamma^{(\Pom \rho \rho)}_{\rho_{1} \rho_{2} \alpha \beta}(-q_{1},-p_{34})\, 
i\Delta^{(\Pom)\,\alpha \beta, \delta \eta}(s_{2},t_{2}) \,
\bar{u}(p_{2}, \lambda_{2}) 
i\Gamma^{(\Pom pp)}_{\delta \eta}(p_{2},p_{b}) 
u(p_{b}, \lambda_{b}) \,, \nonumber\\
\label{amplitude_gamma_pomeron_a}\\
%\end{eqnarray}
%
%\begin{eqnarray}
&&{\cal M}^{(b)}_{\lambda_{a} \lambda_{b} \to \lambda_{1} \lambda_{2} K^{+}K^{-}} 
= (-i)
\bar{u}(p_{1}, \lambda_{1}) 
i\Gamma^{(\gamma pp)}_{\mu}(p_{1},p_{a}) 
u(p_{a}, \lambda_{a})\,
i\Delta^{(\gamma)\,\mu \nu}(q_{1})\,
i\Gamma^{(\gamma KK)}_{\nu}(p_{4},p_{u}) \nonumber \\
&&\qquad \times  
i\Delta^{(K)}(p_{u}) \, 
i\Gamma^{(\Pom KK)}_{\alpha \beta}(p_{u},-p_{3})\,
%i\Gamma^{(\Pom \rho \rho)}_{\rho_{1} \rho_{2} \alpha \beta}(-q_{1},-p_{34})\, 
i\Delta^{(\Pom)\,\alpha \beta, \delta \eta}(s_{2},t_{2}) \,
\bar{u}(p_{2}, \lambda_{2}) 
i\Gamma^{(\Pom pp)}_{\delta \eta}(p_{2},p_{b}) 
u(p_{b}, \lambda_{b}) \,,  \nonumber\\
\label{amplitude_gamma_pomeron_b} \\
%\end{eqnarray}
%
%\begin{eqnarray}
&&{\cal M}^{(c)}_{\lambda_{a} \lambda_{b} \to \lambda_{1} \lambda_{2} K^{+}K^{-}} 
= (-i)
\bar{u}(p_{1}, \lambda_{1}) 
i\Gamma^{(\gamma pp)}_{\mu}(p_{1},p_{a}) 
u(p_{a}, \lambda_{a}) \,
i\Delta^{(\gamma)\,\mu \nu}(q_{1}) \nonumber \\
&&\qquad \times  
i\Gamma^{(\Pom \gamma KK)}_{\nu, \alpha \beta}(q_{1},p_{4},-p_{3}) \,
i\Delta^{(\Pom)\,\alpha \beta, \delta \eta}(s_{2},t_{2}) \,
\bar{u}(p_{2}, \lambda_{2}) 
i\Gamma^{(\Pom pp)}_{\delta \eta}(p_{2},p_{b}) 
u(p_{b}, \lambda_{b}) \,.
\label{amplitude_gamma_pomeron_c}
\end{eqnarray}

Here the $\gamma$ and $\Pom$ propagators and the $\gamma pp$,
$\Pom pp$ vertices are given in section~3 of \cite{Ewerz:2013kda}.
The $K$ propagator is standard and given after (\ref{amplitude_u}) above.
The $\gamma KK$, $\Pom KK$ and $\Pom \gamma KK$ vertices
are as the corresponding vertices for pions, see appendix~B of \cite{Bolz:2014mya},
but with $\beta_{\Pom \pi \pi}$ replaced by $\beta_{\Pom KK}$ (\ref{Kp_couplings_pom}).

In order to assure gauge invariance and ``proper'' cancellations 
among the three terms (\ref{amplitude_gamma_pomeron_a}) 
to (\ref{amplitude_gamma_pomeron_c}) 
we have introduced, somewhat arbitrarily, one common energy dependence on $s_{2}$
defined as:
\begin{eqnarray}
s_{2} = (p_{b} + q_{1})^{2} = (p_{2} + p_{34})^{2}
\end{eqnarray}
for the pomeron propagator in all three diagrams.
Gauge invariance requires 
\begin{eqnarray}
\lbrace {\cal M}^{(a)} + {\cal M}^{(b)} + {\cal M}^{(c)} \rbrace|_{p_{1} + p_{a} \to q_{1}} = 0 \,.
\label{gauge_invariance}
\end{eqnarray}
This is satisfied as we see easily be replacing in 
(\ref{amplitude_gamma_pomeron_a}) - (\ref{amplitude_gamma_pomeron_c})
$\Gamma^{(\gamma pp)}_{\mu}(p_{1},p_{a})$ by $q_{1 \mu}$.
The formulas (\ref{amplitude_gamma_pomeron_a}) - (\ref{amplitude_gamma_pomeron_c})
do not include hadronic form factors 
for the inner subprocess $\gamma \Pom \to K^{+}K^{-}$.
A possible way to include form factors
for the inner subprocesses is to multiply the amplitude obtained from
(\ref{amplitude_gamma_pomeron_a}) to (\ref{amplitude_gamma_pomeron_c})
with a common factor, 
see \cite{Poppe:1986dq,Szczurek:2002bn,Klusek-Gawenda:2013rtu,Lebiedowicz:2014bea},
\begin{eqnarray}
{\cal M}^{(\gamma \Pom \to K^{+}K^{-})} = 
({\cal M}^{(a)} + {\cal M}^{(b)} + {\cal M}^{(c)}) \,F(p_{t}^{2},p_{u}^{2},p_{34}^{2})\,.
\label{amplitude_gamma_pomeron_abc}
\end{eqnarray}
A common form factor for all three diagrams
is chosen in order to maintain gauge invariance,
and a convenient form is given in \cite{Klusek-Gawenda:2017lgt}
\begin{eqnarray}
F(p_{t}^{2},p_{u}^{2},p_{34}^{2}) = 
\frac{\left[F(p_{t}^{2})\right]^{2} + [F(p_{u}^{2})]^{2}}{1+[\tilde{F}(p_{34}^{2})]^{2}}\,,
\label{ff_Poppe}
\end{eqnarray}
with the exponential parametrizations
\begin{eqnarray}
F(p_{t}^{2}) &=& \exp \left( \frac{p_{t}^{2}-m_{K}^{2}}{\Lambda_{K}^{2}} \right)\,, \\
F(p_{u}^{2}) &=& \exp \left( \frac{p_{u}^{2}-m_{K}^{2}}{\Lambda_{K}^{2}} \right)\,, \\
\tilde{F}(p_{34}^{2}) &=& \exp \left( \frac{-(p_{34}^{2}-4m_{K}^{2})}{\Lambda_{K}^{2}} \right)\,.
\label{ff_Poppe_aux}
\end{eqnarray}
The parameter $\Lambda_{K}$ should be fitted to the experimental data.
%Note that the form factor $F(t)$ is normalized to unity for $t = m_{p}^{2}$.
%with $\Lambda_{K}$ being a free parameter.
We expect it in the range of 0.8 to 1~GeV.

For the $\Pom \gamma$-exchange the amplitude has the same structure
with $p(p_{a}), p(p_{1}) \leftrightarrow p(p_{b}), p(p_{2})$,
$t_{1} \leftrightarrow t_{2}$ and $s_{2} \leftrightarrow s_{1}$.
We shall consider also contributions involving non-leading reggeons.
For the $f_{2 \Reg}$ exchange 
the formulae have the same tensorial structure as for pomeron exchange
and are obtained from (\ref{amplitude_gamma_pomeron_a})
to (\ref{amplitude_gamma_pomeron_c}) with the corresponding
effective $f_{2 \Reg} pp$, $f_{2 \Reg} KK$ and $f_{2 \Reg} \gamma KK$
vertices and the $f_{2 \Reg}$ reggeon propagator;
see \cite{Ewerz:2013kda,Bolz:2014mya}.
Analogous statements hold for the case for the $a_{2\Reg}$ reggeon exchange.
The relevant reggeon-kaon coupling constants are given in Eq.~(\ref{Kp_couplings_reg}).
The contributions involving $C = -1$ reggeon exchanges are different.
We recall that $\Reg_{-} = \omega_{\Reg}, \rho_{\Reg}$ exchanges 
are treated as effective vector exchanges in our model; see Sec.~3 of \cite{Ewerz:2013kda}.
The vertex for $\rho_{\Reg} \gamma KK$ is in analogy 
to the vertex $\rho_{\Reg} \gamma \pi\pi$ given in (B.81) of \cite{Bolz:2014mya}.
The $\omega_{\Reg}$ exchange is treated in a similar way.

%-------------------------------------------------------------------
\subsection{Photoproduction of $\phi$ meson}
\label{sec:section_2}
%-------------------------------------------------------------------

Since the proton contains no valence $s$ quarks we shall assume that
the amplitude for the $\gamma p \to \phi p$ reaction 
at high energies includes only the pomeron exchange contribution.
In contrast, in the amplitudes for the $\gamma p \to \rho^{0} p$ reaction 
\cite{Lebiedowicz:2014bea,Bolz:2014mya}
and for the $\gamma p \to \omega p$ reaction \cite{Cisek:2011vt}
also reggeon exchanges play an important role.

The amplitude for the $\gamma p \to \phi p$ reaction
with the tensor-pomeron exchange can be written 
in complete analogy to $\gamma p \to \rho^{0} p$ (see \cite{Lebiedowicz:2014bea,Bolz:2014mya})
as follows
\begin{eqnarray}
&&\Braket{\phi(p_{\phi},\lambda_{\phi}),p(p_{2},\lambda_{2})
|{\cal T}|
\gamma(q,\lambda_{\gamma}),p(p_{b},\lambda_{b})} 
\equiv  
\nonumber \\  
&& {\cal M}_{\lambda_{\gamma} \lambda_{b} \to \lambda_{\phi} \lambda_{2}}(s,t) =
(-i) \, (\epsilon^{(\phi)\,\mu})^* \,
i\Gamma_{\mu \nu \alpha \beta}^{(\Pom \phi \phi)}(p_{\phi},q) \,  
i\Delta^{(\phi)\,\nu \kappa}(q) \,
i\Gamma^{(\gamma \to \phi)}_{\kappa \sigma}(q)\,
\epsilon^{(\gamma)\,\sigma} 
%c^{(\gamma \to \rho)}\,
%(\Delta_{T}^{(\rho)})^{-1} \,
\nonumber \\ 
&& \qquad \qquad \qquad \qquad \quad \times 
i\Delta^{(\Pom)\,\alpha \beta, \delta \eta}(s,t)\, 
\bar{u}(p_{2}, \lambda_{2}) 
i\Gamma_{\delta \eta}^{(\Pom pp)}(p_{2},p_{b}) 
u(p_{b}, \lambda_{b}) \,,
\label{phip_tot_opt_aux0}
\end{eqnarray}
where $p_{b}$, $p_{2}$ and $\lambda_{b}$, $\lambda_{2} = \pm \frac{1}{2}$ 
denote the four-momenta and helicities of the ingoing and outgoing protons,
$\epsilon^{(\gamma)}$ and $\epsilon^{(\phi)}$ are the polarisation vectors 
for photon and $\phi$ meson with the four-momenta $q$, $p_{\phi}$
and helicities $\lambda_{\gamma} = \pm 1$, $\lambda_{\phi} = \pm 1, 0$, respectively.
We use standard kinematic variables
\begin{eqnarray}
&&s = W_{\gamma p}^{2} = (p_{b} + q)^{2} = (p_{2} + p_{\phi})^{2}\,, \nonumber \\
&&t = (p_{2} - p_{b})^{2} = (p_{\phi} -q)^{2}\,.
\label{2to2_kinematic}
\end{eqnarray}

In (\ref{phip_tot_opt_aux0}) the $\phi$ propagator is of the same
structure as for $\rho^{0}$ and $\omega$ in (3.2) of \cite{Ewerz:2013kda}.
Our ansatz for the $\Pom \phi \phi$ vertex follows the one for the $\Pom \rho \rho$
in (3.47) of \cite{Ewerz:2013kda}.

\begin{figure}[!ht]
\includegraphics[width=0.49\textwidth]{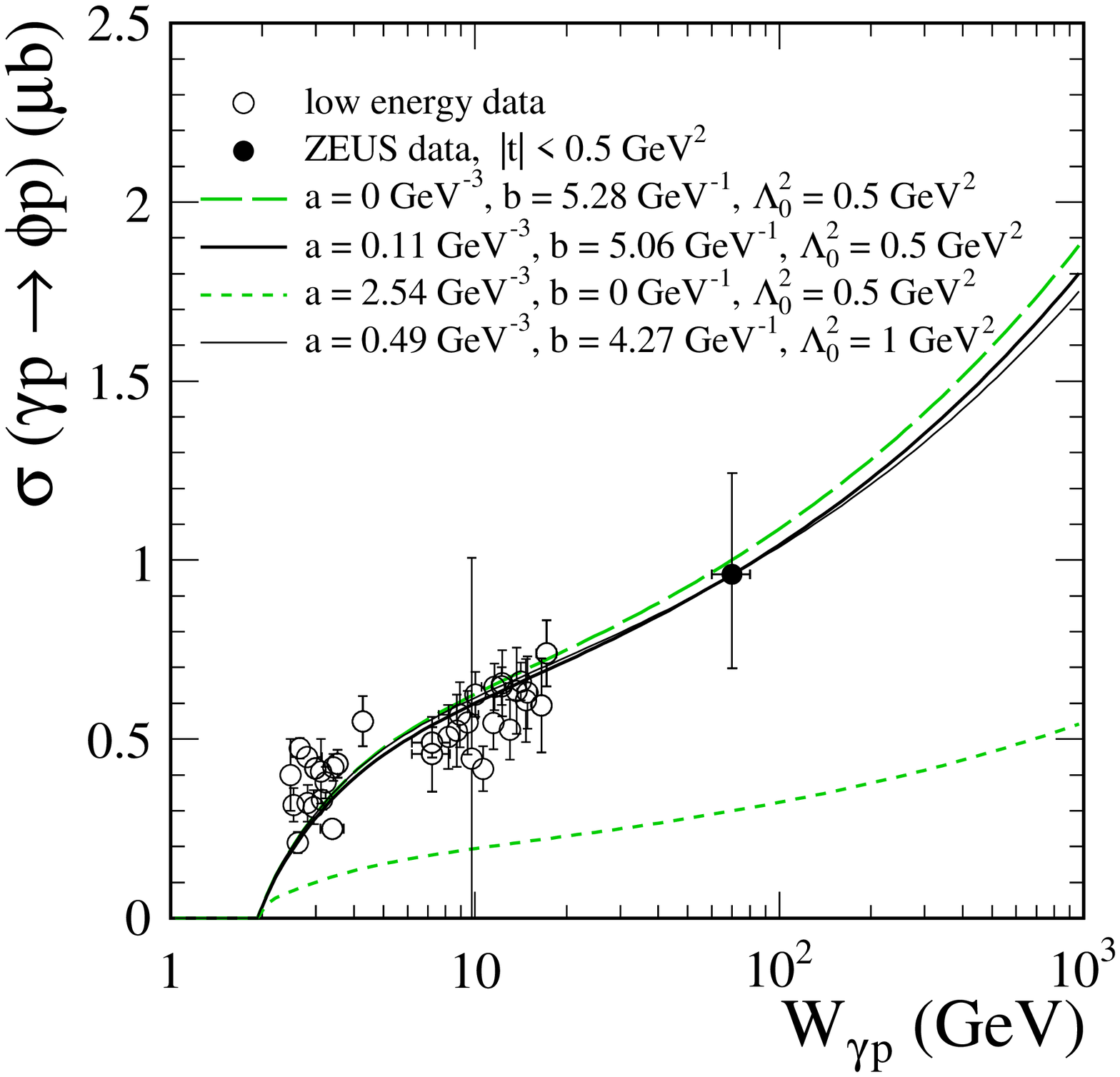}
\includegraphics[width=0.49\textwidth]{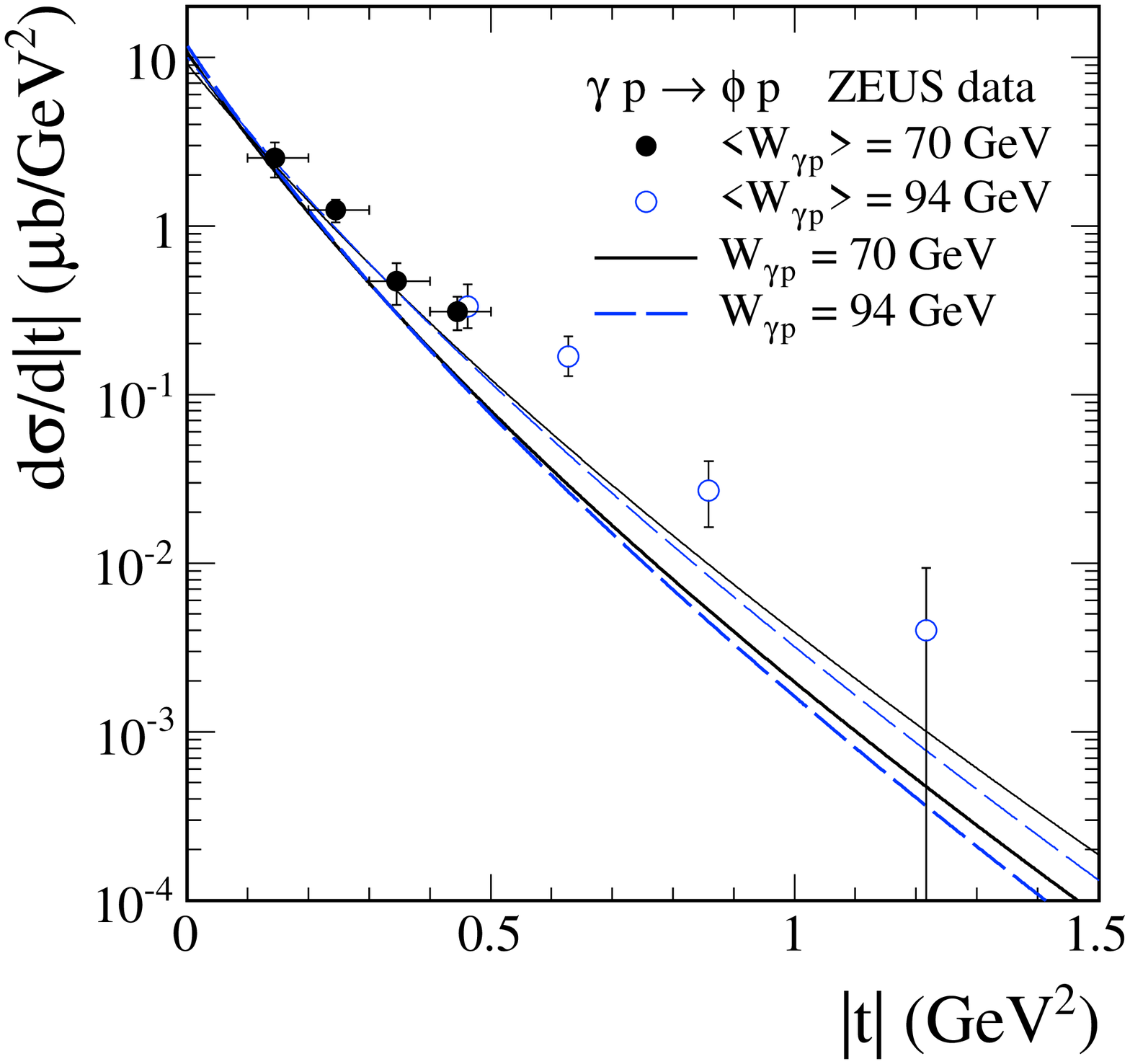}
\caption{\label{fig:photoprod_phi}
\small
Left panel:
The elastic $\phi$ photoproduction cross section
as a function of the center-of-mass energy $W_{\gamma p}$.
Our results are compared with the HERA data
\cite{Derrick:1996af} (solid mark) 
and with a compilation of low energy data from 
\cite{Ballam:1972eq,Fries:1978di,Egloff:1979mg,Aston:1980fm,
Barber:1981fj,Atkinson:1984cs,Busenitz:1989gq,Laget:2000gj} (open circles). 
Right panel: 
The differential cross section $d\sigma/d|t|$ for the 
$\gamma p \to \phi p$ process.
The ZEUS data at low $|t|$ (at $\gamma p$ average center-of-mass 
energy $\langle \sqrt{s} \rangle = 70$~GeV and the squared photon virtuality $Q^{2} = 0$~GeV$^{2}$, 
solid marks, \cite{Derrick:1996af}) 
and at higher $|t|$ (at $\langle \sqrt{s} \rangle = 94$~GeV 
and $Q^{2} < 0.01$~GeV$^{2}$, open circles, \cite{Breitweg:1999jy}) are shown.
The black solid lines represent the results for $\sqrt{s} = 70$~GeV,
while the blue dashed lines represent the results for $\sqrt{s} = 94$~GeV.
The thick lines correspond to results with $\Lambda_{0}^{2} = 0.5$~GeV$^{2}$,
$a_{\Pom \phi\phi} = 0.11$~GeV$^{-3}$, and $b_{\Pom \phi\phi} = 5.06$~GeV$^{-1}$.
The thin lines correspond to results with $\Lambda_{0}^{2} = 1$~GeV$^{2}$,
$a_{\Pom \phi\phi} = 0.49$~GeV$^{-3}$, and $b_{\Pom \phi\phi} = 4.27$~GeV$^{-1}$.
}
\end{figure}

In the high-energy small-angle approximation we get, 
using (D.19) in appendix~D of \cite{Lebiedowicz:2013ika},
\begin{eqnarray}
{\cal M}_{\lambda_{\gamma} \lambda_{b} \to \lambda_{\phi} \lambda_{2}} (s, t) &\cong
%c^{(\gamma \to \rho)}\,
&i e \dfrac{m_{\phi}^{2}}{\gamma_{\phi}}\,
\Delta_{T}^{(\phi)}(0) \,
%i\Delta^{(\rho)}_{\sigma \mu}(0)
%\epsilon_{\gamma}^{\mu} \,
%(\epsilon_{\rho}^{\nu})^* \,
(\epsilon^{(\phi)\, \mu})^*
\epsilon^{(\gamma)\, \nu} 
\left[2 a_{\Pom \phi \phi}\,\Gamma_{\mu \nu \kappa \lambda}^{(0)}(p_{\phi},-q)
- b_{\Pom \phi \phi}\,\Gamma_{\mu \nu \kappa \lambda}^{(2)}(p_{\phi},-q) \right]
%V_{\mu \nu \kappa \lambda}(s,t,q,p_{\phi})
\nonumber \\
&& \times 3 \beta_{\Pom NN}\, \frac{1}{2s} 
(- i s \alpha'_{\Pom})^{\alpha_{\Pom}(t)-1}\,
%\nonumber \\
%&& \times
(p_2+p_b)^{\kappa} (p_2+p_b)^{\lambda} \,
\delta_{\lambda_{2} \lambda_{b}} F_{1}(t) F_{M}(t)\,, \nonumber\\
\label{phip_tot_opt_aux}
\end{eqnarray}
where the explicit tensorial functions $\Gamma_{\mu \nu \kappa \lambda}^{(i)}(p_{\phi},-q)$, 
$i$ = 0, 2,
are given in Ref.~\cite{Ewerz:2013kda}, formulae (3.18) and (3.19), respectively.
In Eq.~(\ref{phip_tot_opt_aux}) 
$4 \pi/ \gamma_{\phi}^{2} =  0.0716$,
$(\Delta_{T}^{(\phi)}(0))^{-1} = -m_{\phi}^{2}$.
The form factors $F_{1}(t)$ and $F_{M}(t)$ are chosen in (\ref{phip_tot_opt_aux})
as the electromagnetic form factors (\ref{Fpion}) only for simplicity.
Here, it seems reasonable to assume rather $\Lambda_{0}^{2} \approx m_{\phi}^{2}$
than $\Lambda_{0}^{2} = 0.5$~GeV$^{2}$ from (\ref{Fpion}).
This will be discussed in Fig.~\ref{fig:photoprod_phi}.
Alternatively, we can take a common form factor
\begin{eqnarray}
F_{\phi p}^{(\Pom)}(t) = \exp \left( B_{\phi p}^{(\Pom)}\, t/2 \right) \,.
\label{Fpion_alternative}
\end{eqnarray}
with the slope parameter $B_{\phi p}^{(\Pom)}$ obtained
from comparison to the experimental data.

%The amplitude (\ref{phip_tot_opt_aux0}) is written in analogy to
%the amplitude for photoproduction of $\rho^{0}$ meson.
%The $\Pom \phi \phi$ vertex is the same as the $\Pom \rho \rho$ vertex 
%given in \cite{Ewerz:2013kda} by formula (3.47).
In order to get estimates for the $\Pom \phi \phi$ coupling constants
$a_{\Pom \phi \phi}$ and $b_{\Pom \phi \phi}$
we make the assumption based on the additive quark model 
\cite{Levin:1965mi,Lipkin:1965fu,Lipkin:1966zzc,Kokkedee:1966wx,Lipkin:1974dv}
(see also chapter II of \cite{Egloff:1979gm}): 
%that at high energies
%the total cross section for transversely polarised $\phi$ mesons
%equals to the relation:
%
\begin{eqnarray}
\sigma_{tot}(\phi(\epsilon^{(m)}), p) = 
\sigma_{tot}(K^{+}, p) + \sigma_{tot}(K^{-}, p) - \sigma_{tot}(\pi^{-}, p)
\label{phip_tot_aqm}
\end{eqnarray}
for transversely polarised $\phi$ mesons ($m = \pm 1$).
In analogy to the $\rho p$ scattering discussed in 
section~7.2 of \cite{Ewerz:2013kda}
the total cross section for the $\phi p$ scattering at high energies
is obtained from (\ref{phip_tot_opt_aux}) as
\begin{eqnarray}
\sigma_{tot}(\phi(\epsilon^{(m)}), p) = 
3 \beta_{\Pom NN} \,
\left[ 2 m_{\phi}^{2}\, a_{\Pom \phi \phi} + (1 - \delta_{m,0}) b_{\Pom \phi \phi} \right]\,
\cos\left[\frac{\pi}{2}(\alpha_{\Pom}(0) -1 )\right]
(s \alpha_{\Pom}' )^{\alpha_{\Pom}(0) -1}\,, \nonumber \\
\label{phip_tot}
\end{eqnarray}
$m = \pm 1, 0$.
With the pomeron parts of the $Kp$ and $\pi p$ 
total cross sections from (\ref{Kp_Kp_tot}) 
above and (7.6) of \cite{Ewerz:2013kda}, respectively,
we obtain from (\ref{phip_tot_aqm}) and (\ref{phip_tot})
\begin{eqnarray}
3 \beta_{\Pom NN} \,
\left( 2 m_{\phi}^{2} \,a_{\Pom \phi \phi} + b_{\Pom \phi \phi} \right)
= 12 \beta_{\Pom NN} (2 \beta_{\Pom KK} - \beta_{\Pom \pi\pi})\,.
\label{phip_tot_comp}
\end{eqnarray}
From (\ref{phip_tot_comp}), (\ref{Kp_couplings_pom}) and 
$\beta_{\Pom \pi \pi} = 1.76$~GeV$^{-1}$ we get
\begin{eqnarray}
2 m_{\phi}^{2} \,a_{\Pom \phi \phi} + b_{\Pom \phi \phi}
= 4 (2 \beta_{\Pom KK} - \beta_{\Pom \pi\pi}) = 5.28\, \mathrm{GeV}^{-1}\,.
\label{phip_tot_comp_aux}
\end{eqnarray}

In the left panel of Fig.~\ref{fig:photoprod_phi} we show
the integrated cross section for the $\gamma p \to \phi p$ reaction,
calculated from (\ref{phip_tot_opt_aux0}),
as a function of the center-of-mass energy together with the experimental data.
The experimental point at $W_{\gamma p} = 70$~GeV
was obtained by extrapolating the differential cross section to
$t=0$ assuming a simple exponential $t$ dependence 
and integrating over the range $|t| < 0.5$~GeV \cite{Derrick:1996af}.
In our calculation we also integrate over the same $t$ range.
We see that our model calculation including only the pomeron exchange
describes the total cross section for the $\gamma p \to \phi p$ reaction fairly well.
\footnote{
At low energies there are other processes contributing, such as
$t$-channel $\pi^{0}$, $\eta$, $\sigma$ meson exchanges; 
see \cite{Titov:2003bk,Kong:2016scm}.
%the $\phi$~bremsstrahlung, baryonic resonances decaying into the $\phi p$ channel 
Thus, the pomeron exchange alone should not be expected 
to fit the low-energy data precisely.
We refer the reader to \cite{Dey:2014tfa,Mizutani:2017wpg}
for measurements of the $\gamma p \to \phi p$ reaction near threshold.
}

The right panel of Fig.~\ref{fig:photoprod_phi} 
shows the differential cross section for elastic $\phi$ photoproduction.
The calculations, performed for two energies, 
$W_{\gamma p} = 70$~GeV and 94~GeV,
are compared with ZEUS data, \cite{Derrick:1996af} and \cite{Breitweg:1999jy}, respectively.
We show results for two parameters of the form factor $F_{M}(t)$,
$\Lambda_{0}^{2} = 0.5$~GeV$^{2}$ and 1~GeV$^{2}$,
represented by the bottom lines and the top lines, respectively.
We can see that the results for $\Lambda_{0}^{2} = 1$~GeV$^{2}$ 
with the relevant values of coupling constants $a$ and $b$
describe more accurately the slope of the $t$ distribution.
%--------------------------------------------------------
%--------------------------------------------------------

%--------------------------------------
\subsection{$\phi(1020)$ meson central production}
\label{sec:photoprod_phi}
%--------------------------------------

The $\phi$ photoproduction is dominated by diffractive
scattering via pomeron exchange.
The amplitude for the $\gamma \Pom$-exchange,
see diagram~(a) in Fig.~\ref{fig:gampom_pomgam_s},
reads as (3.5) of \cite{Lebiedowicz:2014bea} with appropriate modifications:
\begin{eqnarray}
&&{\cal M}^{(\gamma \Pom \to \phi \to K^{+}K^{-})}_{\lambda_{a} \lambda_{b} \to \lambda_{1} \lambda_{2} K^{+}K^{-}} 
= (-i)
\bar{u}(p_{1}, \lambda_{1}) 
i\Gamma^{(\gamma pp)}_{\mu}(p_{1},p_{a}) 
u(p_{a}, \lambda_{a}) \nonumber \\
&&\qquad \times  
i\Delta^{(\gamma)\,\mu \sigma}(q_{1})\, 
i\Gamma^{(\gamma \to \phi)}_{\sigma \nu}(q_{1})\,
%c^{(\gamma \to \rho)} \,
i\Delta^{(\phi)\,\nu \rho_{1}}(q_{1}) \,
i\Delta^{(\phi)\,\rho_{2} \kappa}(p_{34})\,
i\Gamma^{(\phi KK)}_{\kappa}(p_{3},p_{4})
\nonumber \\
&& \qquad \times 
i\Gamma^{(\Pom \phi \phi)}_{\rho_{2} \rho_{1} \alpha \beta}(p_{34},q_{1})\, 
i\Delta^{(\Pom)\,\alpha \beta, \delta \eta}(s_{2},t_{2}) \,
\bar{u}(p_{2}, \lambda_{2}) 
i\Gamma^{(\Pom pp)}_{\delta \eta}(p_{2},p_{b}) 
u(p_{b}, \lambda_{b}) \,.
\label{amplitude_gamma_pomeron}
\end{eqnarray}

Here we use the $\phi$ propagator with the simple Breit-Wigner expression
as defined in (3.7), (3.10) and (3.11) of \cite{Lebiedowicz:2014bea} with
\begin{eqnarray}
\Delta_T^{(\phi)}(s) = 
\frac{1}{s - m_{\phi}^2 + i \sqrt{s} \Gamma_{\phi}(s)} \,.
\label{phi_propagator}
\end{eqnarray}
In (\ref{phi_propagator})
the running (energy-dependent) width is approximately parametrized as
\begin{eqnarray}
\Gamma_{\phi}(s) = \Gamma_{\phi}
\left( \frac{s - 4 m_{K}^2}{m_{\phi}^2 - 4 m_{K}^2} \right)^{3/2}
\frac{m_{\phi}^2}{s} \, \theta(s-4 m_{K}^2)\,.
\end{eqnarray}
A more accurate parametrization of $\Delta_T^{(\phi)}(s)$ and $\Gamma_{\phi}(s)$
must take into account also non $K \overline{K}$ decay channels
of the $\phi$, in particular, the $3 \pi$ decays which amount to
$(15.32 \pm 0.32) \%$ of all decays, see \cite{Olive:2016xmw}.
For such a program one could use the methods explained for the $\rho$ propagator
in \cite{Melikhov:2003hs}; see also \cite{Ewerz:2013kda,Bolz:2014mya}.

For the $\phi KK$ vertex we have
\begin{eqnarray}
i\Gamma^{(\phi KK)}_{\kappa}(p_{3},p_{4}) = -\frac{i}{2} \,g_{\phi K^{+} K^{-}} 
(p_{3} - p_{4})_{\kappa}\,.
\label{Gamma_phiKK}
\end{eqnarray}
The $g_{\phi K^{+} K^{-}}$ coupling constant 
can be determined from the partial decay width $\Gamma(\phi \to K^+ K^-)$,
\begin{eqnarray}
\Gamma(\phi \to K^{+} K^{-}) = \frac{1}{24 \pi} 
\frac{p_{K}^3}{m_{\phi}^{2}} |g_{\phi K^{+} K^{-}}|^2 \,,
\label{partial_decay_width}
\end{eqnarray}
where $p_{K} = \sqrt{m_{\phi}^2/4 - m_{K}^2}$.
With the parameters of Table~\ref{table:table},
assuming $g_{\phi K^{+} K^{-}} > 0$, we get
\begin{eqnarray}
g_{\phi K^{+} K^{-}} = 8.92 \,.
\label{g_phi_1020KK}
\end{eqnarray}

In the diagram of Fig.~\ref{fig:gampom_pomgam_s}
at the $\Pom \phi \phi$ vertex the incoming $\phi$ is always off shell,
the outgoing $\phi$ also may be away from the nominal 
``mass shell'' $p_{34}^{2} = m_{\phi}^{2}$.
As suggested in \cite{Bolz:2014mya}, see (B.82) there,
we insert, therefore, in the $\Pom \phi \phi$ vertex extra form factors.
A convenient form, given in (B.85) of \cite{Bolz:2014mya}
(see also (3.9) of \cite{Lebiedowicz:2014bea}) is
\begin{eqnarray}
\tilde{F}^{(\phi)}(k^{2}) 
%= \left[1 + \dfrac{|k^{2}(k^{2}-m_{\rho}^{2})|}{\Lambda_{\rho}^{4}}  \right]^{-n_{\rho}}
= \left[1 + \dfrac{k^{2}(k^{2}-m_{\phi}^{2})}{\tilde{\Lambda}_{\phi}^{4}} \right]^{-\tilde{n}_{\phi}}
\label{ff_Nachtmann}
\end{eqnarray}
%
%with $\Lambda_{\rho}$ a parameter in the range 1 to 2~GeV and $n_{\rho} > 0$.
with $\tilde{\Lambda}_{\phi}$ a parameter close to 2~GeV and $\tilde{n}_{\phi} > 0$.
%In Fig.~8 of \cite{Lebiedowicz:2014bea}
%it was shown the influence of parameters in (\ref{ff_Nachtmann}) 
%on the two-pion invariant mass distribution for the $\rho(770)$ photoproduction.
In practical calculations we also include in (\ref{Gamma_phiKK})
the form factor
\begin{eqnarray}
F^{(\phi K K)}(p_{34}^{2}) = 
\exp{ \left( \frac{-(p_{34}^{2}-m_{\phi}^{2})^{2}}{\Lambda_{\phi}^{4}} \right)}\,,
\quad \Lambda_{\phi} = 1\;{\rm GeV}\,.
\label{FphiKK_ff}
\end{eqnarray}
%

%--------------------------------------
\section{Results}
\label{sec:results}
%--------------------------------------

%~~~~~~~~~~~~~~~~~~~~~~~~~~~~~~~~~~~~~~~~~~~~~~~~~~~~~~~~~~~~~~~~~~~~~
\begin{table}[!h]
\caption{Some parameters of our model.
The columns indicate the equation numbers where
the parameter is defined and their numerical values used in the calculations.}
\begin{tabular}{|l|l|l|}
\hline
Parameter for			& Equation 	& Value\\ \hline \hline
\textbf{diffractive continuum}	& &  \\ \hline  
$\Lambda_{off,M}$ &	 (\ref{off-shell_form_factors_mon}) & 0.7~GeV \\ 
$\Lambda_{0}^{2}$ & (\ref{Fpion}); (3.34) of \cite{Ewerz:2013kda} & 0.5~GeV$^{2}$ \\	 
$M_{0}$ & (\ref{A6})-(\ref{A7b}), (\ref{Kp_Kp_el})\; et seq. & 1~GeV \\
$M_{-}$ & (\ref{A8}), (\ref{Kp_Kp_el})\; et seq.; (6.63) of \cite{Ewerz:2013kda} & 1.41~GeV \\ \hline \hline
\boldmath{$f_{0}(980)$}	& & 	\\ \hline 
$g'_{\Pom \Pom f_{0}}$ & (\ref{vertex_pompomS}) et seq.; (A.18) of \cite{Lebiedowicz:2013ika} & 0.53 \\ 
$g''_{\Pom \Pom f_{0}}$ & (\ref{vertex_pompomS}) et seq.; (A.20) of \cite{Lebiedowicz:2013ika} & 2.67 \\ 
$g_{f_{0} K^{+}K^{-}}$ & (\ref{gamma_f0KK}), (\ref{g_f0980_KK}) & 3.48 \\
$\Lambda_{f_{0}}$ &	(\ref{Fpompommeson_ff}) 		& 1~GeV \\ \hline \hline
%\boldmath{$f_{0}(1370)$}	& & \\ \hline 
%$g'_{\Pom \Pom f_{0}}$ & (\ref{vertex_pompomS}) et seq.; (A.18) of \cite{Lebiedowicz:2013ika} & 0.81/2.0  \\ 
%$g_{f_{0} K^{+}K^{-}}$ & (\ref{gamma_f0KK}) & 2.47; Eq.~(\ref{g_f01370_KK}) \\ \hline \hline
\boldmath{$f_{0}(1500)$}	& & \\ \hline 
$g'_{\Pom \Pom f_{0}}$ & (\ref{vertex_pompomS}) et seq.; (A.18) of \cite{Lebiedowicz:2013ika} & 0.35 \\ 
$g''_{\Pom \Pom f_{0}}$ & (\ref{vertex_pompomS}) et seq.; (A.20) of \cite{Lebiedowicz:2013ika} & 1.71 \\ 
$g_{f_{0} K^{+}K^{-}}$ & (\ref{gamma_f0KK}), (\ref{g_f01500_KK}) & 0.69 \\
$\Lambda_{f_{0}}$ &	(\ref{Fpompommeson_ff}) 		& 1~GeV \\ \hline \hline
\boldmath{$f_{0}(1710)$}	& & \\ \hline 
$g'_{\Pom \Pom f_{0}}$ & (\ref{vertex_pompomS}) et seq.; (A.18) of \cite{Lebiedowicz:2013ika} & 0.23 	\\ 
$g''_{\Pom \Pom f_{0}}$ & (\ref{vertex_pompomS}) et seq.; (A.20) of \cite{Lebiedowicz:2013ika} & 1.3 	\\ 
$g_{f_{0} K^{+}K^{-}}$ & (\ref{gamma_f0KK}), (\ref{g_f01710_KK}) & 2.36  	\\
$\Lambda_{f_{0}}$ &	(\ref{Fpompommeson_ff}) 		& 1~GeV \\ \hline \hline
\boldmath{$f_{2}(1270)$}	& & 	\\ \hline 
$g_{\Pom \Pom f_{2}}^{(2)}$ & (\ref{vertex_pompomT}) et seq.; (A.13) of \cite{Lebiedowicz:2016ioh} & 9.0 	\\ 
$g_{f_{2} K^{+}K^{-}}$ & (\ref{gamma_f2KK}), (\ref{g_f21270_KK}) & 5.54 	\\
$\Lambda_{f_{2}}$ &	(\ref{Fpompommeson_ff_tensor}) 		& 1~GeV  	\\ \hline \hline
\boldmath{$f'_{2}(1525)$}& & 	\\ \hline 
$g_{\Pom \Pom f_{2}}^{(2)}$ & (\ref{vertex_pompomT}) et seq.; (A.13) of \cite{Lebiedowicz:2016ioh} & 2.0 	\\ 
$g_{f_{2} K^{+}K^{-}}$ & (\ref{gamma_f2KK}), (\ref{g_f21525_KK}) & 7.32 	\\
$\Lambda_{f_{2}}$ &	(\ref{Fpompommeson_ff_tensor}) 		& 1~GeV  	\\ \hline \hline
\textbf{photoproduction continuum}	& &  	\\ \hline 
$\Lambda_{K}$ &	(\ref{ff_Poppe}) - (\ref{ff_Poppe_aux})  & 1~GeV  	\\ \hline \hline
\boldmath{$\phi(1020)$}	& &  	\\ \hline 
$a_{\Pom \phi \phi}$ & (\ref{phip_tot_comp_aux}) et seq. & 0.49~GeV$^{-3}$   \\
$b_{\Pom \phi \phi}$ & (\ref{phip_tot_comp_aux}) et seq. & 4.27~GeV$^{-1}$   \\
$\Lambda_{0}^{2}$ & (\ref{Fpion}) & 1~GeV$^{2}$  	\\
$g_{\phi K^{+}K^{-}}$ & (\ref{Gamma_phiKK})-(\ref{g_phi_1020KK}) & 8.92 	\\ 
$\tilde{\Lambda}_{\phi}$ & (\ref{ff_Nachtmann}); (3.9) of \cite{Lebiedowicz:2014bea} 		& 2~GeV  	\\ 
$\tilde{n}_{\phi}$       & (\ref{ff_Nachtmann}); (3.9) of \cite{Lebiedowicz:2014bea} 		& 0.5  \\ 
$\Lambda_{\phi}$ & (\ref{FphiKK_ff}) 		& 1~GeV  	\\ \hline
\end{tabular}
\label{table:parameters}
\end{table}
%~~~~~~~~~~~~~~~~~~~~~~~~~~~~~~~~~~~~~~~~~~~~~~~~~~~~~~~~~~~~~~~~~~~~~

In this section we present results for 
integrated cross sections of the reaction $pp \to pp K^{+} K^{-}$
and dikaon invariant mass distributions.
For convenience of the reader we collect in Table~\ref{table:parameters}
the numerical values of default parameters of our model used in calculations.
There are also the parameters of pomeron/reggeon-kaon couplings,
see (\ref{Kp_couplings_pom}) and (\ref{Kp_couplings_reg}),
not shown in Table~\ref{table:parameters},
obtained from fits to kaon-nucleon total cross-section
data as discussed in section~\ref{sec:diff_continuum}.
Our attempts to determine the parameters of pomeron-pomeron-meson couplings
as far as possible from experimental data have been presented 
in sections~\ref{sec:diff_f0} and \ref{sec:diff_f2},
and in Refs.~\cite{Lebiedowicz:2013ika,Lebiedowicz:2016ioh}.
Note that we take here somewhat smaller values of the pomeron-pomeron-meson coupling parameters
than in our previous paper \cite{Lebiedowicz:2013ika}
because there they were fixed at the WA102 energy where we expect
also large contributions to the cross sections from the reggeon exchanges.
We have checked for the central $K^{+}K^{-}$ continuum contribution
calculated at $\sqrt{s} = 13$~TeV and 
for three different cuts on pseudorapidities
$|\eta_{K}| < 1$, $|\eta_{K}| < 2.5$, and $2 < \eta_{K} < 4.5$,
that adding the exchange of secondary reggeons
increases the cross section by 2.4 \%, 2.9 \%, and 6.5 \%, respectively.
We expect a similar role of secondary reggeons for production of resonances at $\sqrt{s} = 13$~TeV.
For continuum $K^{+}K^{-}$ photoproduction we find even less effect
on the cross sections from secondary reggeons than for the purely diffractive production above.
Recently, in Ref.~\cite{Lebiedowicz:2018sdt}, we also discussed 
the role of reggeons for the $pp \to pp p \bar{p}$ reaction.

Many of the parameters listed in Table~\ref{table:parameters}
were obtained from fits to available data
but they are still rather uncertain and some are only our educated guess.
Clearly, it would be desirable to experimentally test 
our predictions obtained with our default parameters
and then adjust these if necessary.
Such an adjustment of the model parameters
will be possible with high-energy experimental data 
for the purely exclusive reactions 
$pp \to pp \pi^{+} \pi^{-}$ and $pp \to pp K^{+} K^{-}$ 
which are expected to become available soon. 
The $\mathtt{GenEx}$ Monte Carlo generator \cite{Kycia:2014hea} 
could be used in this context.

\begin{figure}[!ht]
\includegraphics[width=0.49\textwidth]{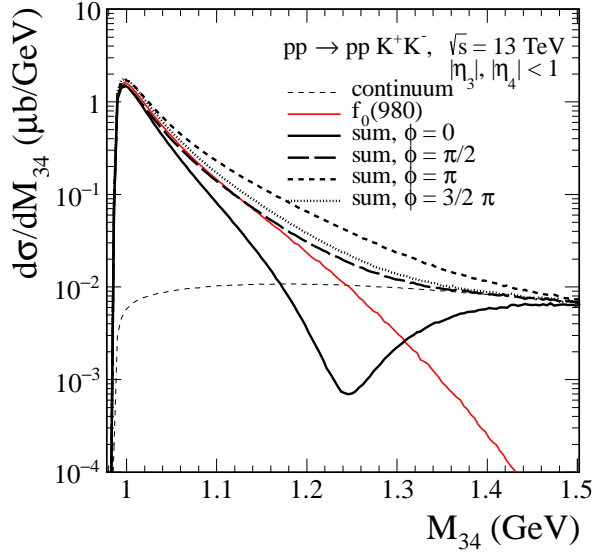}
  \caption{\label{fig:M34_f0980}
  \small
Two-kaon invariant mass distribution for $pp \to pp K^{+} K^{-}$ 
at $\sqrt{s} = 13$~TeV and $|\eta_{K}| < 1$.
Absorption effects were included here by the gap survival factor 
$\langle S^{2} \rangle = 0.1$.
Results for a coherent sum of non-resonant continuum and $f_{0}(980)$ meson 
for different values of the phase factor $\phi_{f_{0}(980)}$
in (\ref{gamma_f0KK_phase_factor}) are shown for illustration.
The short-dashed line represents the non-resonant continuum contribution
obtained for the monopole off-shell kaon form factors (\ref{off-shell_form_factors_mon})
with $\Lambda_{off,M} = 0.7$~GeV.
The red solid line represents the $f_{0}(980)$ resonant term.
}
\end{figure}

In Fig.~\ref{fig:M34_f0980} we present the $K^{+} K^{-}$ invariant mass distribution
at $\sqrt{s} = 13$~TeV and $|\eta_{K}| < 1$.
Here we take into account the non-resonant continuum, 
including both pomeron and reggeon exchanges,
and the scalar $f_{0}(980)$ resonance created here only by the pomeron-pomeron fusion.
We show results for different values of 
the relative phase $\phi_{f_{0}(980)}$ 
in the coupling constant (\ref{g_f0980_KK})
not known \textit{a priori} 
\begin{eqnarray}
g_{f_{0}(980) K^{+} K^{-}} \to g_{f_{0}(980) K^{+} K^{-}} \,e^{i \phi_{f_{0}(980)}} \,.
\label{gamma_f0KK_phase_factor}
\end{eqnarray}
We can see that the complete result indicates a large interference 
effect of the continuum and the $f_{0}(980)$ terms.
It should be recalled that the $f_{0}(980)$ resonance appears 
as a sharp drop around the 1~GeV region in the $\pi^{+} \pi^{-}$ mass spectrum.
The black solid line corresponds to the calculations 
with the phase used for $\pi^{+} \pi^{-}$ exclusive production.
The phase for $K^{+} K^{-}$ does not need to be the same
as the production of $\pi \pi$ and $K \bar{K}$ systems may be
a complicated coupled-channel effect not treated here explicitly.
In some of the following figures
we show predictions for two representative values for this phase,
$\phi_{f_{0}(980)} = 0$ and $\pi/2$.
We must leave it to the experiments to determine this phase from data.
%-------------------------------------------------------
%--------------------------------------------------------

\begin{figure}[!ht]
\includegraphics[width=0.49\textwidth]{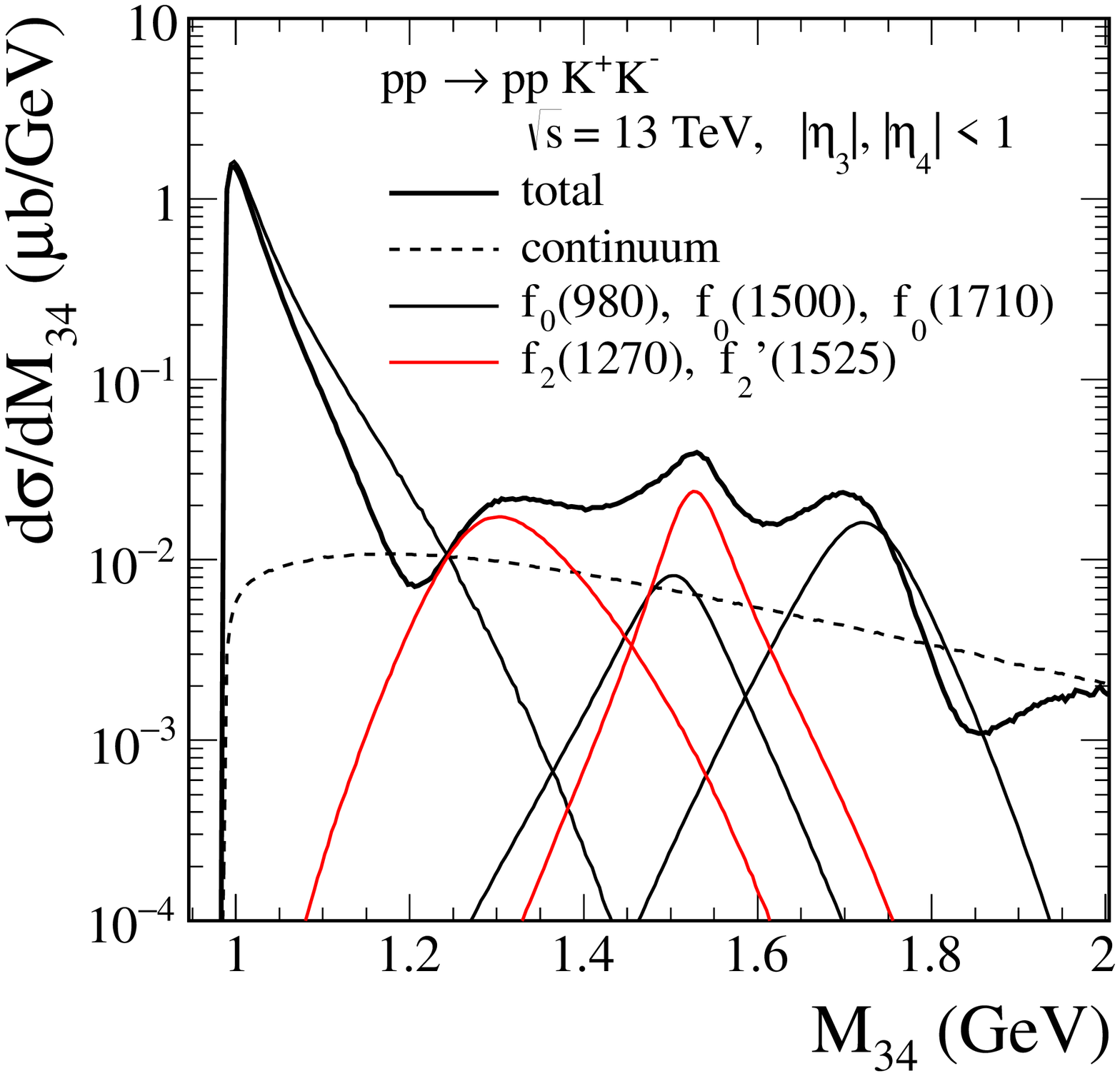}
\includegraphics[width=0.49\textwidth]{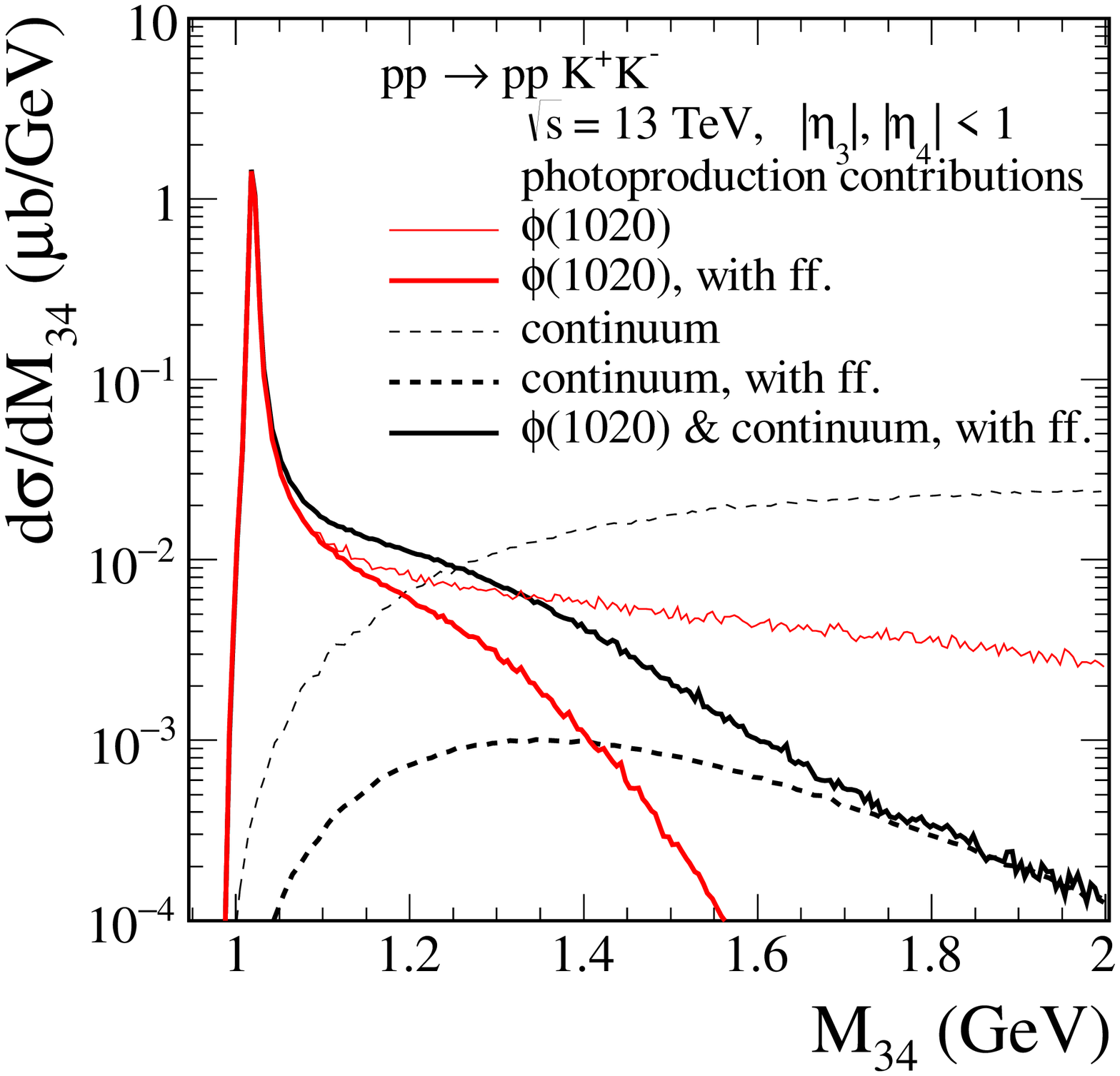}
  \caption{\label{fig:M34_deco}
  \small
Two-kaon invariant mass distributions for $pp \to pp K^{+} K^{-}$ 
at $\sqrt{s} = 13$~TeV and $|\eta_{K}| < 1$.
The left panel shows the purely diffractive contributions and
the right panel shows the diffractive photoproduction results.
In the left panel the black line represents a coherent sum of non-resonant
continuum, and $f_{0}(980)$, $f_{0}(1500)$, $f_{0}(1710)$, $f_{2}(1270)$
and $f'_{2}(1525)$ resonant terms.
For the $f_{0}(980)$ resonant term we take $\phi_{f_{0}(980)} = 0$ 
in (\ref{gamma_f0KK_phase_factor}).
The dashed line represents the non-resonant continuum contribution.
The individual resonance contributions are shown for better understanding.
In the right panel the photoproduction terms are shown without
and with some form factors included in the amplitudes.
The lower lines correspond to results 
for the $\phi(1020)$ photoproduction with form factor (\ref{FphiKK_ff})
and for the non-resonant term with form factors (\ref{ff_Poppe}) - (\ref{ff_Poppe_aux}).
Absorption effects have been included here by the gap survival factors
$\langle S^{2} \rangle = 0.1$ for the diffractive contributions
and $\langle S^{2} \rangle = 0.9$ for the photoproduction contributions.}
\end{figure}

As can be clearly seen from the left panel of Fig.~\ref{fig:M34_deco} 
the resonance contributions generate a highly structured pattern.
In the calculations we include
the non-resonant continuum, and the dominant scalar $f_{0}(980)$, $f_{0}(1500)$, $f_{0}(1710)$,
and tensor $f_{2}(1270)$, $f'_{2}(1525)$, resonances decaying
into the $K^{+} K^{-}$ pairs.
In principle, there may also be a contribution from the broad scalar $f_{0}(1370)$ meson.
%The existence of $f_{0}(1370)$ resonance
%has been considered to be the crucial
%evidence for the supernumery nature of scalars in this mass
%region, and therefore of the existence of a scalar glueball,
%albeit mixed with other scalars.
The right panel of Fig.~\ref{fig:M34_deco} shows the photoproduction contributions
without and with some form factors included in the amplitudes.
The narrow $\phi(1020)$ resonance is visible above the continuum term.
It may in principle also be visible on top of the broader $f_{0}(980)$ resonance.
This will be discussed in Fig.~\ref{fig:dsig_dM34_LHC}.
%-------------------------------------------------------
%--------------------------------------------------------

\begin{figure}[!ht]
\includegraphics[width=0.49\textwidth]{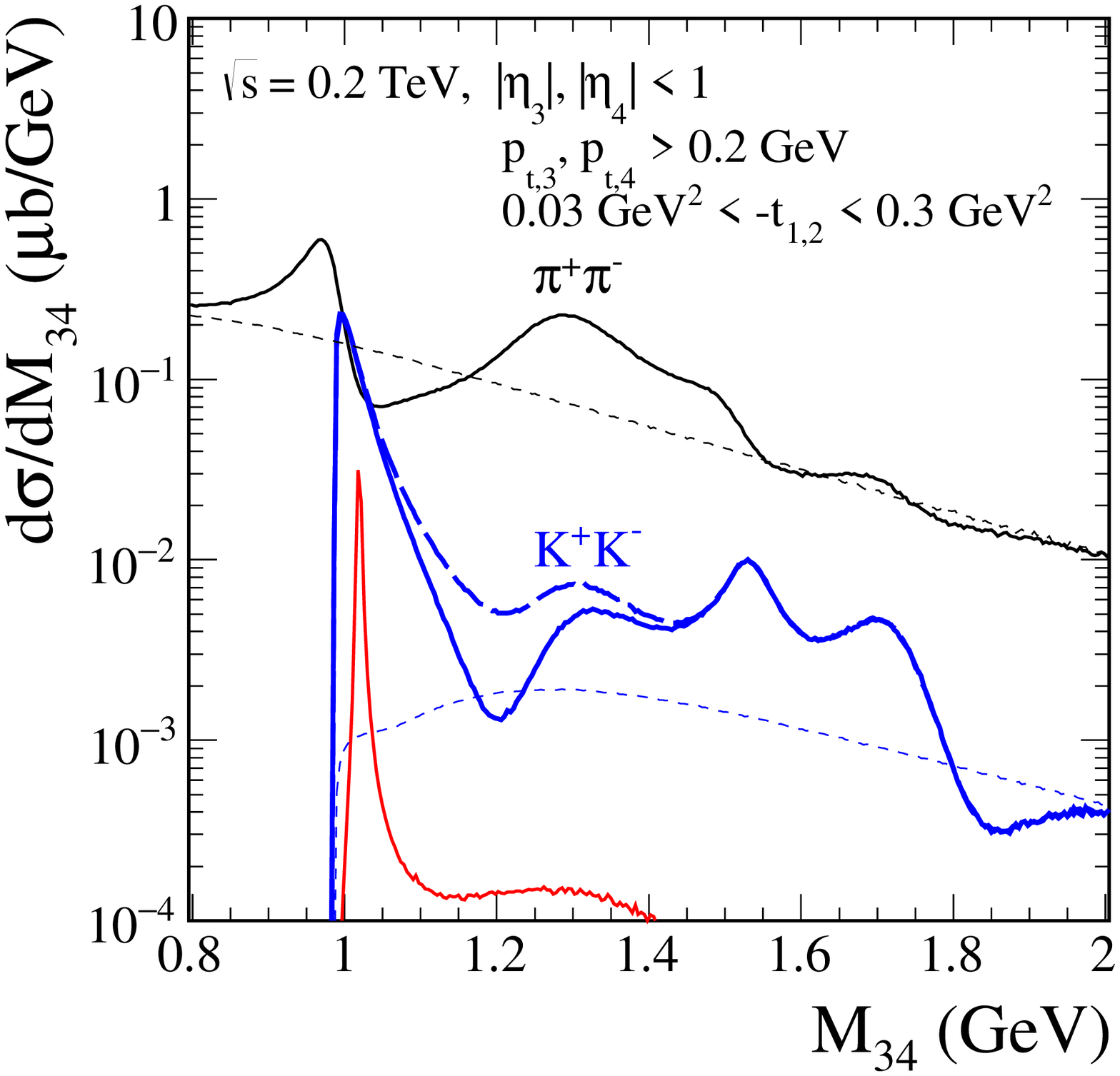}
\includegraphics[width=0.49\textwidth]{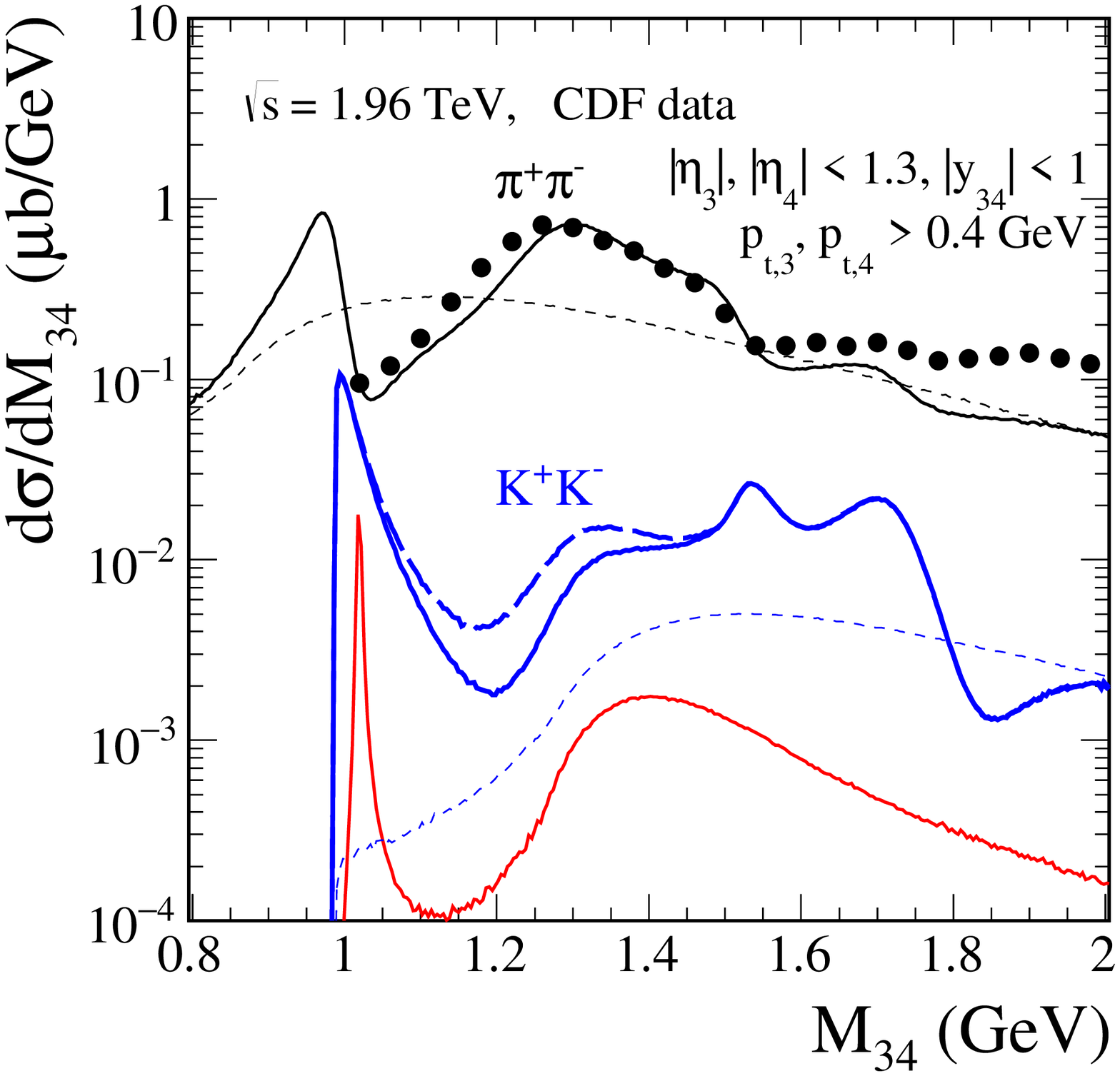}
  \caption{\label{fig:dsig_dM34}
  \small
The invariant mass distributions for
centrally produced $\pi^{+} \pi^{-}$ (the black top lines) 
and $K^{+} K^{-}$ (the blue bottom lines) pairs
with the relevant experimental kinematical cuts specified in the legend.
Results including both the non-resonant continuum and the resonances are presented.
The short-dashed lines represent the purely diffractive continuum term.
The solid and long-dashed blue lines correspond to 
the results for $\phi_{f_{0}(980)} = 0$ and $\pi/2$ in (\ref{gamma_f0KK_phase_factor}), respectively.
The lower red line represents the $\phi(1020)$ meson 
plus continuum photoproduction contribution.
The CDF experimental data from \cite{Aaltonen:2015uva} in the right panel
for the $p\bar{p} \to p\bar{p} \pi^{+} \pi^{-}$ reaction are shown for comparison.
Absorption effects were taken into account effectively by the gap survival factors.}
\end{figure}

\begin{figure}[!ht]
\includegraphics[width=0.49\textwidth]{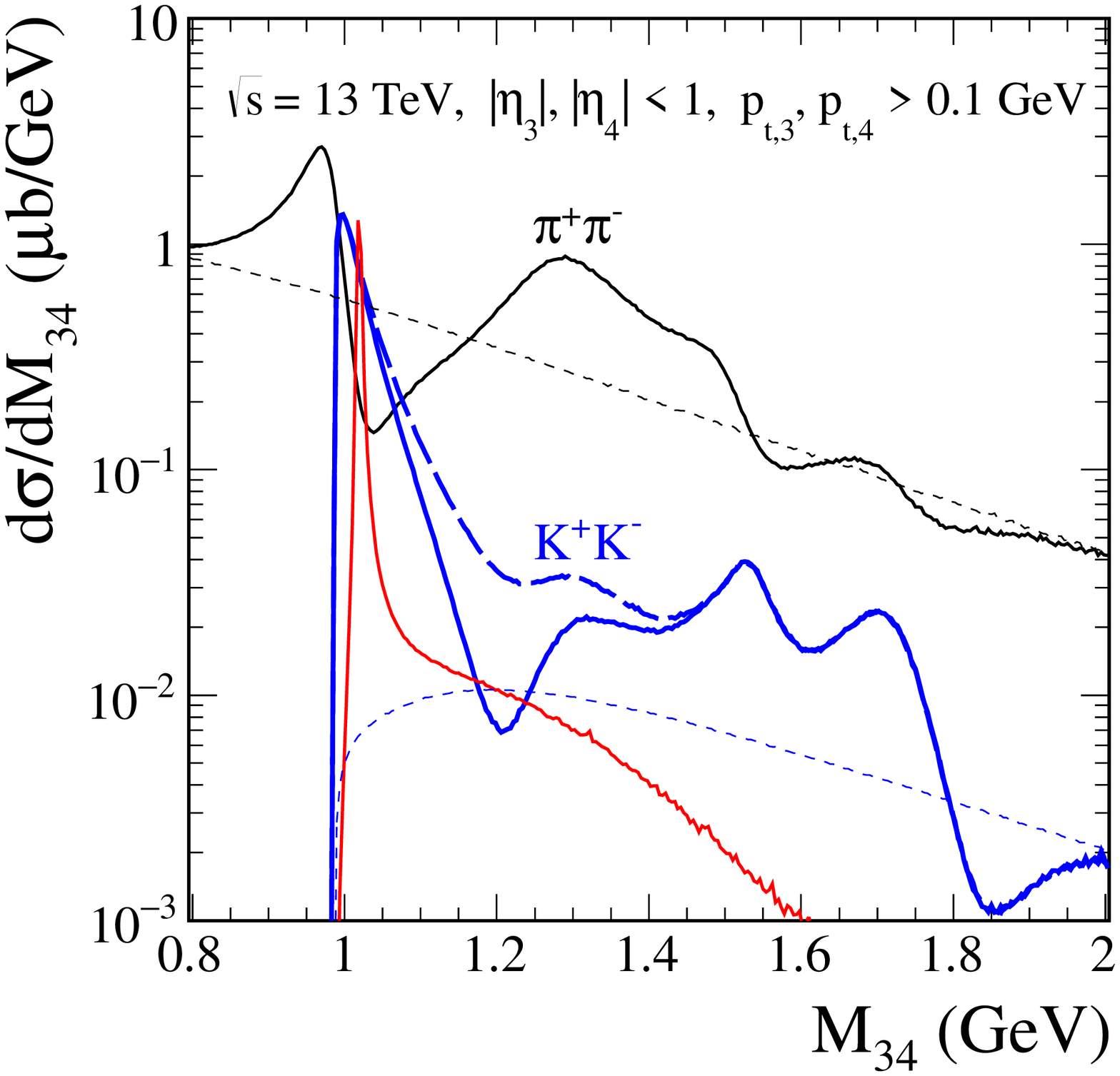}
\includegraphics[width=0.49\textwidth]{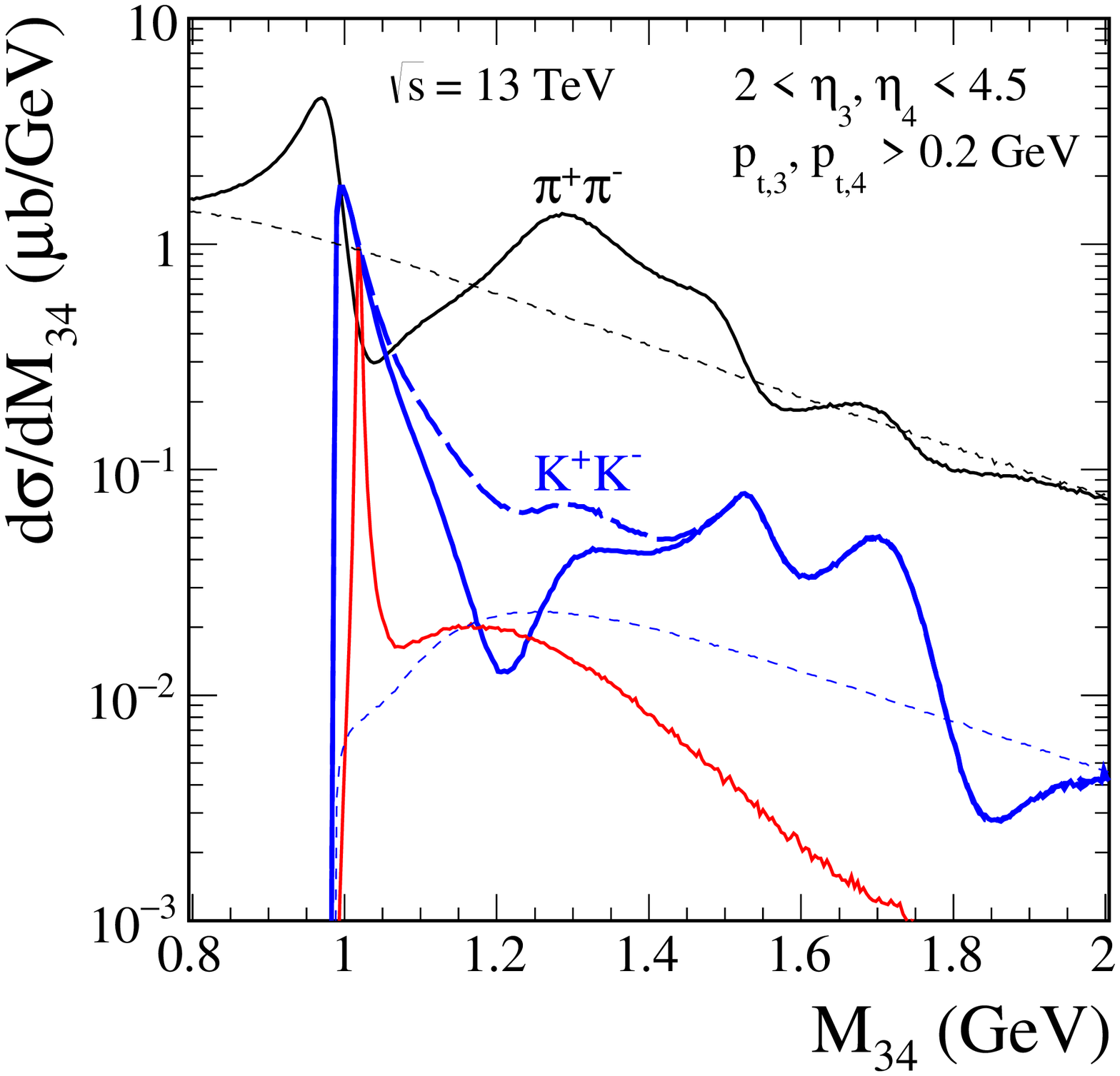}\\
\includegraphics[width=0.49\textwidth]{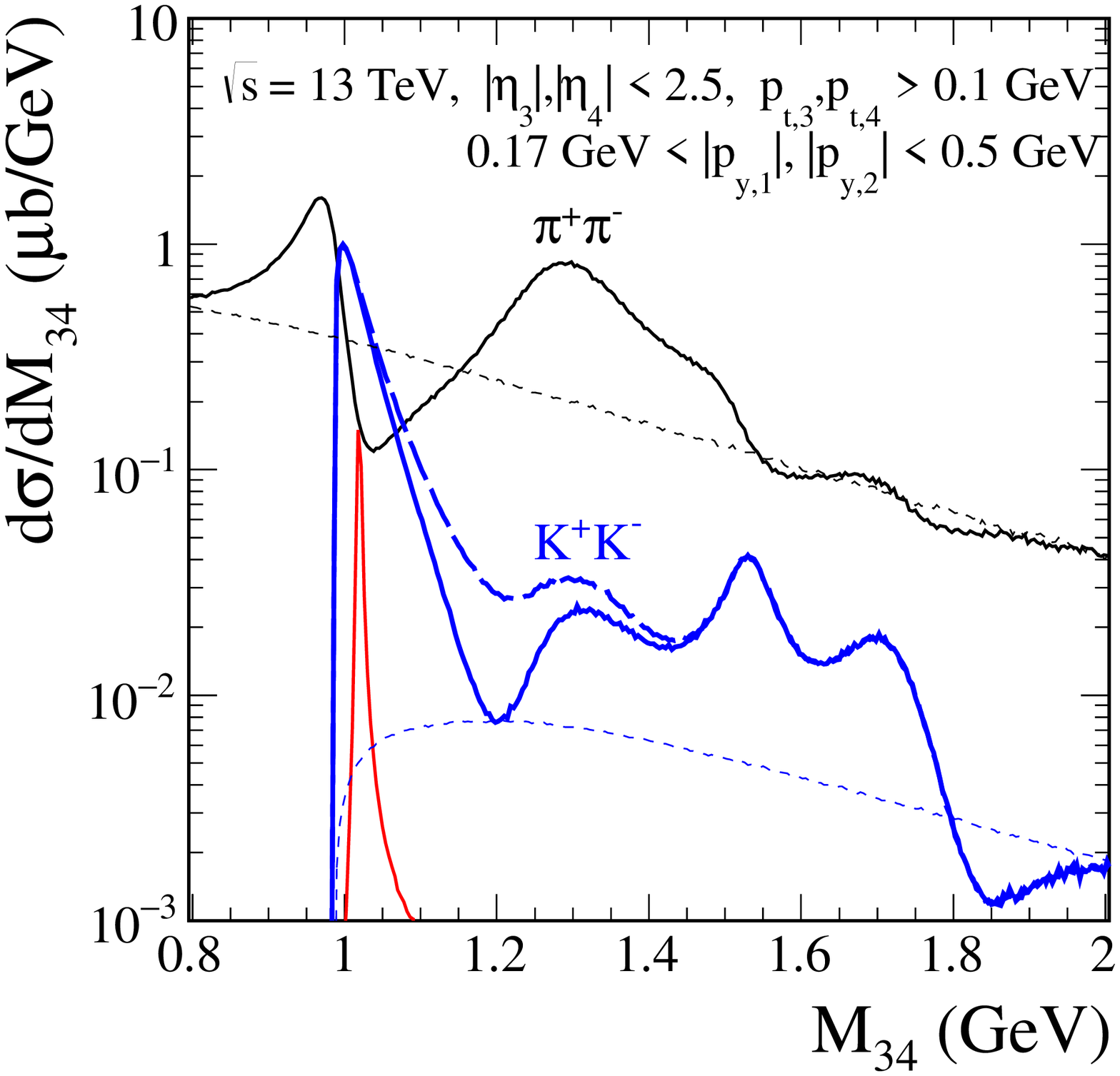}
  \caption{\label{fig:dsig_dM34_LHC}
  \small
The same as in Fig.~\ref{fig:dsig_dM34} but for $\sqrt{s} = 13$~TeV
and different experimental cuts specified in the legend.
Absorption effects were taken into account effectively by the gap survival factors,
$\langle S^{2} \rangle = 0.1$ for the purely diffractive contribution
and $\langle S^{2} \rangle = 0.9$ for the photoproduction contribution.}
\end{figure}

In Figs.~\ref{fig:dsig_dM34} and 
\ref{fig:dsig_dM34_LHC} we show the invariant mass distributions 
for centrally produced $\pi^{+} \pi^{-}$ (the black lines) 
and $K^{+} K^{-}$ (the blue lines) pairs
imposing cuts on pseudorapidities and transverse momenta of produced particles
that will be measured in the RHIC and LHC experiments.
The $pp \to pp \pi^{+} \pi^{-}$ reaction was discussed 
within the tensor-pomeron model in \cite{Lebiedowicz:2016ioh}.
The short-dashed lines represent the purely diffractive continuum term.
The solid lines represent the coherent sum of the diffractive
continuum, and the scalar $f_{0}(980)$, $f_{0}(1500)$, $f_{0}(1710)$,
and tensor $f_{2}(1270)$, $f'_{2}(1525)$ resonances.
For the $pp \to ppK^{+}K^{-}$ reaction we show 
predictions for $\phi_{f_{0}(980)} = 0$ and $\pi/2$
in (\ref{gamma_f0KK_phase_factor}),
see the solid and long-dashed blue lines, respectively.
The $f_{0}(980)$ resonance term in the $pp \to ppK^{+} K^{-}$ reaction
is calculated with the upper limit for the coupling, 
$g_{f_{0}(980) K^{+} K^{-}} = 3.48$; see (\ref{g_f0980_KK}).
The lower red lines show the photoproduction contributions.
The diffractive and photoproduction contributions to 
$K^{+}K^{-}$ production must be added coherently at the amplitude level 
and in principle could interfere.
However, this requires the inclusion of absorption effects (at the amplitude level) 
that are different for both classes of processes, see e.g. \cite{Lebiedowicz:2014bea}.
In \cite{Lebiedowicz:2016ioh} we found that for the reaction $pp \to pp \pi^{+} \pi^{-}$ 
a similar interference effect is below 1\%.
The reader is asked to note different shapes of the $\pi^{+} \pi^{-}$
and $K^{+} K^{-}$ invariant mass distributions for different experimental setups.
In the left panel of Fig.~\ref{fig:dsig_dM34} we show distributions for the STAR experiment.
In the right panel we show results for the CDF experimental conditions 
together with data for the $p\bar{p} \to p\bar{p} \pi^{+} \pi^{-}$ reaction \cite{Aaltonen:2015uva}.
The limited CDF acceptance, in particular the $p_{t} > 0.4$~GeV condition
on centrally produced $K^{+}$ and $K^{-}$ mesons, 
causes a reduction of the cross sections in the region $M_{34} < 1.3$~GeV;
see e.g. the clearly visible minimum for the photoproduction term there.

The calculations were done at Born level and the absorption
corrections were taken into account by multiplying
the cross section for the corresponding collision energy
by a common factor $\langle S^{2} \rangle$
obtained from \cite{Lebiedowicz:2015eka} and \cite{Lebiedowicz:2011tp}.
For the purely diffractive contribution the gap survival factors
$\langle S^{2} \rangle = 0.2$ and 0.1
for $\sqrt{s} = 0.2$~TeV and $1.96$~TeV, respectively, were taken.
For the photoproduction contribution the Born calculation
was multiplied by the factor $\langle S^{2} \rangle = 0.9$; 
see \cite{Lebiedowicz:2014bea}.
The absorption effects lead to a huge damping of the cross section for the purely
diffractive contribution and a relatively small reduction
of the cross section for the $\phi(1020)$ photoproduction contribution.
Therefore we expect that one could observe the $\phi$ resonance term,
especially when no restrictions on the leading protons are included.
This situation is shown in Fig.~\ref{fig:dsig_dM34_LHC},
see the top left and right panels 
for the ALICE and LHCb experimental conditions, respectively.
However, the final answer can only be given considering 
the experimental mass resolution of a given experiment. 
Here, for $\sqrt{s} = 13$~TeV, we take $\langle S^{2} \rangle = 0.1$
for the purely diffractive contribution
and $\langle S^{2} \rangle = 0.9$ for the photoproduction contribution.
In the bottom panel of Fig.~\ref{fig:dsig_dM34_LHC} we show results 
with extra cuts on the leading protons
of 0.17~GeV~$< |p_{y,1}|, |p_{y,2}|<$~0.5~GeV 
as will be the momentum window for ALFA 
on both sides of the ATLAS detector \cite{Adamczyk_priv}.
Here the $\phi(1020)$ resonance is not so-well visible.
%--------------------------------------------------------
%--------------------------------------------------------
%--------------------------------------------------------
%--------------------------------------------------------

\begin{figure}[!ht]
\includegraphics[width=0.45\textwidth]{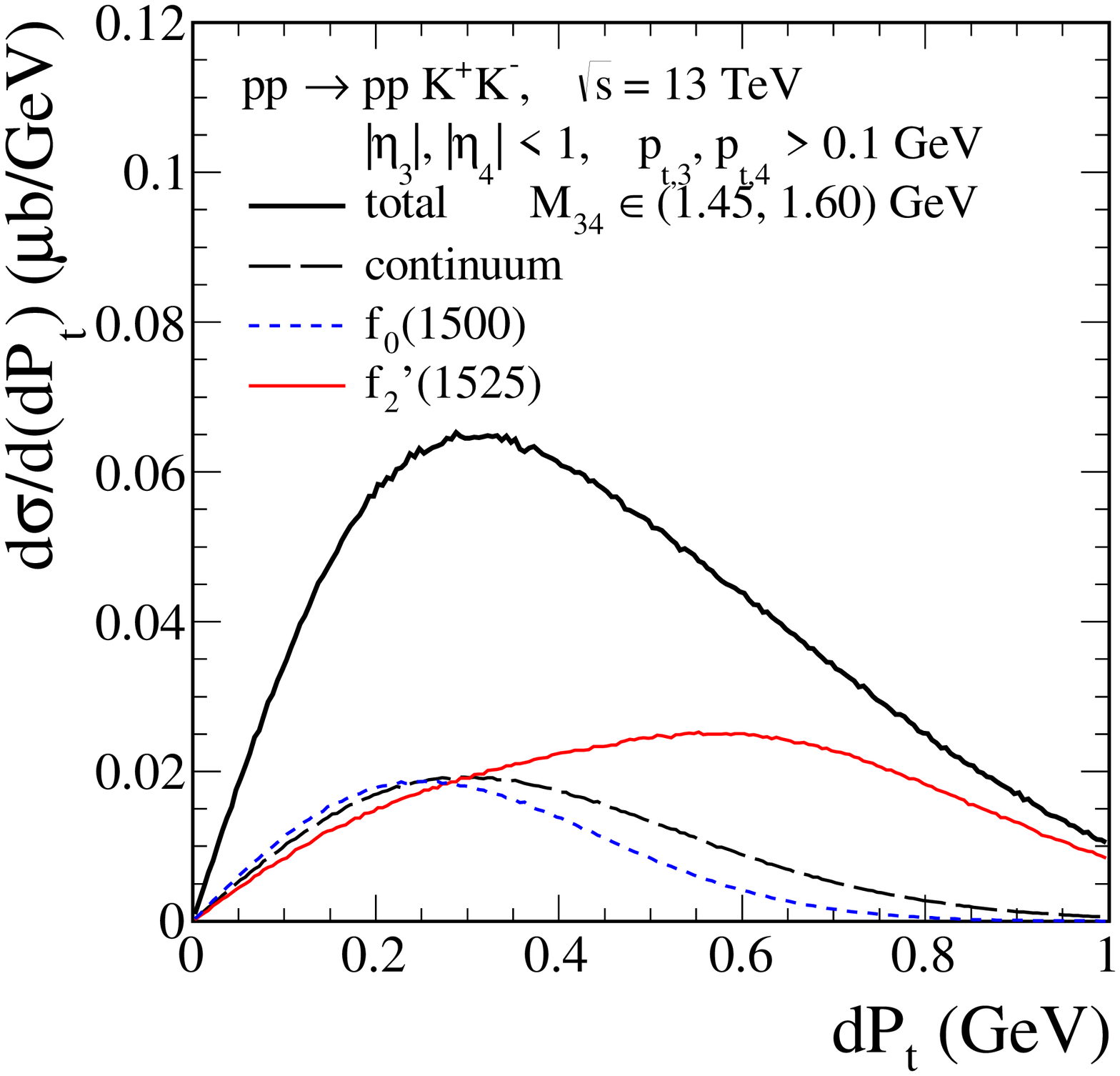}
\includegraphics[width=0.45\textwidth]{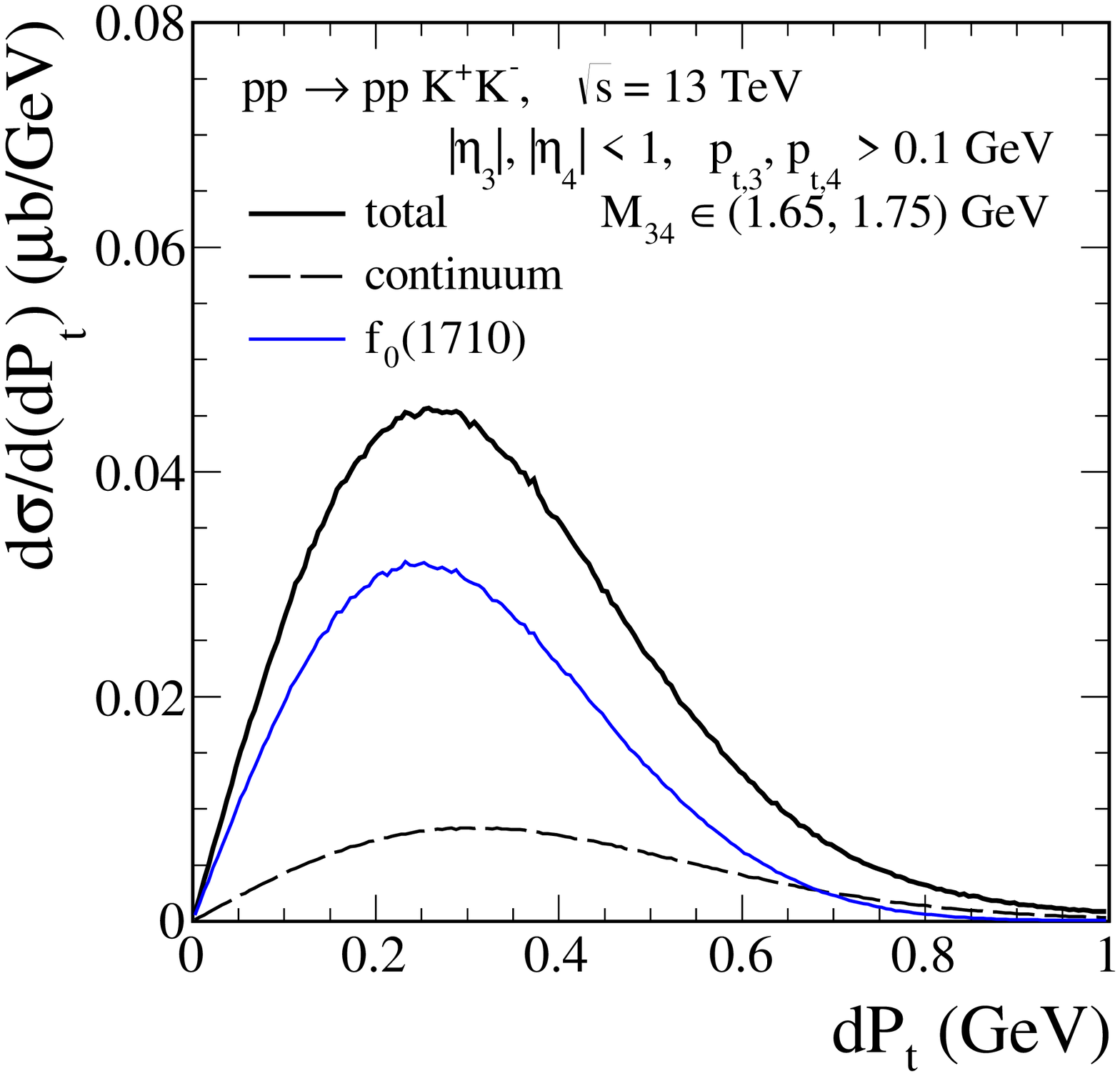}
  \caption{\label{fig:dsig_dptperp}
  \small
The differential cross sections $d\sigma/d(dP_{t})$
as a function of the $dP_{t}$ ``glueball filter'' variable (\ref{dPt_variable})
for the $pp \to pp K^{+} K^{-}$ reaction.
Calculations were done for $\sqrt{s} = 13$~TeV, $|\eta_{K}| < 1$,
$p_{t,K} > 0.1$~GeV, and in two dikaon invariant mass regions,
$M_{34} \in (1.45, 1.60)$~GeV and $M_{34} \in (1.65, 1.75)$~GeV, 
see the left panel and the right panel, respectively.
Both, non-resonant continuum and resonances are included here.
No absorption effects were taken into account here.}
\end{figure}

\begin{figure}[!ht]
\includegraphics[width=0.45\textwidth]{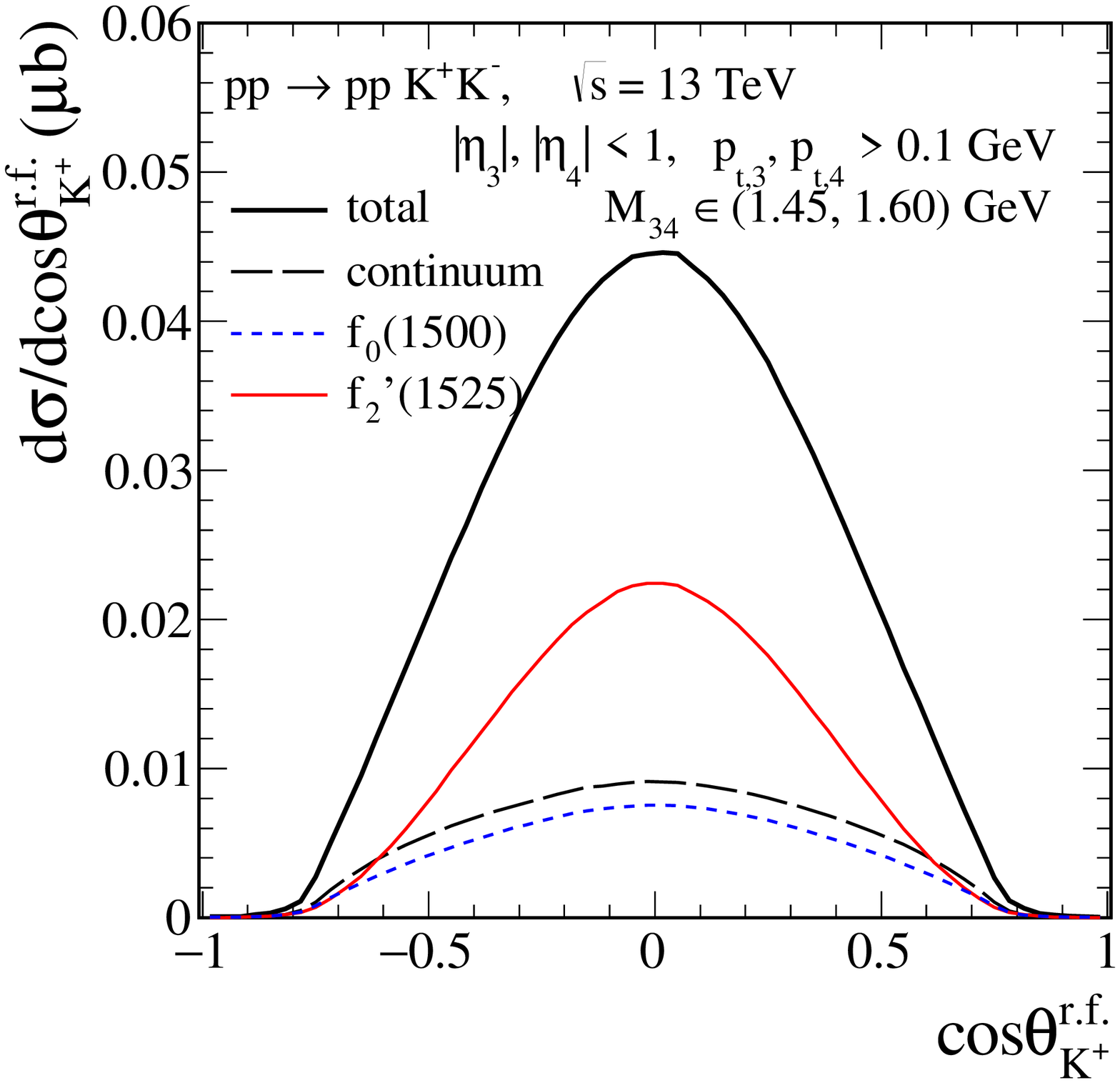}
\includegraphics[width=0.45\textwidth]{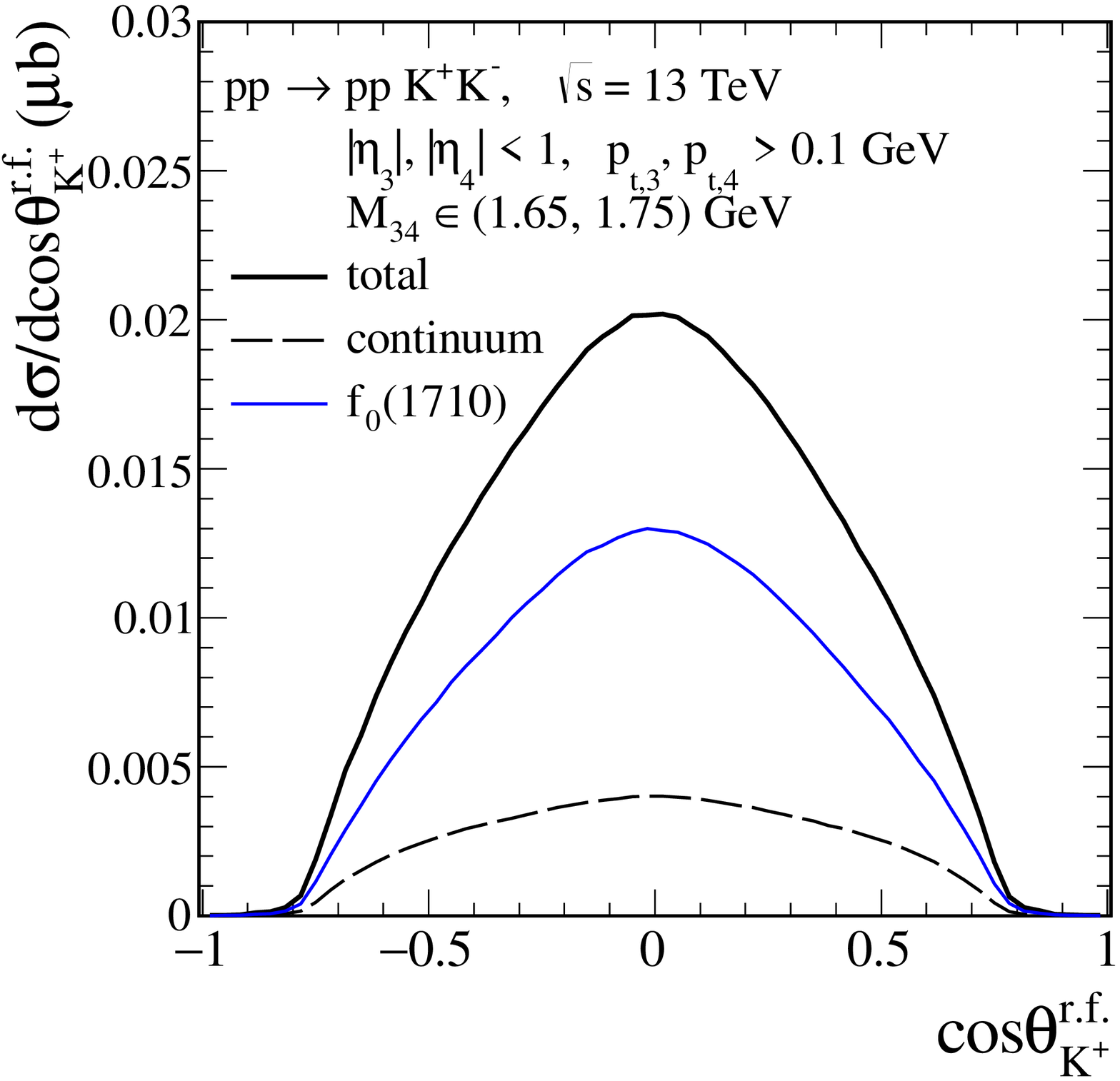}
  \caption{\label{fig:dsig_dz3rf}
  \small
The differential cross sections $d\sigma/dcos\theta_{K^{+}}^{\,r.f.}$
as a function of the cosine of the polar angle $\theta_{K^{+}}^{\,r.f.}$
in the $K^{+} K^{-}$ rest frame for the $pp \to pp K^{+} K^{-}$ reaction
for two different windows of $K^{+} K^{-}$ invariant mass.
The predictions shown correspond to $\sqrt{s} = 13$~TeV
and include cuts $|\eta_{K}| < 1$ and $p_{t,K} > 0.1$~GeV.
The meaning of the lines is the same as in Fig.~\ref{fig:dsig_dptperp}.
No absorption effects were taken into account here.}
\end{figure}

In Figs.~\ref{fig:dsig_dptperp} and \ref{fig:dsig_dz3rf}
we present differential observables for the ALICE kinematics 
($\sqrt{s} = 13$~TeV, $|\eta_{K}| < 1$, $p_{t,K} > 0.1$~GeV)
and for two regions:
$M_{34} \in (1.45, 1.60)$~GeV (the left panels) and 
$M_{34} \in (1.65, 1.75)$~GeV (the right panels).
Fig.~\ref{fig:dsig_dptperp} shows the distributions
in the ``glueball filter'' variable $dP_{t}$; see (\ref{dPt_variable}).
We see that the maximum for the $q \bar{q}$ state $f'_{2}(1525)$
is around of $dP_{t} = 0.6$~GeV. On the other hand,
for the scalar glueball candidates $f_{0}(1500)$ and $f_{0}(1710)$
the maximum is around $dP_{t} = 0.25$~GeV, that is,
at a lower value than for the $f'_{2}(1525)$.
This is in accord with the discussion in section~\ref{sec:diff_f2}
and in Ref.~\cite{Barberis:1996iq}.
%--------------------------------------------------------
%--------------------------------------------------------

Angular distributions in the dimeson rest frame
are often used to study the properties of dimeson resonances.
Fig.~\ref{fig:dsig_dz3rf} shows the distribution
of the cosine of $\theta_{K^{+}}^{\,r.f.}$, 
the polar angle of the $K^{+}$ meson with respect to the beam axis,
in the $K^{+} K^{-}$ rest frame.
\footnote{It is also possible to consider a related observable, 
defined with respect to the pomeron/reggeon exchange three-vector; 
see e.g. \cite{Barberis:1999am,Austregesilo:2013vky}.}
%--------------------------------------------------------
%--------------------------------------------------------

It should be emphasized that our predictions were done 
with our choice of parameters collected in Table~\ref{table:parameters}.
From the partial wave analysis, performed 
by the WA76/102 collaborations \cite{French:1999tm,Barberis:1999am},
more amount of the $S$-wave than of the $D$-wave 
in the mass region around 1.5~GeV was observed.
This observation was confirmed by the E690 experiment \cite{Reyes:1997ei} 
at the Fermilab Tevatron at $\sqrt{s} = 40$~GeV 
in the $pp \to p_{\mathrm{slow}} \,K_{S}^{0}K_{S}^{0} \,p_{\mathrm{fast}}$ reaction.
This would change the behavior of invariant mass distribution
around $M_{KK} = 1.5$~GeV.
Note that the relative phase between
$K^{+}K^{-}$-continuum and $f_{0}(1500)$ amplitudes
is not so well determined and a constructive interference
cannot be excluded.
Here we do not show explicitly the corresponding result.

%--------------------------------------
\section{Conclusions}
\label{sec:conclusions}
%--------------------------------------

We have discussed central exclusive production (CEP) of $K^{+}K^{-}$ pairs 
in proton-proton collisions at high energies. 
We have taken into account purely diffractive and diffractive photoproduction mechanisms.
For the purely diffractive mechanism we have included
the continuum and the dominant scalar $f_{0}(980)$, $f_{0}(1500)$, $f_{0}(1710)$
and tensor $f_{2}(1270)$, $f'_{2}(1525)$ resonances decaying
into $K^{+} K^{-}$ pairs.
The amplitudes have been calculated using Feynman rules within
the tensor-pomeron model \cite{Ewerz:2013kda}.
The effective Lagrangians and the vertices for $\Pom \Pom$ fusion 
into the scalar and tensor mesons were discussed in \cite{Lebiedowicz:2013ika}
and \cite{Lebiedowicz:2016ioh}, respectively.
The model parameters of the pomeron-pomeron-meson couplings
have been roughly adjusted to recent CDF data \cite{Aaltonen:2015uva}
and then used for predictions for the STAR, ALICE, CMS and LHCb experiments.
For the photoproduction of $K^{+} K^{-}$ pairs
we have discussed the dominant $\phi(1020)$ meson contribution 
and the non-resonant (Drell-S\"oding) contribution.
Similar mechanisms were discussed in \cite{Lebiedowicz:2014bea} 
for the $\pi^{+}\pi^{-}$ photoproduction.
The coupling parameters of the tensor pomeron to the $\phi$ meson
have been fixed based on the HERA experimental data for 
the $\gamma p \to \phi p$ reaction \cite{Derrick:1996af,Breitweg:1999jy}.

In the present study we have focused mainly on 
the invariant mass distributions of centrally produced $K^{+}K^{-}$.
In Fig.~\ref{fig:dsig_dM34} we also presented, for comparison,
the purely diffractive contribution previously developed 
in \cite{Lebiedowicz:2016ioh} for the central production of $\pi^{+}\pi^{-}$ pairs.
The pattern of visible structures in the invariant mass distributions
is related to the scalar and tensor isoscalar mesons
and it depends on experimental kinematics.
One can expect, with our default choice of parameters,
that the scalar $f_{0}(980)$, $f_{0}(1500)$, $f_{0}(1710)$
and the tensor $f_{2}(1270)$, $f'_{2}(1525)$ mesons
will be easily identified experimentally in CEP.

The $\phi$-photoproduction and purely diffractive 
contributions have different dependences on the proton transverse momenta.
Furthermore, the absorptive corrections for the $K^{+}K^{-}$ photoproduction processes lead
to a much smaller reduction of the cross section than for the diffractive ones.
It can therefore be expected that the $\phi$-photoproduction 
will be seen in experiments requiring only a very small deflection angle
for at least one of the outgoing protons.
However, we must keep in mind that other processes can contribute in experimental studies 
of exclusive $\phi$ production where only large rapidity gaps 
around the centrally produced $\phi$ meson are checked 
and the forward and backward going protons are not detected.
Recently, experimental results for this kind of processes have been published 
by the CDF \cite{Aaltonen:2015uva} and CMS \cite{Khachatryan:2017xsi} collaborations. 
We refer the reader to Ref.~\cite{Lebiedowicz:2016ryp} in which
$\rho^{0}$ production in $pp$ collisions was studied with one proton
undergoing diffractive excitation to an $n \pi^{+}$ or $p \pi^{0}$ system. 

In addition, we have presented distributions in the so-called
glueball filter variable, $dP_{t}$ (\ref{dPt_variable}), which shows different behavior
in the $K^{+}K^{-}$ invariant mass windows 
around glueball candidates with masses $\sim 1.5$~GeV and $\sim 1.7$~GeV
than in other regions.
Also examples of angular distributions in the $K^{+}K^{-}$ rest frame were shown.
The $dP_{t}$ distribution may help to interpret the relative rates
between the $f_{0}(1500)$ and $f'_{2}(1525)$ resonances
and to resolve the controversial discussion about
the existence of the supernumerous resonances
in the scalar sector \cite{Ochs:2013gi}.

Finally we note that central exclusive $\phi$ production
in $pp$ collisions offers the possibility to search for
effects of the elusive odderon, as was pointed out in \cite{Schafer:1991na}.
The odderon was introduced on theoretical grounds in \cite{Lukaszuk:1973nt,Joynson:1975az}.
For a review of odderon physics see e.g. \cite{Ewerz:2003xi}.
The experimental status of the odderon is still unclear even if there
seems to be some evidence for it
from the recent TOTEM result \cite{Antchev:2017yns}.
For recent discussions of possible odderon effects
in $pp$ elastic scattering at LHC energies see \cite{Martynov:2017zjz,Khoze:2017swe,Khoze:2018bus,Broilo:2018els}.
Using the methods and results of the present paper
it would be straightforward to include also $\phi$ production
by odderon-pomeron fusion and to discuss odderon effects,
e.g. in $K^{+}$ - $K^{-}$ distributions,
in a way analogous to the program presented in \cite{Bolz:2014mya}.
But this is beyond the scope of our present paper.

To summarize: we have given a consistent treatment of 
central exclusive $K^{+}K^{-}$ continuum and resonance production
in an effective field-theoretic approach.
Our studies could help in understanding the production mechanisms
of some light resonances and their properties
in the $pp \to ppK^{+} K^{-}$ reaction.
A rich structure has emerged which will give experimentalists
interesting challenges to check and explore it.

%--------------------
\acknowledgments
%--------------------
The authors are grateful to Leszek Adamczyk and Carlo Ewerz for discussions.
This work was partially supported by
the Polish National Science Centre Grants No. 2014/15/B/ST2/02528
and No. 2015/17/D/ST2/03530
and by the Center for Innovation and Transfer of Natural Sciences 
and Engineering Knowledge in Rzesz\'ow, Republic of Poland/PL.

%------------------------------------------------------------------
{
\begin{small}
\bibliography{refs}
\end{small}
}
%------------------------------------------------------------------

\end{document}